\documentclass[english,12pt,a4paper,pdftex]{article}

\usepackage{aaltothesis}
\usepackage{color}
\definecolor{Blue}{rgb}{0.3,0.3,0.2}
\usepackage[pdftex]{graphicx} 
\usepackage{amssymb}
\usepackage{pdfpages}
\usepackage[toc,page]{appendix}
\usepackage[printonlyused]{acronym}

\usepackage[pdfpagemode=None,colorlinks=true,urlcolor=black,linkcolor=black,citecolor=black,pdfstartview=FitH,backref=page]{hyperref}

\usepackage{url}

\usepackage{algorithm}
\usepackage{algcompatible}

\usepackage{amsfonts,amssymb,amsbsy,amsmath}
\numberwithin{equation}{section}

\usepackage{varwidth}
\usepackage{booktabs}
\usepackage{caption}
\usepackage{color}   
\usepackage{hyperref}
\usepackage{cite}
\hypersetup{
    colorlinks=true, 
    linktoc=all,     
    linkcolor=blue,  
   citecolor=red,
   pdftitle={Repeated Games for Inter-operator Spectrum Sharing},
   pdfauthor={Bikramjit Singh},
   pdfsubject={Radio Communications},
   pdfkeywords={game theory, inter-operator spectrum sharing, noncooperative game, cooperative spectrum sharing, carrier price, surplus}
}

\hypersetup{pdfpagemode=UseOutlines}

\makeatletter
\AtBeginDocument{\renewcommand*{\AC@hyperlink}[2]{#2}}
\makeatother

\renewcommand*{\backref}[1]{}
\renewcommand*{\backrefalt}[4]{
\ifcase #1 (Not cited.)
\or        (Cited on page~#2.)
\else     (Cited on pages~#2.)
\fi}

\usepackage{pifont}

\usepackage{epstopdf}

\usepackage[skip=0cm,list=true,labelfont=it]{subcaption}

\usepackage{chngcntr}
\counterwithin{figure}{section}
\counterwithin{table}{section}

\usepackage{cases}

\usepackage{empheq}

\usepackage{xcolor}

\begin{document}


\university{aalto university}{aalto-yliopisto}
\school{School of Electrical Engineering}{teknillinen korkeakoulu}
\department{School of Electrical Engineering\\
Department of Communications and Networking Technologies}\\
\professorship{Radio Communications}\\
\code{S-72}
\univdegree{MSc}

\author{Singh, Bikramjit}
\thesistitle{Repeated Games For Inter-operator Spectrum Sharing}

\place{Espoo,}
\date{June 16, 2014}
\supervisor{Prof.\ Olav Tirkkonen}{Prof.\ Olav Tirkkonen}
\instructor{D.Sc.\ (Tech.) Konstantinos}{D.Sc.\ (Tech.) Konstantinos Koufos}
\instructor{Koufos}{D.Sc.\ (Tech.) Konstantinos Koufos}
\uselogo{blue}{!}{elec}

\makecoverpage

\newpage



\begin{abstractpage}[english]
\noindent
As wireless communication becomes an ever-more evolving and pervasive part of the existing world, system capacity and \ac{QoS} provisioning are becoming more critically evident. In order to improve system capacity and \ac{QoS}, it is mandatory that we pay closer attention to operational bandwidth efficiency issues. We address this issue for two operators' spectrum sharing in the same geographical area.\\ 

\noindent
We model and analyze interactions between the competitive operators coexisting in the same frequency band as a strategic noncooperative game, where the operators simultaneously share the spectrum dynamically as per their relative requirement. If resources are allocated in a conventional way (static orthogonal allocation), spectrum utilization becomes inefficient when there is load asymmetry between the operators and low inter-operator interference.\\

\noindent
Theoretically, operators can share resources in a cooperative manner, but pragmatically they are reluctant to reveal their network information to competitors. By using game theory, we design a distributed implementation, in which self-interested operators play strategies and contend for the spectrum resources in a noncooperative manner. We have proposed two game theoretic approaches in the thesis, one using a virtual carrier price; and the other based on a mutual history of favors. The former approach takes into account a penalty proportional to spectrum usage in its utility function, whereas in the latter, operators play strategies based on their history of interactions, i.e., how well the other behaved in the past. Finally, based on the simulations, we assess the performance of the proposed game theoretic approaches in comparison to existing conventional allocations.

\end{abstractpage}

\mysection{Acknowledgement}

\noindent
This work was carried out at the Department of Communications and Networking at Aalto University as a part of METIS project, which is partly funded by European Union.\\

\noindent
I would like to express my sincere gratitude to my supervisor Prof. Olav Tirkkonen, for giving me the opportunity to work on this research project, and his guidance and insightful comments during the work. Special thanks are also extended to my instructor D.Sc.\ (Tech.) Konstantinos Koufos for the inspiring discussions and guidance while writing the thesis.\\

\noindent
Finally, I would like to thank my mother who has given me a lifetime of loving support and encouragement.
\\[12\baselineskip]
Espoo, June 16, 2014\\ \\
Bikramjit Singh

\newpage
\phantomsection


\addcontentsline{toc}{section}{Contents}
{\hypersetup{linkcolor=black}
\tableofcontents
}
\clearpage
\phantomsection


\addcontentsline{toc}{section}{List of Acronyms}
\section*{List of Acronyms}
\begin{acronym}

\acrodef{3GPP}[3GPP]{3\textsuperscript{rd} Generation Partnership Project}
\acrodef{BER}[BER]{bit error rate}
\acro{BIM}[BIM]{binary interference matrix}
\acro{BS}[BS]{base station}
\acro{CDF}[CDF]{cumulative distributive function}
\acro{CR}[CR]{cognitive radio}
\acro{CRN}[CRN]{cognitive radio network}
\acro{CSI}[CSI]{channel state information}
\acro{DFS}[DFS]{dynamic frequency selection}
\acro{DRRM}[DRRM]{distributed radio resource management}
\acro{DSA}[DSA]{dynamic spectrum allocation}
\acro{FSA}[FSA]{fixed spectrum allocation}
\acrodef{IMT}[IMT]{International Mobile Telecommunications}
\acrodef{ISM}[ISM]{Industrial, Scientific and Medical}
\acrodef{ITU}[ITU]{International Telecommunication Union}
\acro{LTE}[LTE]{Long Term Evolution}
\acro{MAC}[MAC]{medium access control}
\acro{MMF}[MMF]{max-min fair}
\acro{NE}[NE]{Nash equillibrium}
\acrodef{NGN}[NGN]{Next-generation network}
\acro{OFDMA}[OFDMA]{Orthogonal Frequency Division Multiple Access}
\acro{PCC}[PCC]{primary component carrier}
\acro{PF}[PF]{proportional fair}
\acro{PHY}[PHY]{physical layer}
\acro{PU}[PU]{primary user}
\acro{RAN}[RAN]{radio access network}
\acro{RAT}[RAT]{radio access technology}
\acro{SCC}[SCC]{secondary component carrier}
\acro{SINR}[SINR]{signal-to-interference plus noise ratio}
\acro{SPNE}[SPNE]{subgame perfect Nash equillibrium}
\acro{NBS}[NBS]{Nash bargaining solution}
\acro{SNR}[SNR]{signal-to-noise ratio}
\acro{SU}[SU]{secondary user}
\acro{QoS}[QoS]{Quality of Service}
\acrodef{UDN}[UDN]{ultra-dense network}
\acro{UE}[UE]{user equipment}
\acro{WRAN}[WRAN]{wireless regional area network}
\acro{WLAN}[WLAN]{wireless local area network}

\acrodef{4G}[4G]{4\textsuperscript{th} generation}
\acrodef{IMT}[IMT]{International Mobile Telecommunications}

\acro{SPAI}[SPAI]{Sparse Inverse}
\acrodef{CN}[CN]{Core Network}
\acrodef{GSM}[GSM]{Global System for Mobile}
\acrodef{WCDMA}[WCDMA]{Wideband Code Division Multiple Access}
\acrodef{MSC}[MSC]{Mobile Switching Center}
\acrodef{RNC}[RNC]{Radio Access Controller}
\acrodef{1G}[1G]{1st Generation}
\acrodef{2G}[2G]{2nd generation}
\acrodef{3G}[3G]{3rd generation}
\acrodef{GPRS}[GPRS]{\ac{GSM} First Evolved to General Packet Radio Service}
\acrodef{EDGE}[EDGE]{Enhanced Data Rates for \ac{GSM} Evolution}
\acrodef{IS-95}[IS-95]{Interim Standard 95}
\acrodef{1xRTT}[1xRTT]{1 Times Radio Transmission Technology}
\acrodef{CDMA2000}[CDMA2000]{Code Division Multiple Access 2000 }
\acrodef{IMT-2000}[ITU-2000]{International Mobile Telecommunications-2000}
\acro{WiMAX}[WiMAX]{Worldwide Interoperability for Microwave Access}
\acro{HSPA}[HSPA]{High Speed Packet Access}
\acro{IEEE}[IEEE]{Institute of Electrical and Electronics Engineers}

\acro{RN}[RN]{Relay Nodes}

\acrodef{CDMA}[CDMA]{Code Division Multiple Access}
\acrodef{FDD}[FDD]{Frequency Division Duplex}
\acrodef{TDD}[TDD]{Time Division Duplex}
\acrodef{OFDMA}[OFDMA]{Orthogonal Frequency Division Multiple Access}

\acro{MSE}[MMSE]{Minimising Mean Square Error}
\acro{SLNR}[SLNR]{Signal to Leakage Plus Noise Ration}
\acrodef{ISD}[ISD]{Inter-site Distance}
\end{acronym}

\clearpage
\phantomsection


\addcontentsline{toc}{section}{List of Figures}

{\hypersetup{linkcolor=black}
\listoffigures
}

\clearpage
\phantomsection


\addcontentsline{toc}{section}{List of Tables}

{\hypersetup{linkcolor=black}
\listoftables
}

\cleardoublepage
\storeinipagenumber
\pagenumbering{arabic}

\acresetall
\setcounter{page}{1}


\section{Introduction}


\subsection{Motivation}

\noindent
Radio spectrum is defined as part of the electromagnetic spectrum with frequencies ranging from 3 Hz to 300 GHz. It is used for various wireless communication tasks - data communications, voice communications, video communications, broadcast messaging, command and control communications, emergency response communications, etc. In the past decade, wireless communication services have seen an unprecedented exponential growth~\cite{WP11NSN} and they are expected to grow tremendously in the future as well~\cite{WP13Huawei}. The studies in~\cite{WP11NSN,WP11Ericsson} projected a 1000 times more traffic, and 50 billion connected devices in mobile networks by 2020. According to a study in~\cite{WP13Ericsson}, the development of 4G systems based on \ac{3GPP} \ac{LTE} \ac{RAT} is progressing on a large scale with 55 million users in November 2012 and nearly 1.6 billion users in 2018.\\

\begin{figure}[b]
\centering
\includegraphics[scale=1, trim = 0mm 4mm 0mm 6mm, clip]{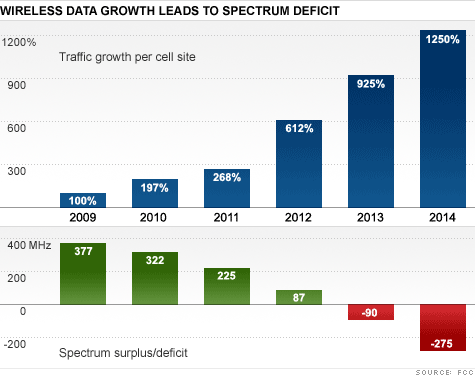}
\caption{U.S. spectrum surplus/deficit situation with growing traffic per cell site~\cite{RP12FCC}}
\label{fig:DataGrowth}
\end{figure}

\begin{figure}[b]
\centering
\includegraphics[scale=0.35, trim = 0mm 0mm 0mm 0mm, clip]{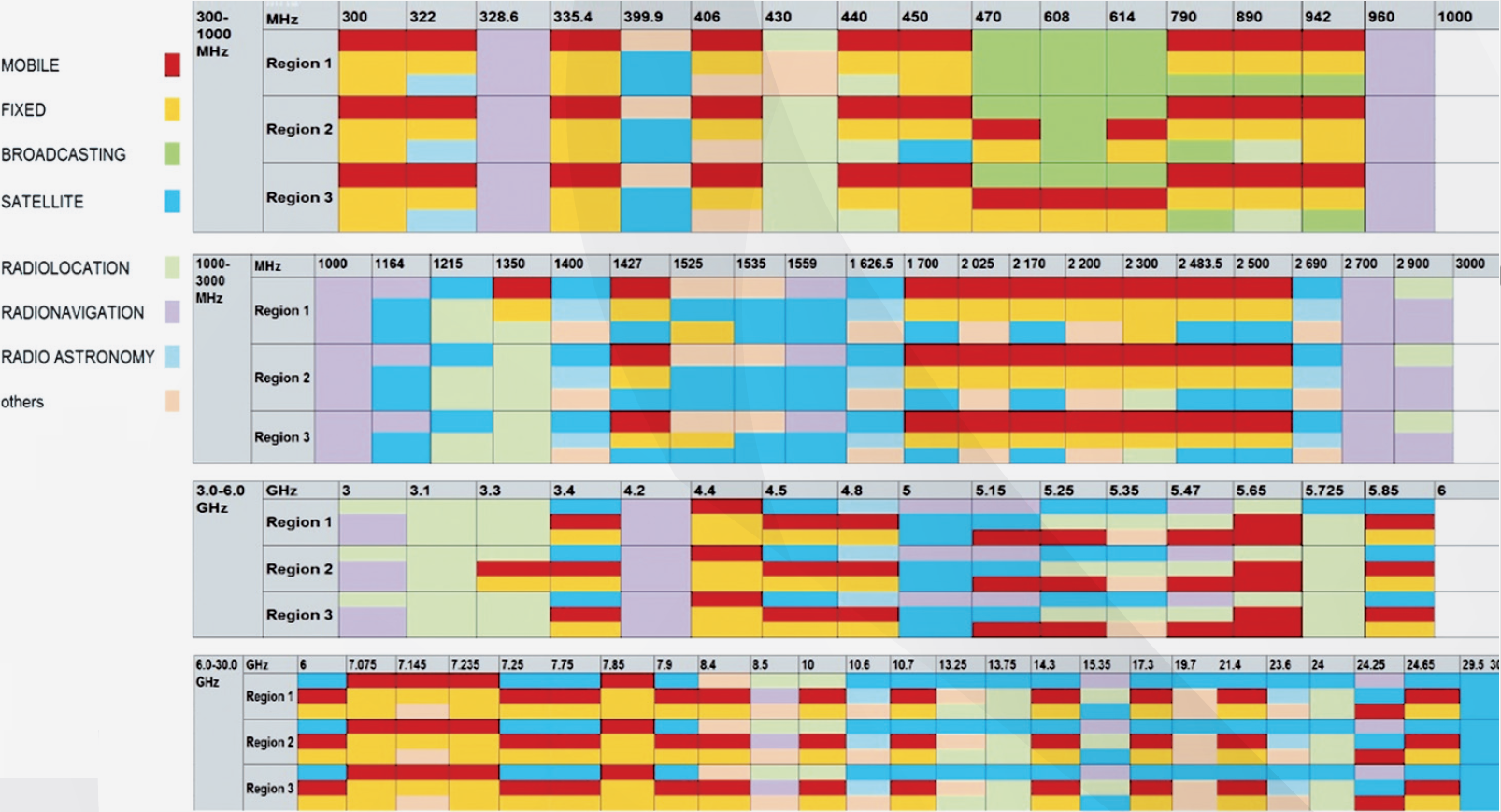}
\caption{Summary of frequency allocation from 0.3 GHz to 30 GHz~\cite{RP11NTIA}}
\label{fig:IMTFreq}
\end{figure}

\noindent
One possible solution to meet the ever-increasing demand is to allocate more spectrum for mobile services, e.g., through spectrum farming. In~\cite{RP12FCC}, quoted in Fig.~\ref{fig:DataGrowth}, it has been shown that by 2014, the mobile traffic per cell site in U.S., will double that of 2012, causing an estimated spectrum deficit of 275 MHz from the surplus of 87 MHz in 2012. Another study by the \ac{ITU} (report ITU-R M.2078~\cite{RP06ITU}) estimates that total spectrum bandwidth requirements for \ac{IMT} services will be up to 1720 MHz by 2020. It will be a challenge to identify such amounts of spectrum and to allocate it exclusively for mobile services.\\

\noindent
Spectrum may be allocated using one of the following authorizations - dedicated, co-primary and unlicensed. Consider the frequency allocation of main \ac{IMT} bands (0.3-30 GHz) by the ITU, shown in Fig.~\ref{fig:IMTFreq}. It is interesting to observe that in co-primary authorization usually more than one service share the same spectrum, e.g., the frequency band of 3.4-4.2 GHz is allocated for both satellite and fixed services. In dedicated authorization, spectrum is exclusively allocated for a single service, e.g., in European region the frequency band of 470-790 MHz is only reserved for broadcasting at the moment. Finally, in unlicensed authorization, unlicensed multi-radio coexists in \ac{ISM} band 2.4 GHz and 5 GHz in which devices like Bluetooth, Wi-Fi, etc. works.\\

\noindent
Though the spectrum map in Fig.~\ref{fig:IMTFreq} looks crowded, it is important to remark that it does not indicate the actual spectrum in use. Based on spectrum usage activity, spectrum utilization in a given geographical area can be regarded as fully utilized, underutilized (sporadically used) or fully unused. The unused or sporadically used spectrum in space and/or time could exist due to many reasons, e.g., the system is idle, or intermittent activity (spectrum holes), or signals are unable to reach the receiver due to heavy losses. One major cause of this underutilization is the static (fixed) allocation of spectrum to the various systems. If a system with a static frequency allocation is not using its assigned spectrum, the resources are wasted. If other systems could utilize the vacant spectrum, spectrum utilization could be improved.\\

\noindent
Various benchmark studies and measurement campaigns have pointed out that a large portion of the allocated spectrum is not actively used in space and time. The FCC Spectrum Policy Task Force in 2002 in their report~\cite{RP02FCC} have reported vast temporal and geographic variations in the allocated spectrum utilization ranging from 15\% to 85\%. \v{C}abri\'{c} \textit{et al.} in~\cite{CP04Cabric} have shown measurements taken in an urban setting revealing a typical utilization of 0.5\% in the 3-4 GHz frequency band, further dropping to 0.3\% in the 4-5 GHz frequency band. In survey~\cite{CP10Valenta} conducted in 2010 globally, it is found that in densely populated areas, less than 20\% of spectrum bands below 3 GHz are used during a working day and the occupation is even lower in rural areas.\\

\noindent
The incongruence between \textquotedblleft spectrum allocation\textquotedblright~and \textquotedblleft spectrum utilization\textquotedblright~suggests that \textquotedblleft spectrum allocation\textquotedblright~is a more significant problem than an actual physical scarcity of spectrum. The fixed spectrum allocation generally worked well in the past because of limited traffic. Nowadays, the pressing demands for more wireless services and the inefficient spectrum utilization necessitate a new communication paradigm to use the existing spectrum opportunistically and more efficiently. Opportunistic use is not necessarily limited to different services but can also be within the same service. For example, multiple operators can share the spectrum resources opportunistically. One promising case could be that operators operating in a shopping mall can use full spectrum resources by localizing themselves to the respective floors instead of a whole shopping mall area and render mobile services on a co-primary basis with negligible inter-operator interference.


\subsection{Overview of Thesis Problem}

\noindent
\acp{NGN} will have higher bandwidth requirements so that they can meet demands of end user capacities and \ac{QoS}. Nowadays, operators are largely following \ac{FSA}. Such, static assignments are disadvantageous because they are time and space invariant, and prevent devices from efficiently utilizing allocated spectrum, resulting in spectrum holes (no devices in the area) and poor utilization~\cite{RP03McHenry}.\\

\noindent
Let us consider multiple \acp{RAN} owned by different operators providing wireless services within and around the small area they control, e.g., offices, restaurants, etc. in a marketplace. Within the same geographical area, there exist different classes of users, as well as different companies/business units, and may have different peak usage times. With orthogonal assignments, the spectrum is underutilized when load conditions of neighbouring operators are subjected to temporal variations. In that scenario, a low load  operator could transfer some of its spectrum resources to a high load operator by using \ac{DSA} and can help it, e.g., in reducing the blocking probability and in avoiding high latency. \ac{DSA} can help operators to adapt to varying channel state conditions and radio frequency environments. If the inter-\ac{RAN} interference is severe, operators tend to share the spectrum with a high degree of orthogonality; if the interference is negligible, operators tend to have a high degree of overlapped carriers (full spread).\\

\noindent
With DSA operators become able to share the spectrum resources as per their relative needs and exercise better performance in their access area. For this, a protocol that coordinates the interaction between multiple operators is needed to achieve improved spectral efficiency by allowing flexible and efficient spectrum use. This is explored in this thesis.


\subsection{Thesis Contribution}

\noindent
In this Thesis, an efficient \ac{DSA} scheme is proposed to improve the operational bandwidth efficiency in a multi-operator scenario. Multiple operators coexist in the same geographical area causing interference to each other. It is assumed that operators' \acp{RAN} have a connection between them. However, the cooperation between operators is on low level. They are unwilling to share their network and operational information due to mutual competition. Also they may send false information to get more advantage from other operators. Operators are thus considered as self-interested entities and will be contending for spectrum resources noncooperatively. By noncooperation, we mean that no operational information is shared amongst the operators. Hence, there is neither need for tight synchronizations (extra overhead) nor new interfaces.\\

\noindent
Game theory provides tools that offer significant insight into the dynamics of noncooperation. It is a promising approach for studying mathematical models of conflict and cooperation between rational decision makers~\cite{BK03Osborne}. It has been recently applied in telecommunication field and has been established as an important tool for modelling interactions and \ac{DSA} techniques for evolving technologies like \ac{CR} or inter-operator spectrum sharing.\\

\noindent
The studied spectrum allocation problem is related to the frequency assignment problem~\cite{CP12Peltomaki}, where a carrier is either used or not. Following the carrier selection approach, two algorithms are developed for dynamic spectrum sharing based on \textit{noncooperative repeated games}. In this, operators adopt an interactive mode of communication and agree upon formulating a policy on how to share carriers amongst them. Due to the fact that operators coexist in the same geographical area for a long time, they interact and build response sequences through a trust game. Interaction is modelled in terms of spectrum usage favors being asked or received by them. The favors are referred to utilization of shared frequency carriers.


\subsection{Thesis Organization}

\noindent
The remainder of this thesis is organized as follows. Chapter \ref{chap:Background} briefly reviews the utility criterion for resource allocation. It also discusses game theory and its models.\\

\noindent
In Chapter \ref{chap:Works}, the related work pertaining to inter-operator spectrum sharing are presented. Besides that, standards closely related to spectrum sharing are also discussed.\\

\noindent
In Chapter \ref{chap:Coop}, inter-operator cooperation has been discussed, and its advantages and challenges are presented. The system model used for the cooperative schemes is reviewed, and the implementation is analyzed mathematically.\\

\noindent
In Chapter \ref{chap:GamePrice}, the proposed DSA scheme based on noncooperative repeated games and virtual carrier pricing is explained. The system model and the utility functions are described alongside its optimization criteria. Finally, an algorithm for distributed dynamic spectrum sharing among the operators is explained.\\

\noindent
In Chapter \ref{chap:GameExpectation}, another distributed noncooperative game theoretic scheme is proposed using mutual history of gains/losses incurred between the participating operators. The system model, utility functions and algorithm are explained. Detailed mathematical analysis is presented to corroborate the algorithm.\\

\noindent
In Chapter \ref{chap:Simulation}, simulation results are presented and analyzed. The simulated scenario, simulation parameters, user distributions, and channel models are explained. The benefits of the proposed \ac{DSA} schemes are then assessed, comparing to static allocation schemes such as orthogonal, full spectrum allocations and a cooperative scheme. Finally, in the last chapter, conclusions are drawn and future work is suggested.


%
%
%

\clearpage



\section{Background} \label{chap:Background}

\noindent
In this chapter, inter-operator spectrum sharing is considered with the aid of the literature discussions. Inter-operator spectrum sharing opens opportunities for the operators to enhance the system level performance. To describe the operator specific performance, we consider performance metric in the form of utility functions. To describe interactions between the operators, we use the theory of games. So, this chapter provides an overview of utility functions and game theory background.\\


\subsection{Utility Criterion for Resource Allocation}

\noindent
Utility-based approaches have recently been widely adopted to quantify the radio resource allocation problems in wireless communication. \textit{Utility function} represents the system's performance level or \ac{QoS}~\cite{JR97Kelly}. The composition of a utility function is strictly non-decreasing (monotonic) and concave function of system parameters. The function describing the system-wide utility and welfare functions studied in economic sciences bears the same characteristics.\\

\begin{figure}[h]
\centering
\includegraphics[scale=1, trim = 0mm 0mm 0mm 0mm, clip]{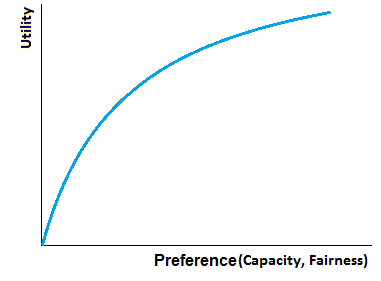}
\caption{Example of utility function behaviour}
\label{UtilityBehavior}
\end{figure}

\noindent
Though utility functions have been used to model various performance parameters such as data traffic (Shannon capacity)~\cite{JR03Sung}, bandwidth allocation~\cite{JR08Wu,CP02Cao}, multiuser diversity~\cite{CH05Navaie}, scheduling/delay-tolerant traffic~\cite{CP01Gao,CP04Liu}, \ac{SNR}/\ac{SINR} improvement, bandwidth pricing applications~\cite{CP02Siris,CP02Marbach,CP02Liu}, fairness~\cite{JR00Bianchi,JR01Liao}, \ac{BER}, energy efficiency~\cite{JR02Saraydar}, sigmoid-like function of \ac{SINR}~\cite{JR07Huang,JR03Xiao} etc. But in the thesis, we focus on the study consisting of fairness in system capacity based utility which can be best described in Fig.~\ref{UtilityBehavior}.

\noindent
Assume a load of $\mathcal{L}=\left\{ 1,2,3,...,n \right\}$ users in the wireless network. The resources are quantified based on user preferences such as \ac{SINR}, or throughput, or allocated bandwidth etc. The associated utility function can be expressed as ${U}\left( {{x}_{i}}\left( t \right) \right)$, where ${{x}_{i}}\left( t \right)$ denotes the user specific quantity, such as experienced user rate of the $i$-th user at time $t$ and $U$ is a function that describes the user satisfaction level, given a quantity. The utility function ${U}\left( {{x}_{i}}\left( t \right) \right)$ is an increasing and strictly concave function representing the decreasing additional benefits with increasing resource allocation.\\

\noindent
From user's satisfaction perspective, networks are interested in optimizing a resource allocation $r$, e.g., carrier allocations, within the resource constraint set $\mathcal{R}$, which maximizes the long term expected aggregated utility,

\begin{equation} \label{eq:UtilityProblemSolution}
\underset{r\in \mathcal{R}}{\mathop{\max }}\,\underset{\mathcal{T}\to \infty }{\mathop{\lim }}\,\frac{1}{\mathcal{T}}{\text{E}_{r}}\left\{ \int\limits_{0}^{\mathcal{T}} \sum\limits_{i\in n} {U} \left( {{x}_{i}}\left( t \right) \right) dt \right\}.
\end{equation}

\noindent
The solution to Eq.~\eqref{eq:UtilityProblemSolution} is called the socially optimal solution. If the \ac{RAN} and the available resources (e.g., a fixed maximum transmit power constraint at the \ac{BS}, available bandwidth etc.) are static, and the user's \ac{QoS} measurements are independent of time (e.g., a user's \ac{SINR}), then Eq.~\eqref{eq:UtilityProblemSolution} can be written as

\begin{equation*}
\underset{r\in \mathcal{R}}{\mathop{\max }}\,\sum\limits_{i\in n}{{{U}}\left( {{x}_{i}} \right)}
\end{equation*}

\noindent
without time $t$.\\

\noindent
A solution that maximizes the sum-throughput utility of all the players might not be practicable, as some of the players might consider it \textit{unfair} in the sense that such a solution is achieved at the expense of some players. In many environments \textit{fairness} might be more important than optimality. Various definitions of fair allocations have been proposed, such as weighted fair~\cite{BK97Keshav}, \ac{MMF}~\cite{JR06HuangACM}, \ac{PF} allocations~\cite{JR97Kelly, JR98Kelly}, etc. Based on various fairness criteria, the utility function can be written as

\begin{subnumcases}{U(x_i) =} \label{eq:fair}
 \frac{w_i}{1-\alpha }x_{i}^{1-\alpha}, &$\text{weighted } \alpha\text{-fairness, }$\label{eq:weighted} \\
&$\alpha > 0,  \text{weights }  (w_i) \geq 0$ \notag \\
{{x}_{i}},&$\text{Max}$\label{eq:max}\\
\lim_{\alpha \to \infty} \frac{1}{1-\alpha }x_{i}^{1-\alpha},&$\text{MMF}$\label{eq:mmf}\\
\text{log}\left( {{x}_{i}} \right),&$\text{PF.}\label{eq:pf}$
\end{subnumcases}

\noindent
The weighted $\alpha$-fair allocations are a parameterized family of fairness criteria. In Eq.~\eqref{eq:weighted}, if $w_i = 1$ and $\alpha \rightarrow 0$ then $\alpha$-fair regains the max-throughput optimization. The max or greedy fairness criterion maximizes the network throughput. The disadvantage of such an allocation is that users with poor channel conditions is starved of resources, which seems somewhat unfair. It would seem fairer, for all users simultaneously have some access to the network s resources. If max-throughput is unfair then, perhaps, \ac{MMF} is the most fair. Amongst all rate allocations, the minimum rate allocated to any flow is maximized over all possible rate allocations, and eventually leading to equal rates for all users. In Eq.~\eqref{eq:weighted}, if $w_i = 1$ and $\alpha \rightarrow \infty$ then the weighted $\alpha$-fair allocations reduces to \ac{MMF} allocations. \ac{PF} is a compromise-based scheduling algorithm. It is based upon maintaining a balance between two competing interests, trying to maximize network throughput while at the same time allowing all users at least a minimal level of service. In Eq.~\eqref{eq:weighted}, for $\alpha = 1$, the weighted $\alpha$-fair objective is not defined, but $\lim_{\alpha \to 1}$ reduces to \ac{PF} allocation. It is important to remark that achieving a fair allocation and achieving a socially optimal allocation do not always conflict with each other, and sometimes both objectives can be achieved by choosing the appropriate utility functions (e.g., ~\cite{JR98Kelly,JR00Mo}).\\

\noindent
With utility-based framework, a network can be modelled using a single function and the network resource allocation problems can be studied in a noncomplex way. The performance of different allocation schemes can be easily compared, e.g., how far they are from the socially optimal solution, or the upper limit of resource usage. It also aids in examining the trade-off between social optimality, and other performance objectives.


\subsection{Introduction to Game Theory}

\noindent
Game theory~\cite{BK03Osborne} is concerned with predicting the outcome of \textit{games of strategies}. Expressed succinctly, game theory is the formal study of decision-making where it analyzes or models the interactions between interdependent decision-making entities that have mutual and possibly conflicting objectives.\\

\noindent
Developed since the first half of the 20\textsuperscript{th} century, it has been used primarily in economics as it is to describing animal behaviour and model competition between companies, and is central to the understandings of various other fields, such as political sciences, psychology, logic and biology. In recent years, telecommunications is one of the new fields that has evolved an emerging interest towards game theory as a tool to analyze conflicts among players, e.g., congestion control, routing, power control, topology control, trust management, dynamic spectrum sharing, etc. The importance of modelling interaction via game-theoretic approach is multifold -

\begin{itemize}

\item Offers a wide range of optimality criteria (e.g., in simultaneous, multistage games),
\item Optimizes problems where no centralized control is present (noncooperative games),
\item Players devise strategies independently and intelligently, and give a power to make decisions locally (noncooperative one-shot games, noncooperative repeated games). With good strategic mechanism players can enforce others to cooperate in noncooperative environment (noncooperative repeated games).

\end{itemize}


\subsubsection{Game Definition}

\noindent
A game is typically formalized as a triple of a set of players, a set of allowable strategies for each player, and a utility function. Utility function represents a player's evaluation of consequences in a game. Players play strategies with the intention to maximize their utilities. Normally the strategies are conflicting, i.e., increasing own utility happens at the expense of other's decreasing utility. So, the players have to be rational while playing strategies as too much greedy approach can harm themselves because of repercussions and too much of trustworthiness can let their exploitations by greedy opponents.\\

\noindent
Representing mathematically, game $\mathcal{G}$,

\begin{equation*}
\mathcal{G}=\left\langle \mathcal{P},\mathcal{S},\left. \mathcal{U} \right\rangle  \right.,
\end{equation*}

\noindent
where

\begin{itemize}
\item $\mathcal{P}$ is a finite set of players, s.t., $\mathcal{P}=\{1,2,3,...,m\}$,
\item $\mathcal{S}$ is an m-tuple of pure strategy sets, one for each player, s.t., $\mathcal{S}=\{{{s}_{1}},{{s}_{2}},{{s}_{3}},...\\,{{s}_{m}}\}$, where ${{s}_{i}}$ is the strategy profile of $i$-th player, s.t., ${{s}_{i}}\in {{S}_{i}}$ and ${{S}_{i}}$ is finite number of allowable strategy set of $i$-th player, ${{S}_{i}}=\{1,...,{{q}_{i}}\}$,
\item $\mathcal{U}$ is the utility function, whose intended interpretation is the award given to a single player at the outcome of the game, s.t., $\mathcal{U}:{{S}_{1}}\times {{S}_{2}}\times ...\times {{S}_{m}}\to \mathbb{R}$.
\end{itemize}


\subsubsection{Game Strategies Type}

\noindent
With different type of strategies, players can obtain various game resolutions - different equilibriums, optimal/suboptimal solution, etc. In this section, we outline the possible strategies describing their behaviour and possible outcome on the game $\mathcal{G}$.

\begin{itemize}

\item Mixed (Randomized) Strategy

\noindent
A mixed strategy for player $i$, with ${{S}_{i}}=\{1,...,{{q}_{i}}\}$ is a probability distribution over ${{S}_{i}}$. In other words, $p_{i}:S_{i} \to [0,1]$, where we have $p_{i}(s_{i}) \ge 0$ for all $s_{i} \in S_{i}$ and $\sum\limits_{{{s}_{i}}\in {{S}_{i}}}{{{p}_{i}}\left( {{s}_{i}} \right)}=1$, i.e.,

\begin{equation*}
{{p}_{i}}\left( 1 \right)+{{p}_{i}}\left( 2 \right)+...+{{p}_{i}}\left( {{q}_{i}} \right)=1.
\end{equation*}

\noindent
We interpret $p_{i}(s_{i})$ as the probability with which player $i$ chooses startegy $s_{i}$.

\item Pure Strategy

\noindent
If in the mixed strategy, the probability associated to ${{s}_{i}}={{s}_{i}}\left( j \right)$ for some $j$ is 1, i.e., $p_{i}(s_{i}(j)) = 1$, where $1 \le s_{i}(j) \le q_{i}$, while for others is 0, then it is called pure strategy.

\item Strictly Dominant Strategy

\noindent
A strategy $s_{i}^{*}\in {{S}_{i}}$ is a strictly dominant strategy to a given startegy $s_{i}^{'}\in {{S}_{i}}$ for player $i$ if $\forall {{s}_{-i}}\in {{S}_{-i}}$, we have,

\begin{equation*}
{{U}_{i}}\left( s_{i}^{*},{{s}_{-i}} \right)>{{U}_{i}}\left( s_{i}^{'},{{s}_{-i}} \right).
\end{equation*}

\noindent
In this case, we say that $s_{i}^{*}$ strictly dominates $s_{i}^{'}$.

\item Weakly Dominant Strategy

\noindent
For any player $i$, a strategy $s_{i}^{*}\in {{S}_{i}}$ weakly dominates another strategy $s_{i}^{'}\in {{S}_{i}}$ if $\forall s_{-i}\in {{S}_{-i}}$,

\begin{equation*}
{{U}_{i}}\left( s_{i}^{*},{{s}_{-i}} \right)\ge {{U}_{i}}\left( s_{i}^{'},{{s}_{-i}} \right).
\end{equation*}

\item Maxmin Strategy 

\noindent
Player $i$ plays strategy ${{s}_{i}}\in {{S}_{i}}$ to the ${{s}_{-i}}\in {{S}_{-i}}$ in order maximize its minimum utility,

\begin{equation*}
\underset{{{s}_{i}}}{\mathop{\max }}\,\underset{{{s}_{-i}}}{\mathop{\min }}\,{{U}_{i}}\left( {{s}_{i}},{{s}_{-i}} \right).
\end{equation*}

\item Best Response

\noindent
A strategy $s_{i}^{*}\in {{S}_{i}}$ is a best response for player $i$ to ${{s}_{-i}}\in {{S}_{-i}}$ if $\forall {{s}_{i}}\in {{S}_{i}}$,

\begin{equation*}
{{U}_{i}}\left( s_{i}^{*},{{s}_{-i}} \right)\ge {{U}_{i}}\left( {{s}_{i}},{{s}_{-i}} \right).
\end{equation*}

\noindent
Note - best response is different from dominant strategy in a way that best response improves utility for a specific strategy ${{s}_{-i}}\in {{S}_{-i}}$ and $\forall {{s}_{i}}\in {{S}_{i}}$, whereas dominant strategy improves utility to a given strategy $s_{i}^{'}\in {{S}_{i}}$ and $\forall {{s}_{-i}}\in {{S}_{-i}}$.

\item Mixed Nash Equilibrium (mixed NE)

\noindent
For a strategic game $\mathcal{G}$, a strategy profile ${{s}^{*}}=\left( s_{1}^{*},s_{2}^{*},s_{3}^{*},...,s_{m}^{*} \right)\in \mathcal{S}$ is a mixed \acs{NE} if for every player $i$, $s_{i}^{*}$ is a best response to $s_{-i}^{*}\in {{S}_{-i}}$. In other words, for every player $i=1,...,m$ and for every mixed strategy ${{s}_{i}}\in {{S}_{i}}$,

\begin{equation} \label{eq:mixedNE}
{{U}_{i}}\left( s_{i}^{*},s_{-i}^{*} \right)\ge {{U}_{i}}\left( {{s}_{i}},s_{-i}^{*} \right).
\end{equation}

\noindent
In other words, no player can improve its own utility by unilaterally deviating from the mixed strategy profile ${{s}^{*}}=\left( s_{1}^{*},s_{2}^{*},s_{3}^{*},...,s_{m}^{*} \right)$.

\item Pure Nash Equilibrium (Pure NE)

\noindent
Strategy profile ${{s}^{*}}$ satisfying Eq.~\eqref{eq:mixedNE} in addition is called a pure \acs{NE} if every $s_{i}^{*}$ is a pure strategy $s_{i}^{*}=s_{i}^{*}\left( j \right)$, for some $j\in {{S}_{i}}$.

\item Subgame Perfect Nash Equillibrium (\acs{SPNE})

\noindent
A strategy profile $s$ is a \acs{SPNE} if it represents a \acs{NE} of every subgame of the original game $\mathcal{G}$. A subgame is a subset of any game that includes an initial node (which has to be independent from any information set) and all its successor nodes.

\item Pareto Optimal

\noindent
A game $\mathcal{G}$ strategy profile $s=\left( {{s}_{1}},{{s}_{2}},{{s}_{3}},...,{{s}_{m}} \right)$ is said to be Pareto optimal if we cannot find another strategy profile $s$ in which it is impossible to make any one player better off without making at least one player worse off. Essentially, it is often treated as a weak efficient solution for the optimization problems beacuse a socially optimal solution is Pareto optimal, but the vice versa is not always true. For example, if each ${{U}_{i}}\left( {{x}_{i}}\left( t \right) \right)$ is monotonic and strictly concave in ${{x}_{i}}\left( t \right)$, and resource constraint set $\mathcal{W}=\{x|\sum\nolimits_{i\in m}{{{x}_{i}}}\le X\}$, where ${{x}_{i}}\left( t \right)$ denotes the \ac{QoS} measurements of the $i$-th user at time $t$, then any resource allocation $w\in \mathcal{W}$ achieves $\sum\nolimits_{i\in m}{{{x}_{i}}}=X$ is Pareto optimal, but there is only one socially optimal solution.

\item Nash Bargaining Solution (\acs{NBS})

\noindent
A pair of utilities $\left( U_{i}^{*},U_{-i}^{*} \right)$ is a \acs{NBS} if it solves the following optimization problem,

\begin{equation*}
\begin{gathered}
\underset{{{U}_{i}},{{U}_{-i}}}{\mathop{\max }}\,\left( {{U}_{i}}-{{d}_{i}} \right)\left( {{U}_{-i}}-{{d}_{-i}} \right) \\ 
\text{subject to }\left( {{U}_{i}},{{U}_{-i}} \right)\in \mathcal{U} \\ 
\left( {{U}_{i}},{{U}_{-i}} \right)\ge \left( {{d}_{i}},{{d}_{-i}} \right), \\ 
\end{gathered}
\end{equation*}

\noindent
where ${{d}_{i}}$ and ${{d}_{-i}}$, are the status quo utilities (i.e., the utility are not meant for bargain with the other player). The \acs{NBS} should satisfy certain axioms:

\begin{itemize}
\item [-] Invariant to affine transformations or Invariant to equivalent utility representations
\item [-] Pareto optimality
\item [-] Independence of irrelevant alternatives
\item [-] Symmetry
\end{itemize}

\item Stackelberg Equilibrium

\noindent
The Stackelberg model can be solved to find the \acs{SPNE}. Assume there are two players, player $i$ act as a leader and player $-i$ its follower. To find the \acs{SPNE} of the game we need to use backward induction, as in any sequential game. Starting from the end (2\textsuperscript{nd} stage), player $-i$ (follower) makes reactive choices depending on the actions of player $i$,

\begin{equation*}
s_{-i}^{f}\left( {{s}_{i}} \right)=\arg \underset{{{s}_{-i}}}{\mathop{\max }}\,{{U}_{-i}}\left( {{s}_{-i}},{{s}_{i}} \right).
\end{equation*}

\noindent
In the 1\textsuperscript{st} stage, player $i$ (leader) always anticipate its rival behaviour initially, makes its strategic choices accordingly,

\begin{equation*}
s_{i}^{l}=\arg \underset{{{s}_{i}}}{\mathop{\max }}\,{{U}_{i}}\left( {{s}_{i}},s_{-i}^{f}\left( {{s}_{i}} \right) \right).
\end{equation*}

\noindent
So, the Stackelberg equilibrium or \acs{SPNE} strategies are $\left( s_{i}^{l},s_{-i}^{f} \right)$.

\end{itemize}


\subsubsection{Game Model Classification}

\noindent
In the game formulation, players act rationally according to their strategies with an objective to maximize their outcome. However, the strategic profile of players is highly influenced by the regulation imposed by the nature of the environment of the game. This led to the proliferation of varieties of the game with possible taxonomies are:\\

\begin{itemize}

\item One-shot vs. Repeated Game

\noindent
One-shot game is also known as non-repeated or single stage game. It is played only once; therefore stakes are high but carries no further repercussions. Here, players may be uninformed about the moves made by other players and might act selfishly to get away with the highest payoff. If a game is not played once, but numerous times then the game is called a repeated game. It allows for a strategy to be contingent on past moves, and have reputation effects and retribution for it. Repeated game is further classified as finitely and infinitely repeated game, but widely studied repeated game is an infinitely repeated game. According to the Folk Theorem, for an infinite repeated game there exists a discount factor $\widehat{\delta }<1$ such that any feasible and individually rational payoff can arise as an equilibrium payoff for any discount factor $\delta \in \left( \widehat{\delta },1 \right)$. Thus, future payoffs are discounted and are less valuable. This is because the consumption in the future is considered less valuable than the present due to time preference (e.g., money). Therefore, the player's total payoff in a repeated game is a discounted sum of each stage payoff. The repeated game holds a variety of equilibrium properties because the threat of retaliation is real due to the repetitive nature of the game and also it has a much bigger strategy space than the one-shot game. Unlike one-shot game, here players can punish hostile players using tit-for-tat strategy~\cite{JR81Axelrod,BK07Axelrod}. A repeated game with perfect monitoring where players' actions are observable is called a multistage game. In this, players announce their strategy publicly and thus, each stage of a multistage game resembles a single stage game.

\item Cooperative vs. Noncooperative Game

\noindent
A cooperative game is defined where group of players enforces cooperative behaviour  on each other. In this, players bargain or negotiate on payoffs and form joint strategies. Cooperative game is pragmatically undesirable because of excessive overhead signalling and trust issues, but it provides a unique Pareto optimal solution for the problem modelling. In the latter, if the competition is between potentially conflicting and self-interested players, then the corresponding game is known as noncooperative game. In a noncooperative game, without centralized control, the players do not cooperate or make deals so that any cooperation among them must be self-enforcing.

\end{itemize}


\subsubsection{Prisoner's Dilemma Example}

\begin{figure}[h]
\centering
\includegraphics[scale=1, trim = 0mm 0mm 0mm 0mm, clip]{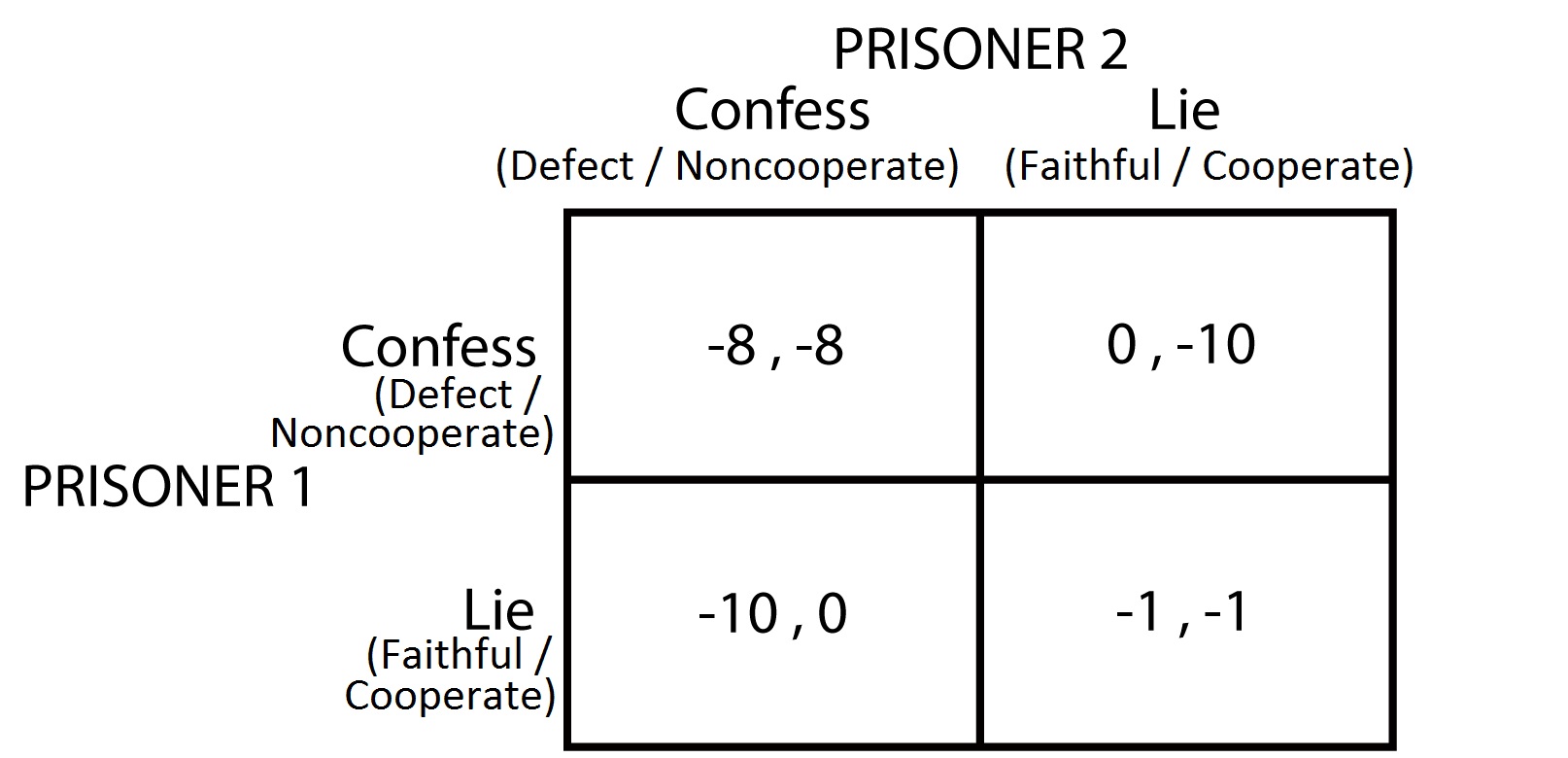}
\caption{Prisoner's dilemma game}
\label{PrisonersDilemma}
\end{figure}

\noindent
The prisoner's dilemma is probably the most widely used game for pedagogical purposes in game theory. Nicknamed in 1950 by Albert W. Tucker, prisoner's dilemma describes a situation where two prisoners are taken into custody in connection with a burglary. However, the authorities possess insufficient evidence to convict them for their crime, only to convict them on the charge of possession of stolen goods. This game is summarized in Fig.~\ref{PrisonersDilemma}.
 
\paragraph{Description}

\noindent\\\\
If prisoners help each other by not confessing the crime, they will both be charged with the lesser prison term of a year each. The authorities will question them in separate interrogation rooms, which mean that the prisoners take decisions simultaneously and they do not know about each other's decision. Thus, the process is noncooperative with imperfect information. The authorities will try to convince each prisoner to confess their crime by offering him an escape clause and to his accomplice a prison term of ten years. If both prisoners defect and confess their crime, they shall be sentenced to eight years. Both prisoners have a common knowledge of the same offered deal along with its consequences, but completely unaware of each other's choices. In this, the prison term can be seen as the respective utilities (negative of prison terms) for each set of choices, and each would like to have largest utility for himself.

\paragraph{Prisoner's Choices}

\noindent\\\\
As the game carries no further repercussions, prisoners are tempted to get away with largest profit (least prison term) and consequently, will defect. Therefore, \textquotedblleft to confess\textquotedblright~is the dominant strategy, thus \textit{(Noncooperation, Noncooperation)} is the \acs{NE} in this one-shot game.\\

\noindent
Let us assume, the two burglars work together over a long time, and repeatedly end up being interrogated by the police, and punished by prison terms. After each round (stage) of interrogation, and based on their choices, they get to know what others did. Again, prisoners would like to have maximum sum-utility (minimal prison terms) for their repeated criminal acts. So, thinking rationally, it makes sense to cooperate over all the stages, to have maximum profit in a repeated sense of game (smaller prison term of one year is given every time for their crimes). Let us say, prisoner $P1$ wants to defect, and no doubt he can get away with zero prison term in that stage, whereas prisoner $P2$ will be awarded ten years of prison term. Beside, prisoner $P2$ will get aware of the last hostile choice made by prisoner $P1$, and later he will defect too, in order to punish prisoner $P1$ for his noncooperation. If they continue with their noncooperative behaviour, then both will always get eight years of prison term for every subsequent crime. Therefore, in repeated games, \textquotedblleft to lie\textquotedblright~is the dominant strategy, thus \textit{(Cooperation, Cooperation)} is the \acs{NE} as it maximizes the profit by getting aware of others' strategies and punishing them with tit-for-tat strategy~\cite{JR81Axelrod,BK07Axelrod} if somebody does not cooperate.

\clearpage


\section{Research Work and Standards for Spectrum Sharing} \label{chap:Works}


\subsection{Related Work} \label{sec:ResWorks}

\noindent
Traditional channel allocation algorithms aim to improve carrier usage for a single system (e.g., cellular systems, femtocells deployments, etc.). In essence, these algorithms realize inter-operator spectrum sharing needs provided that operators with neighbouring \acp{RAN} are willing to work together. The spectrum sharing algorithms can be classified on the knowledge of the domain where inter-operator interference is handled, i.e., frequency, time and/or spatial domain. Further, there are two extremes in regard to the cooperative arrangement between the operators in distributed inter-operator spectrum sharing depending on whether the operators cooperate with each other or not. On one hand, the operators may behave in a totally selfish manner and respond to the opponents by using the sequence of the best responses. On the other hand, the operators may be honest and fully cooperative.

\paragraph{Cooperative Games}

\noindent\\\\
In~\cite{JR09Garcia}, autonomous component carrier selection scheme is proposed, a distributed solution to the cross-tier interference management for HetNet. The \acp{BS} exchange \ac{BIM} entries representing a single value over the entire bandwidth encapsulating the information of outgoing and incoming inter-cell interference reduction. \acp{BS} select the \acp{PCC} based on their coverage (maximum path loss), whereas \acp{SCC} are selected based on \acp{BIM} exchanged. In~\cite{CP11Prasad}, \acp{BS} select carriers if the corresponding capacity gain on its served \acp{UE} is greater than the losses of the neighbouring \acp{BS}. However, the network suffers from many carrier re-selections. In~\cite{CP13Amin}, a dynamic carrier selection scheme is proposed using \ac{BIM} per component carrier instead of full bandwidth (as in~\cite{JR09Garcia}), and avoids carrier re-selections by estimating capacity gains and losses. In~\cite{CP12Ahmed}, the interference condition is communicated in the form of interference prices instead of \acsp{BIM}. An upper bound on the sum-capacity of two operators is identified in~\cite{CP12Anchora} assuming that operators exchange their user-specific channel quality indicators over all shared channels. In~\cite{CP07Niyato}, a repeated game Cournot model for the cooperative spectrum sensing in \ac{CR} is presented in which multiple \acp{SU} share spectrum with a \ac{PU} through a bidding process using a determined spectrum pricing functions set by \ac{PU}.\\

\noindent
In time domain spectrum sharing, the operators could, for instance, trade time slots~\cite{CP06Middleton,CP07Bennis}. Operators with a low load could borrow their time resources to heavily loaded operators helping them to reduce the blocking probability and frame delay~\cite{CP07Bennis}. This scheme improves spectrum utilization efficiency at the cost of high signalling overhead and a good time synchronization requirement among operators. For higher efficiency, one could also allow \acp{UE} connecting to the \textquotedblleft best \ac{BS}\textquotedblright~whether it belongs to their home network or not provided that all operators utilize the same \ac{RAT}~\cite{JR09Bennis}.\\

\noindent
Inter-operator spectrum sharing in the spatial domain has been considered in~\cite{CP10Lindblom,CP11Jorsweik}, modelling each operator by a transmitter-receiver link. In~\cite{CP10Lindblom}, operators use inter-operator interference as a bargaining value to enable cooperation, and compute their beam-forming vectors. With cooperative beam-forming, operators increase their system throughput by lowering the overall inter-operator interference. In~\cite{CP11Jorsweik}, operators exchange their CSI, and utilize cooperative transmit beam-forming to steer their beams towards the desired receiver.\\

\noindent
Besides spectrum utilization efficiency, cooperative algorithms for inter-operator spectrum sharing could jointly maximize a function that incorporates the utility function of each operator. In~\cite{JR06HuangIEEE}, the sum-utility is maximized by properly distributing the available power budget across multiple carriers. Joint power control and scheduling (as an operator may want to favor users based on their channel conditions) are identified in~\cite{CP12Ahmed} for maximizing the sum-\ac{PF} utility. In~\cite{CP13Amin}, multiple cells exchange interference prices, and partition the spectrum so that the sum-utility is maximized. Unfortunately, the cooperation mechanism forces each operator/link to reveal its network-specific information, e.g., how much interference it receives in the form of interference prices, or may be the utilities, \ac{CSI} information, load, etc. The method can also be generalized where operators cooperate with each other for maximizing their sum-operator utility. In practice operators are expected to have some elements of selfish behaviour, and may send malicious information to others (e.g., falsified interference prices, erroneous gains/losses over the component carriers, etc.) in order to get a higher share of available resources. Fully-cooperative inter-operator spectrum sharing schemes cannot be used unless there is a mechanism to identify and punish the non-trustworthy parties.

\paragraph{Noncooperative One-shot Games}

\noindent\\\\
The study in~\cite{JR09Bennis} considers one-shot noncooperative games between multiple operators where each operator is modelled by a single transmitter-receiver link. The operators maximize their sum-rate over multiple carriers in a selfish manner. Since the utility function (sum-rate) is concave, equilibrium exists, and the power allocation vector for each link at an equilibrium point is identified. Under certain conditions, there is a unique equilibrium and noncooperative spectrum sharing game becomes predictable. However, the equilibrium point in one-shot noncooperative games could be inefficient for some if not for all of the players~\cite{JR07Etkin}.\\

\noindent
In the literature~\cite{JR09Bennis,JR07Etkin}, strict ways to apply punishment are considered where resource allocations of one-shot noncooperative games are enforced after a player deviates from cooperation. A common assumption in is that the one-shot games are enforced forever so as no player has the incentive to deviate. This kind of strategy is quite strict because it does not incorporate forgiveness and punishes all players even if a single player deviates. Besides, the existing studies consider the power spectral density as the strategy space and optimize the power allocation over multiple carriers for the different players/links/operators.

\paragraph{Noncooperative Repeated Games}

\noindent\\\\
Operators are expected to share spectrum a long time, and have persistent, known identity. As a result, they can learn from each other's behaviour, build reputations and achieve higher utility in comparison with one-shot selfish strategies. In that case, the interaction between operators would rather be modelled by repeated games among selfish players. Besides utility functions and strategy spaces used in the cooperative game model, repeated games of selfish players require also a punishment mechanism when a player deviates from the specified rules or mechanisms to otherwise, specify reciprocity. In~\cite{JR09Wu}, a repeated game approach is used to improve the inefficient \acs{NE} of a one-shot game where it optimizes to eliminate the incentive for deviation in the finite time interval. However, the cooperation strategy is arbitrarily set, i.e., orthogonal spectrum sharing is used, and the networks are assumed to have complete information about each other's parameters. Due to the fact that operators in advance agree on orthogonal spectrum sharing, the finite punishment period introduces forgiveness and alleviates the demand for perfect detection accuracy.\\

\noindent
Noncooperative repeated games have been considered for scenarios where the players do not have equal rights, e.g., primary-secondary property rights model in spectrum sharing. In~\cite{CP05Ileri}, noncooperative repeated games are considered to model and analyze the competition among multiple \acp{SU} to access the \ac{PU}'s channels. The spectrum leasing process identified as a monopoly market in which the \ac{PU} has the full control over leasing process. The market is secondary driven where \acp{SU} act as relays with their demand functions based on the acceptance probability model for the users. \acs{NE} is considered as the solution for this noncooperative game, where each \ac{SU} tries to maximize its payoff function in a selfish manner. The games are based on the assumption that \acp{SU} are honest and will not cheat. Another game spectrum leasing models considered in~\cite{JR08Niyato,JR09Niyato} consisting of multiple strategic \acp{PU} (as against the one in~\cite{CP05Ileri}) and \acp{SU}. In this, the active trading of \acp{PU}' spare radio resources with the \acp{SU}' stochastic demand is modelled using a Markov chain to describe \acp{SU}' buying opportunities from \acp{PU}. Both models behaviour are very similar in nature with an only exception that~\cite{JR08Niyato} does not account the user's wireless details. With the existence of multiple sellers (\acp{PU}) and buyers (\acp{SU}); both groups' participants try to maximize their payoffs selfishly and the problem broken down to two different problems: the buyers' problem of revenue maximization and spectrum pricing, and the sellers' problem of spectrum access. For such active models,~\cite{JR08Niyato,JR09Niyato} use market-equilibrium-based approaches to understand the behavioural economics  behind it. In~\cite{JR09Sengupta}, a competitive spectrum leasing model is presented where a central mediating entity acts as a spectrum broker and distributes spectrum among different competing service providers through auctioning~\cite{BK02Krishna}.\\

\noindent
In~\cite{JR06Sun}, fair resource allocation is realized in time domain using the game-theoretic approach to enable concerned operators (auctioneers) maximize their respective revenues alongside ensuring the maximization of user's payoffs (throughput), as well. The revenue here is defined as user's quantitative measure of reservation preference (of channel) upon time slots. In this approach, the second price auction mechanism (Vickrey auctions) is employed where users bid for a wireless channel in the auction competing for resources (securing time slots) for their throughput maximization. Users with better channel coefficients have a better chance to secure the bid. The bidding here is not related to the willingness to pay for the resources; it only acts as a tool securing the possession of resources for the user with good channel conditions.\\

\noindent
Most studies in noncooperative games~\cite{CP05Ileri,JR08Niyato,JR09Niyato,JR09Sengupta} are bounded to centralized scenarios or \ac{CR}, where \ac{PU} (or auctioneer) has full control over the information exchange about the spectrum utilization state and negotiation on the spectrum allocation. Secondly, market-driven mechanisms have been explored widely as a promising approach for spectrum sharing where \acp{PU} trade unused spectrum to \acp{SU} dynamically. However, operators favor decentralized resource management and, at the same time, are hesitant in adopting market driven sharing schemes as they may not want to touch their revenue model. Also, auction based spectrum access~\cite{JR09Sengupta,JR06Sun} poses larger overhead constraints, beside it might require the government's nod in its adoption. Therefore, operators are reluctant to engage in any kind of monetary payoffs, or auctioning spectrum, or transferring load as discussed in the majority of the noncooperative games and drive us to need for newer policies for modelling the noncooperative game models for spectrum sharing.

\subsection{Related Standards}

\noindent
Regulators, operators, suppliers, users of radio communication services and radio equipment rely on technical standards to ensure that radio systems perform as designed. From the spectrum use perspective, there are already some closely related standards in existence for which we would like to discuss, and point out how they differ from the inter-operator spectrum sharing needs in the following section.

\subsubsection{802.11 for Intra-cell/Inter-cell Transmission}

\noindent
IEEE 802.11~\cite{STD80211} is a set of \ac{MAC} and \ac{PHY} specifications for implementing \ac{WLAN} communication in the 2.4, 3.6, 5 and 60 GHz frequency bands. To cope with special problems of wireless transmission in intra-cell/inter-cell data transmission, IEEE 802.11 \ac{MAC} carries two different access mechanisms, the mandatory distributed coordination function (DCF) which provides distributed channel access based on CSMA/CA (carrier sense multiple access with collision avoidance) and the optional point coordination function (PCF) which provides centrally controlled channel access through polling.\\

\noindent
The 802.11 protocol can be employed to tackle the \textit{intra-cell}/\textit{inter-cell} transmission problems in the low-density setting.
However, the performance of 802.11 to resolute \textit{inter-operator} transmission problems in an \ac{UDN} scenario is no better than the already existing technologies, e.g., LTE small cell deployments. In~\cite{WP11Qualcomm}, an analysis is presented showing that LTE co-channel picocells offer a better user experience and system capacity improvement than Wi-Fi nodes. In addition, Wi-Fi nodes also lack better support for mobility/handoff, \ac{QoS}, security and self-organized networks. So, there is a need of a set of protocols or policies that cope with wireless transmission issues in the multi-operator \ac{UDN} system.

\subsubsection{802.11h for Spectrum Management in 5 GHz Band}

\noindent
With an advent of the IEEE 802.11 \ac{WLAN} standard, the persistent thrust to open up spectrum for unlicensed use created a need for \ac{DFS}. \ac{DFS} is supported by the novel IEEE 802.11h~\cite{STD80211h} \ac{WLAN} standard which allows 5 GHz capable IEEE 802.11 devices to share spectrum with radar (or satellite) devices without causing interference to the radar operation. The concept of \ac{DFS} is to have the unlicensed IEEE 802.11h device monitors the presence of a radar/satellite on the channel it is using and, if the level of the radar is above a certain threshold, the device vacates the existing channel, and then monitors and selects another channel if no radar was detected.\\

\noindent
This standard is different from the inter-operator spectrum sharing requirements in the sense that \textit{only IEEE 802.11h} device adjusts its spectrum needs in 5GHz band via \ac{DFS} in order to avoid co-channel operation with radar systems. Whereas in the inter-operator spectrum sharing, \textit{multiple players} share the spectrum dynamically, and have equal rights on the same frequency bands at the same time.

\subsubsection{802.16h for Improved Coexisting Mechanism}

\noindent
The task of inter-operator spectrum sharing necessitates the networks for peaceful coexistence in the geographical area and formulates policies on the development. The IEEE 802.16h~\cite{STD80216h} License-Exempt Task Group, a unit of IEEE 802.16 Broadband Wireless Access Standards Committee realizes improved mechanisms for time domain spectrum sharing under the coordinated coexistence mode. It develops standards, e.g., \ac{MAC} enhancements or policies, and recommended practices enabling coexistence between license-exempt systems in wireless MAN. It also focuses in hierarchical sharing spectrum applications of coexisting systems with primary radio systems. The operation is not limited to license-exempt bands, but extending to all bands where 802.16-2004 is applicable.\\

\noindent
For the execution of spectrum sharing policies, a distributed architecture for radio resource management is suggested (IEEE C802.16h-05/004) that enables communication and exchange of parameters between multiple networks formed by one 802.16 \ac{BS} and its associated \acp{UE}. Each \ac{BS} has a \ac{DRRM} entity and build up a database for sharing information related to actual and intended future usage of radio spectrum. 802.16h protocol realizes all necessary functions required for spectrum sharing amongst coexisting systems, e.g., detecting the co-located \ac{RAN} topology, registering to the \ac{DRRM} database, or negotiation for radio spectrum sharing. While interacting with \ac{MAC} or \ac{PHY}, the \ac{DRRM} uses the coexistence protocol to communicate with other \acp{BS} and regional license-exempt databases. In this manner, using the inter-system communication, IEEE 802.16 protocol helps to achieve harmonious sharing of an unlicensed or shared radio spectrum.\\

\noindent
The IEEE 802.16h coexistence protocol works in \textit{time domain}. However, because of the tight requirements for a good time synchronicity between the operators; the coexistence is wished in \textit{frequency domain}, and in that case the standard IEEE 802.16h fails to serve the purpose.

\subsubsection{802.22 for using White Spaces in the TV Frequency Spectrum}

\noindent
The development of the IEEE 802.22~\cite{STD80222} \ac{WRAN} standard aims at opportunistic use of white spaces using \ac{CR} techniques. White spaces refer to geographically unused spectrum made available for use at locations where spectrum is not being used by TV broadcasting services. IEEE 802.22 \acp{WRAN} are designed to operate in the VHF and UHF TV broadcast bands on a non-interfering basis, and to bring broadband access to hard-to-reach low population density areas, for e.g., rural environments up to 100 km from the transmitter. Each \ac{WRAN} will boost up a connection speed up to 22 Mbps per channel with no harmful interference to the existing TV broadcast stations. Therefore, it has timely potential for a worldwide applicability.\\

\noindent
The IEEE 802.22 standard is intended for \textit{centralized} inter-network resource sharing and targets typical centralized scenario that enables \acp{SU} to reuse unused spectrum of \ac{PU}. On the contrary, in non-centralized scenarios without a central co-ordinator, this standard is not well established and push for the need of more versatile inter-operator spectrum sharing standards for \textit{non-centralized} distributed implementations.

\clearpage


\section{Cooperative Spectrum Sharing} \label{chap:Coop}

\begin{figure}[b]
\centering
\includegraphics[scale=.52, trim = 0mm 0mm 0mm 0mm, clip]{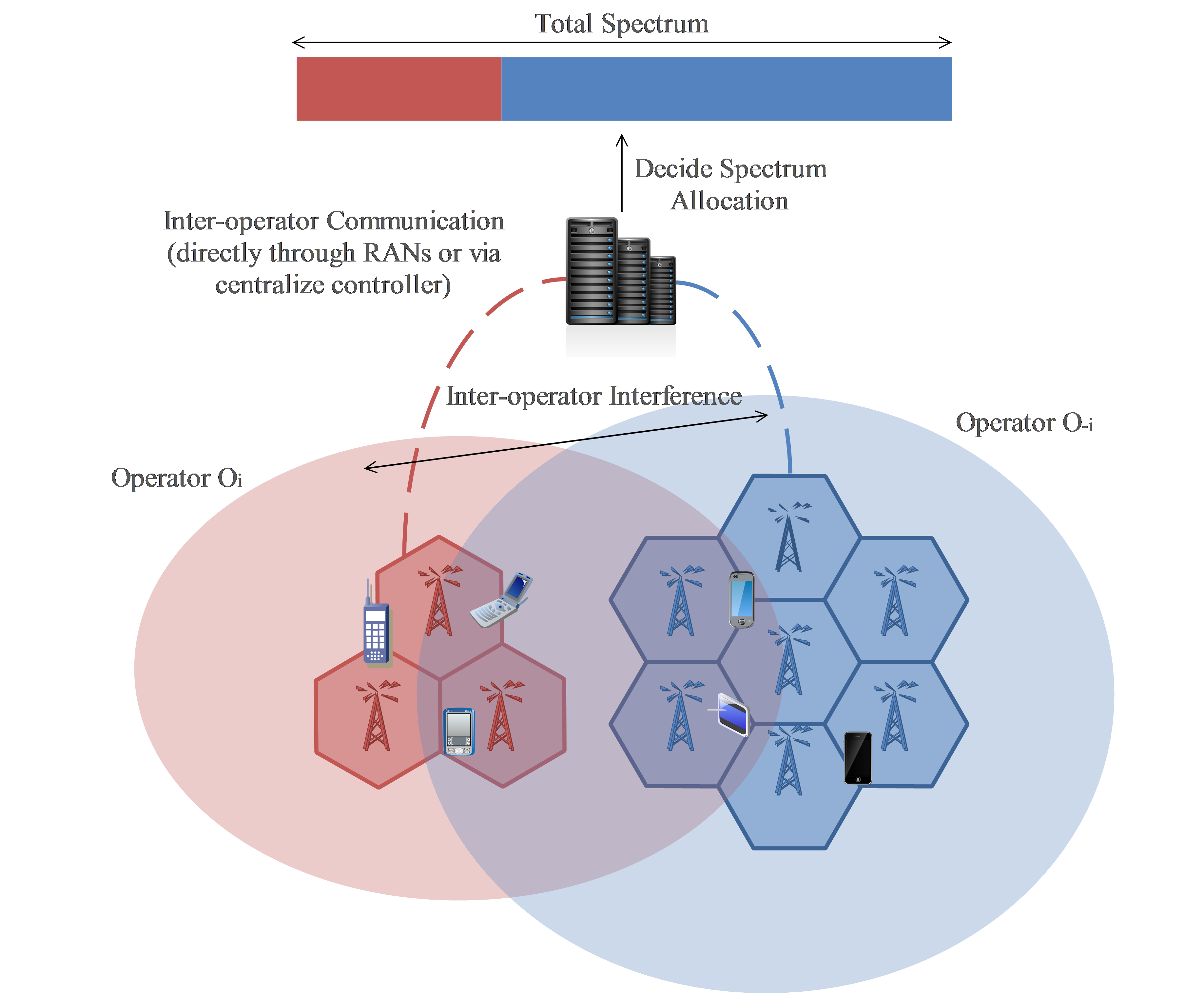}
\caption{Operators share spectrum in a cooperative manner}
\label{fig:Coop}
\end{figure}

\noindent
Cooperative spectrum sharing can effectively improve spectrum efficiency, and thus mitigate the network congestion or the wasteful usage of spectrum resources. Cooperative schemes are largely desirable in the trustworthy networks, and where extra signalling overheads are not an issue. The network resources can be effectively utilized by implementing cooperative schemes in the cellular systems within an operator. In this chapter, we discuss the notion behind cooperative spectrum sharing and investigate it mathematically as it provides us benchmark studies for henceforth discussed noncooperative spectrum sharing techniques for non-trusted and self-interested network operators.\\

\noindent
The \textit{cooperative communications} amongst players can be realized in many ways - in time~\cite{CP06Middleton,CP07Bennis,JR08Simeone,JR08Leshem}, frequency~\cite{CP13Amin,CP12Ahmed,JR09Garcia,CP12Anchora,CP07Niyato,JR06HuangIEEE,JR12Su,JR08Leshem,JR09Suris,JR10Yang}, code~\cite{JR11Liu}, or spatial~\cite{CP11Jorsweik,CP12Zhai,JR04Laneman,JR11Saad,JR08Huang} domain. In~\cite{JR04Laneman}, the network players exploit space diversity, and whenever a player encounters poor channel access conditions, it relays its data to the other player's network which acts as cooperative relays. In~\cite{JR11Saad} the benefit of MIMO communications has been studied by using cooperative players as relay nodes. In~\cite{JR08Leshem,JR11Liu}, cooperative game theory has been used to analyze the interference channel. Problem of opportunistic spectrum access has been addressed in~\cite{JR09Suris} using cooperative game theory. The author showed that the \acs{NBS} achieves the best trade-off between fairness and optimality in spectrum allocation. Distributed power control for \ac{CRN} has been analyzed in~\cite{JR10Yang} based on cooperative game theory. In~\cite{JR08Huang}, a cooperative auction based algorithm is introduced for relay assignment and allocate relay transmit power among bidding users, for which the unique \acs{NE} is achieved distributively via globally updating the best response bid in a completely asynchronous manner\footnote{For additional information on related work, please refer Chapter \ref{chap:Works} Section \ref{sec:ResWorks}.}.\\

\noindent
With a brief review of cooperation algorithms in the field of telecommunication, we henceforth model and investigate the cooperation phenomena in spectrum sharing, in detail. The motivation behind the following investigative study is~\cite{CP13Amin,CP11Prasad}, in which operators exchange interference prices and distribute spectrum resources amongst them by estimating their utility gain/loss and jointly making the decisions so that their sum-utility is maximized.


\subsection{System Model}

\noindent
We consider operators $O_{i}$ and $O_{-i}$ sharing same spectrum, e.g., in a shopping mall scenario. Assume that each operator can construct a number that characterizes the level of service enjoyed by the users served by the operator. Such a number is here called a network utility. It may, e.g., be defined in terms of the distribution of the service provided to users, such as a suitable linear combination of average cell throughput and cell edge throughput. In a cooperative scheme, it is crucial that all operators have the same utility function. For simplicity, we will assume utility functions that are directly formed from the throughputs enjoyed by the user. Then, the operators schedule their users to the available carriers and jointly maximize, e.g., the sum-\ac{PF} rate function with weights reflecting the portion of time a user is multiplexed onto a carrier. An algorithm to approximate the cooperative optimal solution is detailed in~\cite{CP13Amin,CP11Prasad} and it is shortly summarized next. Fig.~\ref{fig:Coop} illustrates the given cooperative scenario.\\

\noindent
Let us assume that each user can measure the interference levels due to transmissions originated from other operator and report them to its serving \ac{BS}. As the operators here are assumed to use the same spectrum, the ability to measure the interference levels originating from the \ac{BS} of another operator would be a straight forward generalization of \ac{LTE} handover measurements. By aggregating such measurements performed by the users, operator $O_{i}$ can form an approximation of the level of interference caused by a \ac{BS} of operator $O_{-i}$, following example the principles outlined in~\cite{JR09Garcia}. The \ac{BS} of an operator, e.g., operator $O_{i}$, asks its users to conduct spectrum measurements over all the carriers utilized by the operator. On receiving the measurement information, operator $O_{i}$ computes for each carrier its utility gain (if the users of operator $O_{-i}$ currently use the carrier, they stop using it) or its utility loss (if the users of operator $O_{-i}$ start using the carrier) and communicates gain/loss to operator $O_{-i}$. Operator $O_{-i}$ in its turn, selects randomly a carrier: (i) if it uses the carrier, it compares its loss (by removing the carrier) to the gain operator $O_{i}$ would achieve (ii) if it does not use the carrier, it compares its gain (by start using the carrier) to the loss operator $O_{i}$ would experience. Operator $O_{-i}$ makes the decision that increases the sum of the utilities of the two operators (a.k.a. cooperative utility). For the new carrier allocation, operator $O_{-i}$ computes its own utility gain/loss, communicates them to operator $O_{i}$, and the interaction continues until the cooperative utility cannot further increase by changing the carrier allocation between the operators. Note that the identification of spectrum allocation and user multiplexing weights across the carriers is a mixed integer programming problem. Approximations to the optimal solution can also be achieved in a centralized manner, but the computational complexity grows quickly for an increasing number of carriers and users. Natural solutions for mixed integer programming problems are iterative, and the protocol depicted above uses a natural distribution of these iterations to the independent decision makers in the problem here, i.e., the operator networks.


\subsection{Cooperative Algorithm}

The discussed algorithm in the form of pseudocode is summarised as below,

\begin{algorithm} [H]
\renewcommand\thealgorithm{}
\caption{Cooperative Spectrum Sharing}
\begin{algorithmic}[1]
\STATE Operator ${O_i}$, ${i \in \mathcal{I}}$ considers carrier ${k \in {K}}$ for the dynamic selection.
\STATE Add unused carrier $k$ by operator $O_{i}$.
\STATEx Calculates utility gain $G_{i,k}$ for added carrier $k$.
\STATEx Compares it with utility loss $L_{-i,k}$ of other operator $O_{-i}$.
\STATEx \hskip\algorithmicindent \textbf{if} {$G_{i,k} > L_{-i,k}$} \textbf{then}
\STATEx \hskip \algorithmicindent \hskip\algorithmicindent \textbf{do} \textit{START} using carrier $k$.
\STATEx \hskip\algorithmicindent \textbf{end if}
\STATE Remove used carrier $k$ by operator $O_{i}$.
\STATEx Calculates utility loss $L_{i,k}$ for removed carrier $k$.
\STATEx Compares it with utility gain $G_{-i,k}$ of other operator $O_{-i}$.
\STATEx \hskip\algorithmicindent \textbf{if} {$L_{i,k} < G_{-i,k}$} \textbf{then}
\STATEx \hskip\algorithmicindent \hskip\algorithmicindent \textbf{do} \textit{STOP} using carrier $k$.
\STATEx \hskip\algorithmicindent \textbf{end if}
\STATE \textbf{go to} 1, and repeat until convergence is achieved.
\end{algorithmic}
\end{algorithm}


\subsection{Mathematical Analysis} \label{sec:CoopMath}

\noindent
In this section, we will present the mathematical analysis dealing with cooperative game between the operators with varying interference conditions and load factor. For the sake of simplicity, we construct the following assumptions in order to get an insight of the behaviour of the algorithm -

\begin{itemize}
\item There are two operators $O_{a}$ and $O_{b}$, each have a \ac{BS}, and a load ${{N}_{a}}$ and ${{N}_{b}}$ respectively,

  \item All users within the \ac{BS}'s access area experience the same \ac{SINR}/\ac{SNR} over the carrier components. With this, scheduling weights are same ($w_{n,k}=1/\text{load}$),

\item No shadowing has been considered; therefore, user rates are function of only distance-dependent path loss,

  \item Operators follow \ac{PF} utility measure ($\sum\nolimits_{n}{\log {{r}_{n}}}$, where $r_{n}$ is the user rate of $n$-th user in the operator's access area).

\end{itemize}

\noindent
We assume that both operators start with orthogonal sharing, and we show that under -

\begin{itemize}
  \item Low Interference; both operators tend to share full spectrum,
  \item High Interference, and with asymmetric loads; both operators tend to share the spectrum orthogonally with high load operator utilizing more component carriers than low load operator,
  \item High Interference, and with symmetric loads; both operators tend to share the spectrum orthogonally with an equal carrier allocation.
\end{itemize}


\subsubsection{Orthogonal Spectrum Sharing} \label{sec:CoopMathOrtho}

\noindent
Let operators $O_{a}$ and $O_{b}$ have respective load $N_{a}$ and $N_{b}$. Both operators were initially sharing the spectrum orthogonally, each were  having equal and non-overlapping $K/2$ carriers allocation and thus no inter-operator interference was generated. For \ac{PF} measure, the utility for operator $O_{a}$ with orthogonal carrier allocation, $U_{o,a}$ can be read as

\begin{equation*}
{{U}_{o,a}}=\sum\limits_{n=1}^{{{N}_{a}}}{\log }\left( \sum\limits_{k=1}^{{{K}_{a}}}{{{w}_{n,k}}{{\log }_{2}}\left( 1+{{\gamma }_{n,k}} \right)} \right).
\end{equation*}

\noindent
As per assumptions and initial conditions, $U_{o,a}$ can be written as

\begin{equation} \label{eq:CoopOrthoUtilA}
{{U}_{o,a}}={{N}_{a}}\log \left( \frac{K}{2}\frac{1}{{{N}_{a}}}{{\log }_{2}}\left( 1+{{\gamma }_{a}} \right) \right),
\end{equation}

\noindent
where ${\gamma }_{a}$ is the \ac{SNR}\footnote{As each operator has a single \ac{BS}, and according to the assumption, \ac{SNR} is same for all component carriers for all users within an operator.} of the users in operator $O_{a}$.\\

\noindent
Similarly, for operator $O_{b}$, utility $U_{o,b}$ is,\\

\begin{equation} \label{eq:CoopOrthoUtilB}
{{U}_{o,b}}={{N}_{b}}\log \left( \frac{K}{2}\frac{1}{{{N}_{b}}}{{\log }_{2}}\left( 1+{{\gamma }_{b}} \right) \right).
\end{equation}


\subsubsection{Cooperative Spectrum Sharing}

\noindent
Operator $O_{a}$ has higher load than operator $O_{b}$, ${{N}_{a}}\gg {{N}_{b}}$. Under asymmetric load it is beneficial (in a cooperative sense) that operator $O_{a}$ uses more carriers than operator $O_{b}$. If high load operator $O_{a}$ cooperatively agreeing with operator $O_{b}$ to switch on one of its unused carriers, the utility for operator $O_{a}$, i.e., $U_{c,a}$, becomes,

\begin{equation*}
U_{c,a}={{N}_{a}}\log \left( \frac{K}{2}\frac{1}{{{N}_{a}}}{{\log }_{2}}\left( 1+{{\gamma }_{a}} \right)+\frac{1}{{{N}_{a}}}{{\log }_{2}}\left( 1+\gamma _{a}^{'} \right) \right),
\end{equation*}

\noindent
where $\gamma^{'}_{a}$ is the new \ac{SINR} corresponding to $k_{i}$-th carrier which both operators use at the same time in the same vicinity and generate inter-operator interference to each other, and all the other symbols have usual meanings. As a result the utility gain for operator $O_{a}$ is, $G_{a} = U_{c,a} - U_{o,a}$,

\begin{equation*}
{{G}_{a}}={{N}_{a}}\log \left( \frac{\left( \frac{K}{2} \right)\log_{2} \left( 1+{{\gamma }_{a}} \right)+\log_{2} \left( 1+\gamma _{a}^{'} \right)}{\left( \frac{K}{2} \right)\log_{2} \left( 1+{{\gamma }_{a}} \right)} \right).
\end{equation*}

\noindent
Similarly, for operator $O_{b}$, utility $U_{c,b}$ is,

\begin{equation*}
U_{c,b}={{N}_{b}}\log \left( \left( \frac{K}{2}-1 \right)\frac{1}{{{N}_{b}}}{{\log }_{2}}\left( 1+{{\gamma }_{b}} \right)+\frac{1}{{{N}_{b}}}{{\log }_{2}}\left( 1+\gamma _{b}^{'} \right) \right),
\end{equation*}

\noindent
and the respective utility loss $L_{b} = U_{o,b} - U_{c,b}$,

\begin{equation*}
{{L}_{b}}={{N}_{b}}\log \left( \frac{\left( \frac{K}{2} \right)\log_{2} \left( 1+{{\gamma }_{b}} \right)}{\left( \frac{K}{2}-1 \right)\log_{2} \left( 1+{{\gamma }_{b}} \right)+\log_{2} \left( 1+\gamma _{b}^{'} \right)} \right).
\end{equation*}

\noindent
Let us define the ratio of rates with and without interference, $R=\frac{\log_{2} \left( 1+{\gamma }' \right)}{\log_{2} \left( 1+\gamma  \right)},\,R<1$ (because SNR $\gamma$ $>$ SINR $\gamma^{'}$ and signal power $>$ noise power, i.e., $\gamma>1$) and re-write the gain/loss as

\begin{equation*}
{{G}_{a}}={{N}_{a}}\log \left( 1+2\frac{{{R}_{a}}}{K} \right),
\end{equation*}

\begin{equation*}
{{L}_{b}}={{N}_{b}}\log {{\left( 1+2\frac{\left( {{R}_{b}}-1 \right)}{K} \right)}^{-1}}.
\end{equation*}

\noindent
The cooperative utility (sum of the operators' utilities) increases, if ${{G}_{a}}>{{L}_{b}}$. The necessary condition is,

\begin{equation*}
{{N}_{a}}\log \left( 1+2\frac{{{R}_{a}}}{K} \right)>{{N}_{b}}\log {{\left( 1+2\frac{\left( {{R}_{b}}-1 \right)}{K} \right)}^{-1}},
\end{equation*}

\begin{equation} \label{eq:CoopHighLoadCond}
{{{\left( 1+\frac{2{{R}_{a}}}{K} \right)}^{{{N}_{a}}}}{{\left( 1+\frac{2\left( {{R}_{b}}-1 \right)}{K} \right)}^{{{N}_{b}}}}>1}.
\end{equation}

\noindent
Similarly, in an unideal case, if low load operator $O_{b}$ switches on one of its unused carriers from the initial equal orthogonal carrier allocation, then the condition is,

\begin{equation} \label{eq:CoopLowLoadCond}
{{{\left( 1+\frac{2\left( {{R}_{a}}-1 \right)}{K} \right)}^{{{N}_{a}}}}{{\left( 1+\frac{2{{R}_{b}}}{K} \right)}^{{{N}_{b}}}}>1}.
\end{equation}

\begin{enumerate}
  
\item Under low interference, both $R_{a}$ and $R_{b}$ approaches one because SINR $\gamma^{'}$ tends to be SNR $\gamma$. Therefore, both Eq.~\eqref{eq:CoopHighLoadCond} and~\eqref{eq:CoopLowLoadCond} are satisfied and both operators start using the unused carriers and share the full spectrum.
  
\item Under high interference, interference power becomes significant and tends to lie closer to the signal power. Thus, SINR ${{\gamma }^{'}}$ tends to be 1, and consequently, ${{\text{R}}_{a}}\to 1/{\log_{2} \left( 1+{{\gamma }_{a}} \right)}$ and ${{\text{R}}_{b}}\to 1/{\log_{2} \left( 1+{{\gamma }_{b}} \right)}$ (because $\log_{2}(1+\gamma ')\approx {{\log }_{2}}2$), which implies that, ${{\text{R}}_{a}}<1$ and ${{\text{R}}_{b}}<1$. Therefore, in Eq.~\eqref{eq:CoopHighLoadCond}, the component $\left( 1+{2{{R}_{a}}}/{K} \right) > 1$, whereas $\left( 1+{2{({R}_{b}-1)}}/{K} \right)<1$, and with ${{N}_{a}}>>{{N}_{b}}$, the left hand side of Eq.~\eqref{eq:CoopHighLoadCond} becomes greater than one and the inequality satisfies. On the other hand, in Eq.~\eqref{eq:CoopLowLoadCond}, $\left( 1+{2{({R}_{a}-1)}}/{K} \right)<1$ and $\left( 1+{2{{R}_{b}}}/{K} \right) > 1$, and with ${{N}_{a}}<<{{N}_{b}}$, the left hand side of Eq.~\eqref{eq:CoopLowLoadCond} becomes less than one and the inequality does not satisfy. It implies that under high interference high load operator gets more orthogonal carriers than low load operator as long as their sum-operator utility increases.
  
\item With equal load, i.e., ${{N}_{a}}\approx {{N}_{b}}=N$, we can assume ${{R}_{a}}\approx {{R}_{b}}=R$, and on that account, both Eq.~\eqref{eq:CoopHighLoadCond} and~\eqref{eq:CoopLowLoadCond} describing the spectrum sharing conditions reduce to a single conditon, and is independent of load, accordingly,

\begin{equation}
{{\left( 1+2\frac{R}{K} \right)\left( 1+2\frac{\left( R-1 \right)}{K} \right) }}>1.
\label{eq:CoopEqLoadHighIntf}
\end{equation}

\noindent
Under high interference (i.e., $R=1/{\log_{2} \left( 1+\gamma  \right)}$, $R<1$), there must be no transference of carriers because loads are equal and the operators must remain as it is as they started initially with an equal orthogonal carrier allocation. Therefore, from Eq.~\eqref{eq:CoopEqLoadHighIntf}, the condition to remain in orthogonal sharing can be obtained as

\begin{equation*}
\left( 1+2\frac{R}{K} \right)\left( 1+2\frac{\left( R-1 \right)}{K} \right)<1.
\end{equation*}

\noindent
So, the condition to remain in equal orthogonal share at high interference is,

\begin{equation} \label{eq:CoopEqLoadHighIntfCond}
\frac{2R\left( R-1 \right)}{K}+2R-1<0.
\end{equation}

\noindent
For a large number of carriers ($K$), the necessary condition obtained from Eq.~\eqref{eq:CoopEqLoadHighIntfCond} is, $R<0.5$, becomes $\gamma >$ 3 or 4.77 dB. In the limiting case with $K=2$  and the necessary condition ($R<0.618$) becomes $\gamma >$ 2.07 or 3.15 dB.

\end{enumerate}

\noindent
As a result, with high inter-operator interference, and for any number of channels, operator with a low load abandons carriers provided that the \ac{SNR} is higher than 4.77 dB.

\clearpage


\section{Repeated Games using Virtual Carrier Price for Spectrum Sharing} \label{chap:GamePrice}

\noindent
Operators may not be willing to share their performance over the different parts of the spectrum, nor willing to decide invariably in favor of cooperative utility. In this sense, a cooperative game model does not describe the interactions between operators in a realistic manner. Instead, interaction between operators could be modelled as noncooperative games. The operators are assumed to interact for long periods of time, and have a well-defined and publicly known identity. Accordingly, an appropriate framework is that of \textit{repeated games}. We assume that each operator has a carrier allocation strategy where it acts based on predetermined rules. One-shot games are not considered in our study as they can result in poor performance for some if not for all of the players~\cite{JR07Etkin}.\\

\noindent
In this chapter, we model \textit{noncooperative repeated games} using \textit{virtual carrier pricing} based utility for inter-operator spectrum sharing. In this, operators estimate their utility gain distributively for a new carrier allocation strategy. For instance, in downlink transmission, the operator can ask its \acp{UE} to measure the carrier utilization and interference levels, and report them to the home \ac{BS}. The operator uses this information to analyze its carrier allocation strategy in which it uses its own unused carrier, or may ask the opponent to stop using it. The operators interact and approve each other carrier allocation strategies if they see utility improvement for themselves. The operators are self-interested; therefore, they scrutinize their mutual interaction in terms of spectrum usage favors given to each other. Finally, the latter chapter presents the simulation results for the proposed model and assesses its performance against the traditional allocation schemes and cooperative algorithm.


\subsection{System Description} \label{sec:GamePriceSysDescription}

\noindent
We propose a dynamic spectrum sharing method in which the operators actively attempt to share its spectrum with the other operators in the downlink based on some policy. Obviously, leasing would mean that the lessee operator system will have to pay certain compensation to the lessor (owner) operator for this additional spectrum. However, instead of monetary compensation for the gained spectrum usage, operators keep track of their mutual spectrum transactions and can ask each other for their fair due based on their mutual history in demanding situations. As operators' identities are publicly known, they strive to behave honestly.\\

\noindent
We consider a geographical area served by the number of operators, with their \acp{RAN} having a connection with each other. For the discussion, the operators are considered as a single cell operator. The set of operators is denoted by, $\mathcal{I}=\{1,2,3.....I\}$. The \ac{BS} distributes the $K$ carriers amongst the $J$ users according to their \ac{CSI} and fairness.\\

\begin{figure}[h]
\centering
\includegraphics[scale=.52, trim = 13mm 0mm 0mm 0mm, clip]{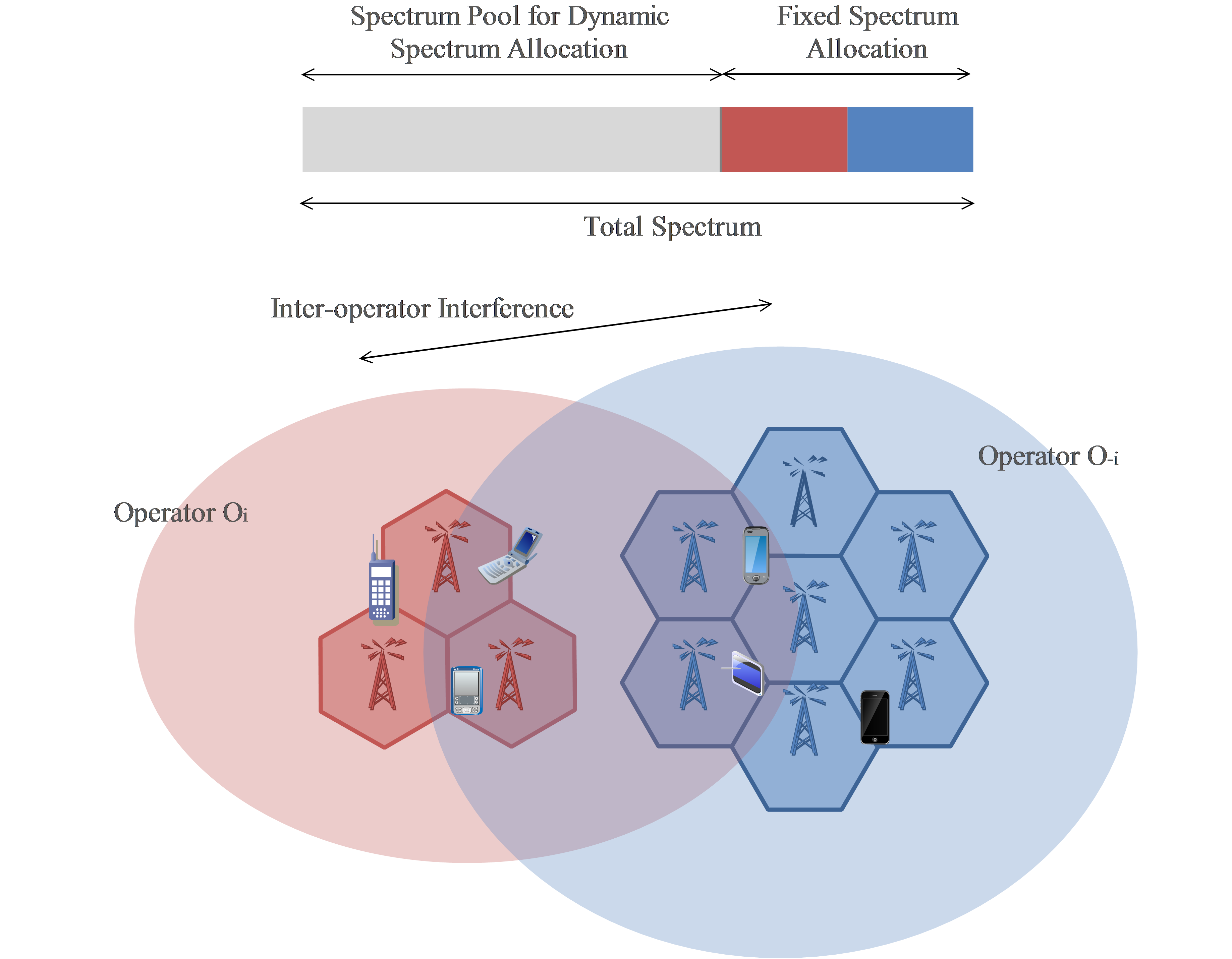}
\caption{Operators contend for spectrum within the common spectrum pool}
\label{SpectrumPool}
\end{figure}

\noindent
The total available spectrum ($K$ carriers) in the given geographical area is divided into two different allocations, namely, (i) Fixed spectrum allocation (FSA), and (ii) Dynamic spectrum allocation (DSA). In \ac{FSA}, each operator has its own independent spectrum usage rights (or \acp{PCC}); therefore, no frequency overlapping occurs, nor it generates any inter-operator interference. However, on the other hand, there exists a common pool of spectrum (or \acp{SCC}), for which operators contend for the spectrum usage rights based on some established policies, and is termed as \ac{DSA}. As multiple operators possess the right to access the spectrum in \ac{DSA} and there is no direct mechanism to control interference between the operators, inter-operator interference is generated as depicted in Fig.~\ref{SpectrumPool}.

\noindent
In the following, we denote by ${{r}_{i,j}}({{k}_{i}},{{k}_{-i}})$ the rate of the $j$-th user of the $i$-th operator. ${k}_{i}$, ${k}_{-i}$ are the carrier allocation of operators ${O}_{i}$ and ${O}_{-i}$ respectively, where the ${O}_{-i}$ signal may interfere with the $O_i$ signal. We estimate throughput ${{T}_{i}}\left( {{k}_{i}},{{k}_{-i}} \right)$ for a single cell operator ${O}_{i}$ serving $J_i$ by the Shannon capacity as

\begin{equation*}
{{T}_{i}}=\sum\limits_{j=1}^{J_i}{{{r}_{i,j}}({{k}_{i}},{{k}_{-i}})},
\end{equation*}

\begin{equation*} \label{eq:GamPricThroughput}
{{T}_{i}}=\sum\limits_{j=1}^{J_i}{\sum\limits_{k=1}^{K}{{{w}_{i,j,k}}({{k}_{i}},{{k}_{-i}})\text{lo}{{\text{g}}_{2}}(1+{\text{SINR}}_{i,j,k}({{k}_{i}},{{k}_{-i}})}}),
\end{equation*}

\noindent
where ${\text{SINR}}_{i,j,k}({k}_{i}, {k}_{-i})$ and ${w}_{i,j,k}$ are the downlink user \ac{SINR} and time scheduling weight of the $k$-th carrier of the $j$-th user in the $i$-th operator, respectively. The ${\text{SINR}}_{i,j,k}({k}_{i}, {k}_{-i})$ is defined as

\begin{equation*}
{{\text{SINR}}_{i,j,k}}=\frac{P_{i}\left( {{k}_{i}} \right){{\mathcal{C}}_{i,j}}}{\left( \sum\nolimits_{q=1,q\ne i}^{I}{P_{q}\left( {{k}_{q}} \right)}{{\mathcal{C}}_{q,j}} \right)+{N_{o}}},
\end{equation*}

\noindent 
where $P_{i}\left( {{k}_{i}} \right)$ is the signal power of the $k$-th carrier, s.t., total power budget $P$ is uniformly distributed over $K_i$ active component carriers out of total $K$ carriers of the $i$-th operator, accordingly, $P/K_i$, ${\mathcal{C}}_{i,j}$  is the channel gain of the $j$-th user within the $i$-th operator, $N_o$ is the power density of the background noise and $\sum\nolimits_{q=1,q\ne i}^{I}{P_q\left( {{k}_{q}} \right)}{{\mathcal{C}}_{q,j}}$ represents the total interference power perceived by the $i$-th user in the $k$-th carrier, which is engendered by the other operators while sharing the same carrier frequency in \ac{DSA}. We define  scheduling weight ${w}_{i,j,k}$ where the $i$-th user is scheduled over the $k$-th component carrier for a fraction ${w}_{i,j,k}$ of time in such a manner that it maximizes throughput $T_{i}$,

\begin{equation*}
\begin{aligned}
& \underset{w_{i,j,k}}{\text{max}}
& & T_{i} \\
& \text{s.t.}
& & \sum\limits_{j=1}^{J_i}{{{w}_{i,j,k}}}=1\forall k \\
& & & {{w}_{i,j,k}}\ge 0\forall \{j,k\}.
\end{aligned}
\end{equation*}


\subsection{System Model}


\subsubsection{Distributed Game Model}

\noindent
Briefly reviewing game theory, for a finite set of $\mathcal{I}$ operators, a game $\mathcal{G}$ in strategic form game can be described as

\begin{equation*}
\mathcal{G}= \left\langle {{S}_{i}},{{U}_{i}} \right\rangle,
\end{equation*}

\noindent
with the following ingredients -
\begin{itemize}
\item ${S}_{i}$ represents the set of strategies (or actions) for each operator $O_{i}$, $i\in \mathcal{I}$ that are feasible during the game $\mathcal{G}$,
\item ${U}_{i}$ is the utility function (or objective function), on the basis of which game $\mathcal{G}$ is played amongst the operators by applying strategies or actions $s_i \in {S}_i$ independently in an effort to fetch the best utility for its own.
\end{itemize}

\noindent
A key concept in noncooperative game theory is the \acf{NE}, which provides a benchmark for investigating how purely rational decision makers would behave~\cite{JR99Myerson}. \ac{NE} is a profile of strategies ($S_{i}$, $S_{-i}$) such that each intelligent operator has knowledge of its environment and thereby rationally acts to maximize its own utility function $U_{i}$, depending on not only its own actions, but also other's actions. Mathematically, a \ac{NE} is defined as

\begin{equation*}
{{U}_{i}}({{s}^{*}_{i}},{{s}_{-i}})\ge {{U}_{i}}(s_{i}^{'},{{s}_{-i}}),
\end{equation*}

\noindent
where $s_{i}^{*}$ is a strict \ac{NE} strategy given the \ac{NE} strategy $s_{-i}$ if, for all $s_{i}^{'}\in {{S}_{i}}$ and ${{s}^{*}_{i}}\subset s_{i}^{'}$.\\

\noindent
Fundamentally, it is assumed that the operators' strategies are independent and chosen at their own will intelligently. However, the game formulation extending to our work contains contingent strategies where acceptance of a strategy requires cooperation from the opponent operator, e.g., one may request the other to switch off an interfering carrier, and the other accedes to the request only if it sees a utility gain in doing it.\\

\noindent
To play such a game $\mathcal{G}$, operator $O_{i}$ evaluates its carrier allocation strategy $s_{i}^{*}$ and checks for its utility\footnote{Here, utility is a function of cell throughput and carrier price component. Refer to Section~\ref{sec:GamePriceUtil} of this chapter for the detailed description.} gain, accordingly,

\begin{equation} \label{eq:GameOwnUtilIneq}
{{U}_{i}}(s_{i}^{*},{{s}_{-i}})>{{U}_{i}}({{s}_{i}},{{s}_{-i}}),
\end{equation}

\noindent
where $s_{i}$ and $s_{-i}$ are the existing strategy profiles of operators $O_{i}$ and $O_{-i}$ respectively. While there could be many viable strategies, however, the operator likely to adopt a strategy that fetches it a highest possible utility gain. If Eq.~\eqref{eq:GameOwnUtilIneq} is satisfied, operator $O_{i}$ requests operator $O_{-i}$ for the fulfilment of its strategy $s_{i}^{*}$. Now, operator $O_{-i} $ analyzes its utility function $U_{-i}$, accordingly,

\begin{equation}  \label{eq:GameOtherUtilIneq}
{{U}_{-i}}(s_{i}^{*},{{s}_{-i}})>{{U}_{-i}}({{s}_{i}},{{s}_{-i}}).
\end{equation}
 
\noindent
If the above given inequality is satisfied, the evaluated strategy $s_{i}^{*}$ is confirmed. After mutual agreement, the new yielded outcome strategy $s_{i}^{*}={{({{s}_{i}})}_{i\in \mathcal{I}}}$ comes into existence, $s_{i}^{*}\to {{s}_{i}}$ and eventually, the strategy profile of the operators converges to \ac{NE}. It has to be noted that the described game model is noncooperative even though the strategies are contingent, but they never reveal any of its utility related information to the other. Besides, the decisions are made locally, unlike in~\cite{CP13Amin} where operators compare their utility gains/losses with each other and jointly make their decisions.


\subsubsection{Utility Function} \label{sec:GamePriceUtil}

\noindent
The utility function is a performance metric, whose design is considered as a bottleneck factor by which then, the given operator tries to optimize this function every time whenever a strategy is played by the other, and plays its own strategy afterwards. Normally, cell throughput or its variant, e.g., \ac{MMF}, \ac{PF}, mean-rate, weighted fair utility~\cite{BK97Keshav,JR98Kelly} are regarded as the true measure of user satisfaction and is the usual choice for the utility function. Operators playing noncooperative games do not have to maintain same utility nor be aware of the utility of other operator. The utility function is occasionally defined as

\begin{equation} \label{eq:UtilThroughput}
{{U}_{i}}=f\left( {{T}_{i}}\left( {{k}_{i}},{{k}_{-i}} \right) \right),
\end{equation}

\noindent 
where $f$ represents the fairness criteria, as described in Eq.~\eqref{eq:fair} to~\eqref{eq:mmf}. Operators play strategies involving carrier allocation $(k_{i},k_{-i})$ and aim to maximize their utility function incessantly,

\begin{equation} \label{eq:UtilMax}
\underset{{{k}_{i}},{{k}_{-i}}}{\mathop{\text{max}}}\,{{U}_{i}}\text{  s}\text{.t}\text{. }{{k}_{i}},{{k}_{-i}}\in K.
\end{equation}

\noindent 
However, the operators are interested in maximizing their throughput over time horizon $\mathcal{T}$ of the repeated games rather than every time instant owing to the fact that the sacrificing operator (with a low load factor) tends to lose a small amount of throughput in order to gain larger throughput benefits during its peak conditions, s.t.,

\begin{equation} \label{eq:GamePriceThTime}
\underset{{{k}_{i}}\left( t \right),{{k}_{-i}}\left( t \right)}{\mathop{\text{max}}}\,\underset{\mathcal{T}\to \infty }{\mathop{\lim }}\,\frac{1}{\mathcal{T}}\int\limits_{0}^{\mathcal{T}}{{{T}_{i}}\left( {{k}_{i}}\left( t \right),{{k}_{-i}}\left( t \right) \right)dt}\text{  s}\text{.t}\text{. }{{k}_{i}},{{k}_{-i}}\in K.
\end{equation}

\noindent
According to Eq.~\eqref{eq:GamePriceThTime} operators cannot maximize their utility at all time instants if the utility function is chosen based throughput alone or its variants (like \ac{PF}, \ac{MMF}, etc. in Eq.~\eqref{eq:UtilThroughput}), which is in contradiction to what it has been stipulated in Eq.~\eqref{eq:UtilMax}. The reason is that, if the utility function chosen based on throughput alone, then the operator, which is sacrificing resources, will always have immediate lower utility. Therefore, either of the Eq.~\eqref{eq:GameOwnUtilIneq} and~\eqref{eq:GameOtherUtilIneq} will never satisfy and thus operators will never share resources.\\

\noindent
However, if it is desirable to have game outcomes that are closer to a social optimum, one may change the utility by a virtual carrier price,

\begin{equation} \label{UtilThroughputPrice}
{{U}_{i}}=f({{T}_{i}})-{{\lambda }_{i}},
\end{equation}

\noindent 
where $\lambda_{i}$ is the virtual carrier price. There is a lot of literature available on spectrum pricing (e.g.,~\cite{CP07Gandhi,CP06Ryan,CP05Ileri,CP04Huang}). Here, the selection for the carrier pricing function is kept simple and precise, instead of being modelled in terms of market based forces as discussed in most of the literature. Therefore, we select

\begin{equation} \label{eq:VirCarPriFun}
{{\lambda }_{i}}=p_1\left( {{e}^{p_2 \frac{\sum\nolimits_{k=1}^{K}{c_{i,k}}}{K_i} }}-1 \right),
\end{equation}

\noindent 
where $p_1$ and $p_2$ are the pricing constants, ${c}_{i,k}$ is the carrier utilization of the $k$-th carrier and $K_i$ is the sum of the active component carriers out of total $K$ carriers  in the $i$-th operator. For a  particular case with two operators in a given geographical area, the carrier utilization $c(k)$ can be set to,

\begin{equation*} 
c\left( k \right)=\left\{ \begin{matrix}
   1,  \\
   0.5,  \\
   0,  \\
\end{matrix}\text{   }\begin{matrix}
   k\in&\text{full carrier}  \\
   k\in&\text{shared carrier}  \\
   k\in&\text{unused carrier}.  \\
\end{matrix}\text{ } \right.
\end{equation*}

\noindent \\
\textit{Virtual Carrier Price} $\lambda$ in utility function $U$ penalises the operators for their carrier usage. In the game, operators aim to maximize their utility function at every game sequence, shown by Eq.~\eqref{eq:UtilMax}. With increased carrier utilization, the heavily loaded operators can have a larger throughput component in comparison to the negative carrier pricing component in their utility function, and in succession their utility increases. This progresses to the heavily loaded operators to afford more spectrum resources, which is not the usual case for the sparsely loaded operators. The selection of an  exponential function for the carrier pricing component in Eq.~\eqref{eq:VirCarPriFun} is due to the fact that it penalises the operators for their increased carrier usage while ensuring the minimal carrier utilization requirement for every operator with negligible price. The label \textquoteleft Virtual\textquoteright~signifies that the price is not measured in monetary terms, rather it is a virtual measure or tool by which operators share the spectrum according to their demands. In this manner, operators become able to share the spectrum resources opportunistically and can maximize their sum-throughput noncooperatively.


\subsubsection{Spectrum Usage Favors} \label{sec:GamePriceFavors}

\noindent
Operators are always motivated by self-interest; therefore, they model negotiations for carriers in terms of spectrum usage favors. The favors are referred to carrier component utilization. It is assumed that the opponent operator cooperates provided that both operators have so far fulfilled about the same number of favors. To do that, for instance, each operator maintains a bookkeeping system listing the number of times each operator has been cooperative. The operators grant favors to the opponents if they see the cooperative spirit. This kind of strategy resembles a tit-for-tat~\cite{JR81Axelrod,BK07Axelrod} strategy in a sense that it is forgiving and avoids immediate punishment. Note that this idea can also be extended to more general cases where operators can grant higher number of favors to the opponent provided they receive some sort of compensation. In this study, though, we do not consider inter-operator communication exterior to radio access and monetary transactions between the different entities.\\

\noindent
Let us assume that operator $O_{i}$ selects randomly a carrier and constructs the possible favors for the different courses of strategic actions -

\begin{itemize}

\item {Both operators utilize carrier $k$. Operator $O_{i}$ asks operator $O_{-i}$ to stop using the carrier in case if it sees the utility gain, accordingly~\eqref{eq:GameOwnUtilIneq}. Operator $O_{-i}$ does the favor only if its own utility gain is positive too as per Eq.~\eqref{eq:GameOtherUtilIneq}.}
\item {Operator $O_{i}$ does not use carrier $k$, but operator $O_{-i}$ does. Operator $O_{i}$ adopts a strategy using carrier $k$. If both operators see the utility gain according to the game (Eq.~\eqref{eq:GameOwnUtilIneq} and~\eqref{eq:GameOtherUtilIneq}), the new strategy is agreed and regarded as a favor as it causes destructive interference to operator $O_{-i}$.}
\item {No operator utilizes the carrier and thus, operator $O_{i}$ can start using it.}
\item {When only operator $O_{i}$ utilizes the carrier, there is no interaction between the operators.}

\end{itemize}

\noindent
The given below Tab.~\ref{tab:FavorsClassification} summarises the strategic actions involving own carrier ($k_{i}$) and interfering  carrier ($k_{-i}$) allocations that defines a favor given to operator $O_{i}$ by operator $O_{-i}$.

\begin{table}[h]
\centering
\caption{Favors Classification}
{\begin{tabular}{ p{7cm}p{3cm}}
\hline
\textbf{Operator} ${{O}_{i}}:\left( k_{i}^{t},k_{-i}^{t} \right)\to \left( k_{i}^{t+1},k_{-i}^{t+1} \right)$ & \textbf{Favor} \\
\hline
$\left( \text{on,on} \right)\to \left( \text{on,off} \right)$ & Yes \\
$\left( \text{off,on} \right)\to \left( \text{on,on} \right)$ & Yes \\
$\left( \text{off,off} \right)\to \left( \text{on,off} \right)$ & No \\
$\left( \text{on,off} \right)\to \left( \text{on,off} \right)$ & No \\
\hline
\end{tabular}}
\label{tab:FavorsClassification}
\end{table}

\noindent
To mitigate the selfish behaviour of the opponents, operators limit the number of outstanding favors. The operators incorporate a hard check stopping criterion for the game where they model the utility function based on Eq.~\eqref{UtilThroughputPrice} and grant spectrum usage favors to each other as long as their outstanding favors are less than surplus limit $S$,

\begin{equation*}
\begin{aligned}
{{O}_{i}}~:&{{h}_{-i}}-{{h}_{i}}\le S,\\
{{O}_{-i}}:&{{h}_{i}}-{{h}_{-i}}\le S.
\end{aligned}
\end{equation*}

\noindent
In contrast, say, if operator $O_{i}$ has received more favors amounting $S$ than it has given to operator $O_{-i}$, operator $O_{-i}$ will not review its requests further anymore unless operator $O_{i}$ starts accepting the favors and bring down the outstanding favors of operator $O_{-i}$ lower than $S$.\\

\noindent
The surplus limit $S$ controls the width of the outstanding favors window; therefore its construction requires an appropriate care. Choosing small values can subdue the game, whereas large values can polarize the game benefiting the particular operators. With a selection of fitting value for the surplus limit, the operators are able to trade resources with fairness; simultaneously keeping check on the operators' unauthorized requests for the unfair gains.


\subsection{Proposed Algorithm I}

\noindent
The proposed algorithm in the form of pseudocode is summarised as below,

\setcounter{algorithm}{0}
\begin{algorithm} [H]
\renewcommand\thealgorithm{}
\caption{Repeated Games Model using Virtual Carrier Price for Inter-operator Spectrum Sharing}
\begin{algorithmic}[1]
\STATE Operator $O_{i}$, where $i \in \mathcal{I}$, analyses strategy $s$ by switching on carrier $k_{i}$ or removing interfering carrier $k_{-i}$. Calculates new utility $U_{i,s}$ and compares it with present utility $U_{i}$.
\STATEx\textbf{if} {$U_{i,s} > U_{i}$} \textbf{then}
\STATE \quad Operator $O_{-i}$ compares its outstanding favors with surplus $S$.
\STATEx\quad\textbf{if} {$h_{-i}-h_{i} \le S$} \textbf{then}
\STATE \quad\quad\begin{varwidth}[t]{\linewidth} Operator $O_{-i}$ compares new utility $U_{-i,s}$ for strategy $s$ with present utility\\ $U_{-i}$. \end{varwidth}
\STATEx \quad\quad\textbf{if} {$U_{-i,s} > U_{-i}$} \textbf{then}
\STATE \quad\quad\quad Strategy s is accepted.
\STATE \quad\quad\quad Favors are updated: ${{h}_{-i}}\to {{h}_{-i}}+1$.
\STATE \quad\quad\textbf{end if}
\STATE \quad\textbf{end if}
\STATE \textbf{end if}
\end{algorithmic}
\end{algorithm}


\subsection{Mathematical Analysis}

\noindent
In this section, we analyze the algorithm mathematically, and provide theoretical results for the optimization of pricing constants. For the analysis, we consider the same assumptions made in Section~\ref {sec:CoopMath}.


\subsubsection{Orthogonal Spectrum Sharing}

\noindent
Referring to Section~\ref{sec:CoopMathOrtho}, the \ac{PF} throughput of operator $O_{a}$ with orthogonal carrier allocation, $T_{o,a}$ is given by Eq.~\eqref{eq:CoopOrthoUtilA} and the same for operator $O_{b}$, $T_{o,b}$ is given by Eq.~\eqref{eq:CoopOrthoUtilB}.\\

\noindent
The sum-\ac{PF} throughput of both operators, $T_{o}$  is, ${{T}_{o}}={{T}_{o,a}}+{{T}_{o,b}}$, i.e.,

\begin{equation} 
{{T}_{o}}={{N}_{a}}\log \left( \frac{K}{2}\frac{1}{{{N}_{a}}}{{\log }_{2}}\left( 1+{{\gamma }_{a}} \right) \right)+{{N}_{b}}\log \left( \frac{K}{2}\frac{1}{{{N}_{b}}}{{\log }_{2}}\left( 1+{{\gamma }_{b}} \right) \right).
\label{eq:GamePriceOrtho}
\end{equation}


\subsubsection{Repeated Games based Spectrum Sharing} \label{sec:GamPriceOpt}

\noindent
Let us assume, ${{N}_{a}}>{{N}_{b}}$, and being $O_{a}$ is the high load operator, so it is more appropriate that operator $O_a$ gets more resources in order to avoid the congestion and blocking probability\cite{CP13Amin,CP11Prasad}. Assume, $\Delta K$ amount of carriers are transferred to high load operator $O_{a}$ by low load operator $O_{b}$ at the end of the sequence of a game from the initial orthogonal allocation (where $\Delta K\in \left( 0,{K}/{2} \right)$). Then the new respective \ac{PF} throughputs of the operators during the game, $T_{g,a}$ and $T_{g,b}$ are,

\begin{equation*} 
{{T}_{g,a}}={{N}_{a}}\log \left( \left( \frac{K}{2}+\Delta K \right)\frac{1}{{{N}_{a}}}{{\log }_{2}}\left( 1+{{\gamma }_{a}} \right) \right),
\end{equation*}

\begin{equation*} 
{{T}_{g,b}}={{N}_{b}}\log \left( \left( \frac{K}{2}-\Delta K \right)\frac{1}{{{N}_{b}}}{{\log }_{2}}\left( 1+{{\gamma }_{b}} \right) \right).
\end{equation*}

\noindent
The sum-\ac{PF} throughput of both operators, ${{T}_{g}}$ is, ${{T}_{g}}={{T}_{g,a}}+{{T}_{g,b}}$,

\begin{equation*} 
{{T}_{g}}={{N}_{a}}\log \left( \left( \frac{K}{2}+\Delta K \right)\frac{1}{{{N}_{a}}}{{\log }_{2}}\left( 1+{{\gamma }_{a}} \right) \right)+{{N}_{b}}\log \left( \left( \frac{K}{2}-\Delta K \right)\frac{1}{{{N}_{b}}}{{\log }_{2}}\left( 1+{{\gamma }_{b}} \right) \right).
\end{equation*}

\noindent
The game is only beneficial if sum-throughput of game is more than the sum-throughput in case of orthogonal sharing (Eq.~\eqref{eq:GamePriceOrtho}), i.e., ${{T}_{g}}>{{T}_{o}}$. Thus,

\begin{equation}  \label{eq:GamPriGamOrt}
{{N}_{a}}\log \left( \frac{K}{2}+\Delta K \right)+{{N}_{b}}\log \left( \frac{K}{2}-\Delta K \right)>{{N}_{a}}\log \left( \frac{K}{2} \right)+{{N}_{b}}\log \left( \frac{K}{2} \right).
\end{equation}

\noindent
Further simplifying,

\begin{equation} \label{eq:GamPriGamOrt2}
\left( 1+2\frac{\Delta K}{K} \right)^{N_{a}} \left( 1-2\frac{\Delta K}{K} \right)^{N_{b}} > 1.
\end{equation}

\noindent
Besides, as per algorithm the tranfer of $\Delta K$ spectrum resources will occur only if the game based utilities, $U_{g}$  are satisified at the operators' end, accordingly, ${{U}_{g}}\left( k^{t+1} \right)>{{U}_{g}}\left( {k}^{t} \right)$, where $k^{t+1}$ and $k^{t}$ are the present and past carrier allocations.\\

\noindent
So, operator $O_{a}$ checks its utility, $U_{g,a}$, accordingly, ${{U}_{g,a}}\left( \frac{K}{2}+\Delta K \right)>{{U}_{g,a}}\left( \frac{K}{2} \right)$, i.e.,

\begin{equation*}
\begin{gathered}
{{N}_{a}}\log \left( \left( \frac{K}{2}+\Delta K \right)\frac{1}{{{N}_{a}}}{{\log }_{2}}\left( 1+{{\gamma }_{a}} \right) \right)-p_1\left( {{e}^{p_2\left( \frac{\frac{K}{2}+\Delta K}{K} \right)}}-1 \right)>\\
{{N}_{a}}\log \left( \frac{K}{2} \frac{1}{{{N}_{a}}}{{\log }_{2}}\left( 1+{{\gamma }_{a}} \right) \right)-p_1\left( {{e}^{p_2\left( \frac{\frac{K}{2}}{K} \right)}}-1 \right),
\end{gathered}
\end{equation*}

\begin{equation}  \label{eq:GamPriHigLoaUtiCon}
{{N}_{a}}\log \left( \frac{K}{2}+\Delta K \right)-{{N}_{a}}\log \left( \frac{K}{2} \right)>p_1{{e}^{p_2\left( \frac{\frac{K}{2}+\Delta K}{K} \right)}}-p_1{{e}^{p_2\left( \frac{\frac{K}{2}}{K} \right)}}.
\end{equation}

\noindent
Similarly, operator $O_{b}$ checks its utility $U_{g,b}$, accordingly, ${{U}_{g,b}}\left( \frac{K}{2}-\Delta K \right)>{{U}_{g,b}}\left( \frac{K}{2} \right)$, i.e.,

\begin{equation}  \label{eq:GamPriLowLoaUtiCon}
{{N}_{b}}\log \left( \frac{K}{2}-\Delta K \right)-{{N}_{b}}\log \left( \frac{K}{2} \right)>p_1{{e}^{p_2\left( \frac{\frac{K}{2}-\Delta K}{K} \right)}}-p_1{{e}^{p_2\left( \frac{\frac{K}{2}}{K} \right)}}.
\end{equation}

\paragraph{Optimization of Pricing Constants}

\noindent\\ \\
With the conditions in Eq.~\eqref{eq:GamPriGamOrt},~\eqref{eq:GamPriHigLoaUtiCon} and~\eqref{eq:GamPriLowLoaUtiCon}, the operators can perform optimization over the spaces, $p_1$, $p_2$ and  $\Delta K$. However, operators strive to maximize the cooperative gain, which can be leveraged by the maximization of the left hand side of Eq.~\eqref{eq:GamPriGamOrt2}. Therefore, the value of $\Delta K$ such that it returns best sum-throughput of the system is accordingly,

\begin{equation} \label{eq:GamePriceMaxProb}
\begin{aligned}
& \underset{\Delta K}{\text{max}}
& & f(\Delta K) \\
& \text{s.t.}
& & f(\Delta K)>0,
\end{aligned}
\end{equation}

\noindent
where from Eq.~\eqref{eq:GamPriGamOrt2}, $f(\Delta K)= \left( 1+2{\Delta K}/{K} \right)^{N_{a}} \left( 1-2{\Delta K}/{K} \right)^{N_{b}}- 1$. The solution to the given maximization probem can be achieved by $df(\Delta K)/d(\Delta K)=0$, assuming $\Delta K$ is a continous resource. Let the obtained solution be $\Delta {{K}_{limit}}$.\\

\noindent
Adding Eq.~\eqref{eq:GamPriHigLoaUtiCon} and~\eqref{eq:GamPriLowLoaUtiCon}, we get,

\begin{equation}  \label{eq:GamPriHigLowSumCon}
\begin{gathered}
{{N}_{a}}\log \left( \frac{K}{2}+\Delta K \right)+{{N}_{b}}\log \left( \frac{K}{2}-\Delta K \right)-{{N}_{a}}\log \left( \frac{K}{2} \right)-{{N}_{b}}\log \left( \frac{K}{2} \right)>\\
p_1\left( {{e}^{p_2\left( \frac{\frac{K}{2}+\Delta K}{K} \right)}}+{{e}^{p_2\left( \frac{\frac{K}{2}-\Delta K}{K} \right)}}-2{{e}^{p_2\frac{1}{2}}} \right).
\end{gathered}
\end{equation}

\noindent
Comparing Eq.~\eqref{eq:GamPriGamOrt} and~\eqref{eq:GamPriHigLowSumCon}, it can be inferred,

\begin{equation}  \label{eq:GamPriPQCon}
p_1\left( {{e}^{p_2\left( \frac{\frac{K}{2}+\Delta K}{K} \right)}}+{{e}^{p_2\left( \frac{\frac{K}{2}-\Delta K}{K} \right)}}-2{{e}^{p_2\frac{1}{2}}} \right)>0.
\end{equation}

\noindent
From Eq.~\eqref{eq:GamPriHigLoaUtiCon},~\eqref{eq:GamPriLowLoaUtiCon} and~\eqref{eq:GamPriPQCon}, pricing constants ($p_1$ and $p_2$) can be obtained accordingly,

\begin{subequations} \label{eq:GamPricEq1}
\begin{align}
&\text{~~~}p_1 e^{\frac{p_2}{2}} \left(\text{cosh} \left(p_2\frac{\Delta K}{K}\right)-1\right)>0,\\
&{{e}^{\frac{p_2}{2}\left( 1+2\frac{\Delta K}{K} \right)}}-{{e}^{\frac{p_2}{2}}}<\log {{\left( 1+2\frac{\Delta K}{K} \right)}^{\frac{{{N}_{a}}}{p_1}}},\\
&{{e}^{\frac{p_2}{2}\left( 1-2\frac{\Delta K}{K} \right)}}-{{e}^{\frac{p_2}{2}}}<\log {{\left( 1-2\frac{\Delta K}{K} \right)}^{\frac{{{N}_{b}}}{p_1}}},
\end{align}
 \end{subequations}

\noindent
where $0<\Delta K<\frac{K}{2}$ and ${{N}_{a}}>{{N}_{b}}$.\\

\noindent
$N_{a}$ and $N_{b}$ represent the overall load conditions in the network, and they do not refer to instantaneous load values. Subtituting, $\Delta K=\Delta K_{limit}$, ${{N}_{a}}={{\widehat{N}}_{high}}$, and ${{N}_{b}}={{\widehat{N}}_{low}}$ in Eq.~\eqref{eq:GamPricEq1}, the optimization equations can be rewritten as

\begin{subequations} \label{eq:GamPricEq2}
\begin{align}
p_1&>0 \label{eq:GamPricEq2a},\\
p_2&\ne0 \label{eq:GamPricEq2b}, \\
{{e}^{\frac{p_2}{2}\left( 1+2\frac{\Delta K_{limit}}{K} \right)}}-{{e}^{\frac{p_2}{2}}}&<\log {{\left( 1+2\frac{\Delta K_{limit}}{K} \right)}^{\frac{{{\widehat{N}}_{high}}}{p_1}}} \label{eq:GamPricEq2c}, \\
{{e}^{\frac{p_2}{2}\left( 1-2\frac{\Delta K_{limit}}{K} \right)}}-{{e}^{\frac{p_2}{2}}}&<\log {{\left( 1-2\frac{\Delta K_{limit}}{K} \right)}^{\frac{{{\widehat{N}}_{low}}}{p_1}}} \label{eq:GamPricEq2d}.
\end{align}
\end{subequations}

\begin{figure}[h]
\begin{subfigure}{.5\linewidth}
  \centering
  \includegraphics[scale=1, trim = 24mm 7mm 31mm 12mm, clip]{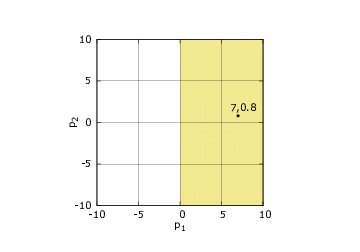}
  \caption{Graphical portrayal of Eq.~\eqref{eq:GamPricEq2a}}
 \label{fig:OptPricConsa}
\end{subfigure}
\begin{subfigure}{.5\linewidth} 
  \centering
  \includegraphics[scale=1, trim = 24mm 7mm 31mm 12mm, clip]{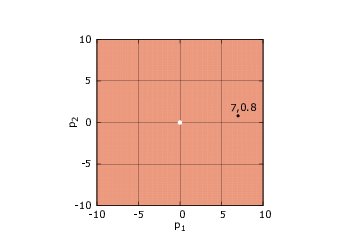}
  \caption{Graphical portrayal of Eq.~\eqref{eq:GamPricEq2b}}
\label{fig:OptPricConsb}
\end{subfigure}\\
\begin{subfigure}{.5\linewidth} 
  \centering
  \includegraphics[scale=1, trim = 24mm 7mm 31mm 12mm, clip]{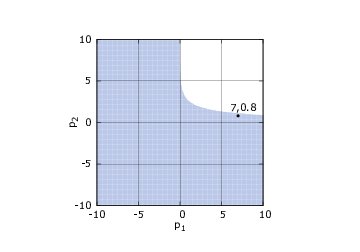}
  \caption{Graphical portrayal of Eq.~\eqref{eq:GamPricEq2c}}
\label{fig:OptPricConsc}
\end{subfigure}
\begin{subfigure}{.5\linewidth}
  \centering
  \includegraphics[scale=1, trim = 24mm 7mm 31mm 12mm, clip]{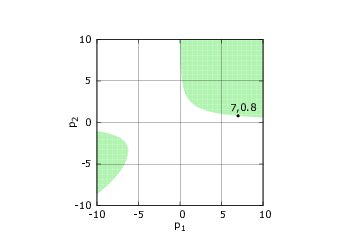}
  \caption{Graphical portrayal of Eq.~\eqref{eq:GamPricEq2d}}
\label{fig:OptPricConsd}
\end{subfigure}\\
\begin{subfigure}{1\linewidth}
  \centering
  \includegraphics[scale=1, trim = 24mm 7mm 31mm 12mm, clip]{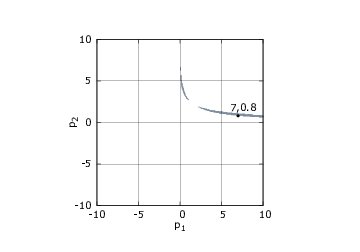}
  \caption{Intersection region of Fig.~\subref{fig:OptPricConsa}-\subref{fig:OptPricConsd}}
 \label{fig:OptPricConse}
\end{subfigure}
\caption{Intersection region in Fig.~\subref{fig:OptPricConse} delineating pricing constants $p_1$ and $p_2$, obtained from Eq.~\eqref{eq:GamPricEq2} for the parameters, $\widehat{N}_{high}=25$, $\widehat{N}_{low}=5$ and $\Delta K_{limit}=2.7.$}
\label{fig:OptPricCons}
\end{figure}

\noindent
In the analysis, $\Delta K_{limit}$ represents the maximum additional carriers allowed to transfer by a low load operator to a high load operator from their initial equal orthogonal carrier allocations. For $\widehat{N}_{high}=25$, $\widehat{N}_{low}=5$, we obtain $\Delta K_{limit}=2.7$ (using Eq.~\eqref{eq:GamePriceMaxProb}), then the optimal values of pricing constants $p_1$ and $p_2$ can be chosen from the depicted region shown in Fig.~\ref{fig:OptPricConse}. In the simulation (Section~\ref{sec:Algo1Analysis}) where many of the assumptions are disregarded, we have fixed the parameters, $p_1=7$ and $p_2=0.8$ in unison with the theoretical solution set (see Fig.~\ref{fig:OptPricConse}). We have observed, out of total carriers $K=8$, the maximum carrier utilization of 5.97 carriers in the case of a high load operator ($N_{a} = 25$) and the minimum carrier utilization of 2.03 carriers in the case of a low load operator ($N_{b} = 5$) (see Fig.~\ref{fig:Algo1HighIntfSur2Sur4}). It shows that observed $\Delta K_{limit} \approx 2$\footnote{From the simulation, $\Delta K_{limit}$ is calculated by using initial and final carrier utilizations of operator $O_a$ as $\left| 8/2 - 5.97\right|$, i.e., 1.97 (or using operator $O_b$'s carrier utilizations, $\left| 8/2 - 2.03\right|$).} for the given pricing constants is in close approximation with the theoretical results.

\clearpage


\section{Repeated Games using Mutual History for Spectrum Sharing} \label{chap:GameExpectation}

\noindent
In this chapter, we propose another coordination protocol than that discussed in Chapter~\ref{chap:GamePrice}. In that chapter, we addressed the issue using a carrier price based utility function. However, there might be a situation where operators are reluctant to entertain the carrier pricing factor in their utility function as it penalises them on their carrier usage. Hence, in this chapter, we address the same issue using a throughput-based utility function and devise a new strategic mechanism for contending spectrum resources amongst the operators.\\

\noindent
We propose a \textit{noncooperative repeated games} model based on \textit{mutual history} of utility gains/losses. Operators estimate their utility gains/losses and compare with their expected gains/losses from the past games. In order to estimate gains/losses, for instance, in downlink transmissions, the operator can ask its \acp{UE} to measure on the carrier utilization and interference levels from the opponent \ac{BS} and report it to the serving \ac{BS}, which is a simple extension of \ac{LTE} handover measurements. Assuming that operators use the same \ac{RAT}, this should be possible. As a result, an operator may select a carrier depending on (i) the carrier utilization by the operator and the opponent, as well as (ii) the history of interactions (i.e., previous games). An operator may ask the opponent to do a favor, e.g., to not to use a carrier. The operators keep track of history of favors (utility gains) provided to each other, and they use it as a check from the unfair game treatment. Finally, in Chapter~\ref{chap:Simulation}, the benefits of the algorithm is assessed in terms of rate distribution by comparing with the static allocation schemes and cooperative solution.


\subsection{System Description}

\noindent
The system description is similar to the one described in Section~\ref{sec:GamePriceSysDescription}.


\subsection{System Model}



\noindent
To begin with, we consider a simple scenario, where the operators have equal rights to the shared part of the spectrum. It is in the benefit of an operator, to demand spectrum resources when it witnesses high load in order to avoid congestion and blocking probability. The demands of an additional spectrum resources are considered as spectrum usage favors. Favors are granted for a single time slot. Prior to asking a favor, the operator evaluates its utility, and if the granted favor results in a utility gain, it goes ahead with its request for that favor.\\

\begin{table}[h]
\centering
\caption{Carrier Utilization}
{\begin{tabular}{ c c}
\hline
\textbf{Operator} ${O}_{i}: k_{i}$ & \textbf{Operator} ${O}_{-i}: k_{-i}$ \\
\hline
1 & 1 \\
0 & 1 \\
0 & 0 \\
1 & 0 \\
\hline
\end{tabular}}
\label{tab:CarrUtil}
\end{table}

\noindent
The carrier utilization between the two operators $O_{i}$ and $O_{-i}$ has four possible outcomes, shown in Tab.~\ref{tab:CarrUtil}. Based on the carrier utilization, the operators are always motivated to draw a favor in which they gain a carrier or may ask the other to stop using it. However, the operators cannot selfishly switch on their unused carriers, or dictate the others to switch off their interfering carriers. The operators follow a proper mechanism to come to a decision allow them to pry for favors based on their relative requirements according to the game model. Unlike Chapter~\ref{chap:GamePrice}, the following game model behaves in a very different manner as it accounts the history of utility gains/losses for evaluation of carrier allocation strategies, as against the game model presented in Chapter~\ref{chap:GamePrice}, where strategies depend on the present utilities only.


\subsubsection{Utility Function}

\noindent
The utility function has already been discussed in Section~\ref{sec:GamePriceUtil} and the operators construct their utility function according to Eq.~\eqref{eq:UtilThroughput}.


\subsubsection{Evaluation of Strategy: Decision Making Process} \label{sec:GameHisDecis}

\noindent
\textit{Strategic Decision Making} is a cognitive process, which produces a final outcome from the selection of a course of action among several alternative scenarios. Several factors can influence the decision making process, such as past experiences~\cite{JR05Juliusson}, cognitive biases~\cite{JR08Stanovich}, etc. However, the fundamental idea behind the decision making process (or selection of a strategy) is based on the players' rational outlook, available information and their past experiences. This theory is known as rational expectations~\cite{BK94Snowdon}, a widely studied hypothesis in the spheres of economics. In this algorithm, we aim to exploit the rational behaviour of the operators by devising a strategic mechanism based on their past experiences and state-of-the-art, which is described henceforth.\\

\noindent
Operator $O_{i}$ evaluates its new carrier allocation strategy $s^{*}_{i}$ and the corresponding increase in the utility referred to as the immediate gain $G$, which is defined as

\begin{equation*} 
{{G}_{i}}={{U}_{i}}\left( s_{i}^{*},{{s}_{-i}} \right)-{{U}_{i}}\left( {{s}_{i}},{{s}_{-i}} \right),
\end{equation*}

\noindent
where $s_{i}$ and $s_{-i}$ are the existing carrier allocation strategy profiles of operators $O_{i}$ and $O_{-i}$ respectively. Operator $O_{i}$ compares gain $G_{i}$ with its expected loss $\widehat L_{i}$ over the previous games. The expected loss is an estimation of future losses that operator $O_{i}$ will incur by sharing its spectrum resources with the other. The expected loss is calculated by averaging the past losses it had incurred by giving favors (say, $h_{-i}$ favors) to the other operator $O_{-i}$ during the game $\mathcal{G}$,

\begin{equation*} 
\widehat {{L}_{i}}=\frac{\sum\limits_{h=1}^{{{h}_{-i}}}{{{L}_{i,h}}}}{{{h}_{-i}}},
\end{equation*}

\noindent
where $\sum L_{i}$ is the totality of past losses operator $O_{i}$ had met at the hands of operator $O_{-i}$ by giving it $h_{-i}$ spectrum usage favors through their mutual interactions. If operator $O_{i}$ finds its immediate gain $G_i$ larger than its expected loss $\widehat {{L}_{i}}$, i.e.,

\begin{equation*}
{{G}_{i}}>\widehat {{L}_{i}},
\end{equation*}

\noindent
operator $O_{i}$ asks a favor from operator $O_{-i}$ to pursue to its new evaluated carrier allocation strategy $s^{*}_{i}$. Now, operator $O_{-i}$ estimates its new immediate loss $L_{-i}$ for the asked carrier allocation strategy $s^{*}_{i}$, which is calculated as

\begin{equation*}
{{L}_{-i}}={{U}_{-i}}\left( s_{i}^{*},{{s}_{-i}} \right)-{{U}_{-i}}\left( {{s}_{i}},{{s}_{-i}} \right),
\end{equation*}

\noindent
and compares with its expected gain $\widehat G_{-i}$ over the previous games. The expected gain is an estimation of future gains operator $O_{-i}$ will witness by getting the spectrum usage favors from the other. The expected gain is calculated by averaging the past gains it had collected during the game $\mathcal{G}$,

\begin{equation*}
\widehat G_{-i}=\frac{\sum\limits_{h=1}^{{{h}_{-i}}}{{{G}_{-i,h}}}}{{{h}_{-i}}},
\end{equation*}

\noindent
where $\sum G_{-i}$ is the totality of past gains operator $O_{-i}$ had collected in $h_{-i}$ number of the spectrum usage favors received from operator $O_{i}$ through their mutual interactions. If operator $O_{-i}$ finds its immediate loss $L_{-i}$ smaller than its expected gain $\widehat G_{-i}$, i.e.,

\begin{equation*}
{{L}_{-i}}<\widehat G_{-i},
\end{equation*}
 
\noindent
operator $O_{-i}$ grants the favor and strategy $s^{*}_{i}$ comes into existence, demonstrating equilibria of the appropriate type, i.e., $s^{*}_{i} \to s_{i}$. The history is updated at the both ends enlisting their respective immediate gain/loss. The game is played in repeated bounds, and in the next time slot, the operator(s) again try to contest for resources randomly. In every case, a decision is conferred based on the present gain/loss and history over the previous games. Further, the operators agree on a limit which restraints the maximum allowable spectrum usage favors (or surplus limit $S$), and helps mediating favors by normative pressure to reciprocate.


\subsubsection{Constraints: Initialization and Spectrum Usage Favors} \label{sec:GameExpectationCons}

\paragraph{Initialization}

\noindent\\ \\
The operation of the distributive algorithm requires an initialization. With initialization, the operators at the beginning of the game contend for spectrum resources having a small load factor and register approximate small gains/losses $\delta$. It is necessary because herewith the operator would be more liberistic in asking favors and in contrast equally conservative in granting them. This unfolding behavior enables operators to respond to asymmetric load conditions. This is illustrated in the mathematical analysis of Section~\ref{sec:GamHisRepeatedGame}.

\begin{equation*} 
\widehat L_{i}\approx \widehat G_{i}\approx \widehat L_{-i}\approx \widehat G_{-i} \approx \delta \text{    at time }t\to 0.
\end{equation*}

\noindent
\paragraph{Spectrum Usage Favors}

\noindent\\ \\
The discussion related to the spectrum usage favors and surplus limit behavior has been posted in Section~\ref{sec:GamePriceFavors}.


\subsection{Proposed Algorithm II}

The proposed algorithm in the form of pseudocode is summarised as below,

\begin{algorithm} [H]
\renewcommand\thealgorithm{}
\caption{Repeated Games Model using Mutual History for Inter-operator Spectrum Sharing}
\begin{algorithmic}[1]
\STATE Operator $O_{i}$, where $i \in \mathcal{I}$, analyses strategy $s$ by switching on carrier $k_{i}$ or removing interfering carrier $k_{-i}$. Calculates immediate utility gain $G_{i,s}$ and compares it with expected utility loss $\widehat L_{i} =\frac{\sum\limits_{h=1}^{{{h}_{-i}}}{{{L}_{i,h}}}}{{{h}_{-i}}} $.
\STATEx \textbf{if} {$G_{i,s} > \widehat L_{i}$} \textbf{then}
\STATE \quad Operator $O_{-i}$ compares its outstanding favors with surplus $S$.
\STATEx \quad\textbf{if} {$h_{-i}-h_{i} \le S$} \textbf{then}
\STATE \quad\quad\begin{varwidth}[t]{\linewidth} Operator $O_{-i}$ compares immediate utility loss $L_{-i,s}$ for strategy $s$ with ex- \\pected utility gain $\widehat G_{-i} =\frac{\sum\limits_{h=1}^{{{h}_{-i}}}{I{{G}_{-i,h}}}}{{{h}_{-i}}} $. \end{varwidth}
\STATEx \quad\quad\textbf{if} {$L_{-i,s} < \widehat G_{-i}$} \textbf{then}
\STATE \quad\quad\quad Strategy s is accepted.
\STATE \quad\quad\quad Operator $O_{i}$ records $G_{i,s}$.
\STATE \quad\quad\quad Operator $O_{-i}$ records $L_{-i,s}$.
\STATE \quad\quad\quad Favors are updated: ${{h}_{-i}}\to {{h}_{-i}}+1$.
\STATE \quad\quad\textbf{end if}
\STATE \quad\textbf{end if}
\STATE \textbf{end if}
\end{algorithmic}
\end{algorithm}



\subsection{Mathematical Analysis} \label{sec:GameExpectMath}

\noindent
In this section, we analyze the algorithm mathematically, and provide the theoretical results. For the analysis, we consider the same assumptions made in Section~\ref {sec:CoopMath}.

\subsubsection{Orthogonal Spectrum Sharing}

\noindent
Initially, the operators were sharing the spectrum in an orthogonal manner, i.e., each operator had allocated non-overlapping fixed number of $K/2$ carriers. The orthogonal utility of operators $O_{a}$ and $O_{b}$, i.e., $U_{o,a}$ and $U_{o,b}$ are described in Eq.~\eqref{eq:CoopOrthoUtilA} and~\eqref{eq:CoopOrthoUtilB} respectively. The Operators play repeated games in order to improve their current under-achieved utilities as described in the next section.
%
%

\subsubsection{Repeated Games based Spectrum Sharing} \label{sec:GamHisRepeatedGame}

We begin the analysis by defining the following symbols for $i$ = $a$, $b$,\\

\noindent
$G_{i}^{t}$, immediate gain of operator $O_{i}$ at time slot $t$,\\
$L_{i}^{t}$, immediate loss of operator $O_{i}$ at time slot $t$,\\
$\widehat G_{i}^{t}$, expected gain of operator $O_{i}$ till time slot $t$,\\
$\widehat L_{i}^{t}$, immediate loss of operator $O_{i}$ till time slot $t$.

\paragraph{Game Initialization}

\noindent\\\\
The game requires to be initialized with a small load. Assume, operators $O_{a}$ and $O_{b}$ have begun with a load of one user initially. With equal loads, both operators play the game for some time and generate approximately equal and small gains/losses. Thus, their expected gains/losses at time slot $t$ are assumed,

\begin{equation} \label{eq:GainLosInit}
\widehat{L}_{a}^{t}\approx \widehat{G}_{a}^{t}\approx \widehat{L}_{b}^{t}\approx \widehat{G}_{b}^{t}\approx \delta.
\end{equation}

\paragraph{Game Beginning}

\noindent\\\\
Consider Operator $O_{a}$ is experiencing a high load while operator $O_{b}$ has a low load, i.e., ${{N}_{a}}>{{N}_{b}}$. Now, two cases arise, either high load operator $O_{a}$ receives more spectrum usage favors in the form of component carriers from low load operator $O_{b}$ or vice-versa over the next time slot $t+1$. Let us calculate the probability of the occurrence for the mentioned cases.\\

\noindent
\textit{Case I}: High load operator $O_{a}$ receives more spectrum usage favor(s) from low load operator $O_{b}$ at time slot $t+1$,

\begin{subequations} \label{eq:BeginGameCaseI}
\begin{align}
G_{a}^{t+1} & > \widehat L_{a}^{t},\\
L_{b}^{t+1} & <\widehat G_{b}^{t}.
\end{align}
\end{subequations}

\noindent
\textit{Case II}: Low load operator $O_{b}$ gets more favor(s) from high load operator $O_{a}$ at time slot $t+1$,

\begin{subequations} \label{eq:BeginGameCaseII}
\begin{align}
G_{b}^{t+1} & >\widehat  L_{b}^{t},\\
L_{a}^{t+1} & < \widehat G_{a}^{t}.
\end{align}
\end{subequations}

%

\noindent
In order to know the probability of occurence for the mentioned cases, we need to know how possibly $G_{a}$ or $G_{b}$ increases and $L_{a}$ or $L_{b}$ decreases in comparison to the expected gain/loss, i.e., $\delta$ (from Eq.~\eqref{eq:GainLosInit}). For this, let us compute the generalized equations for both immediate gain ($G$) and immediate loss ($L$).\\

\noindent
Let us assume, an operator gets spectrum resources from another operator\footnote{We assume complete carrier transference is equivalent to even number of favors.}, then the immediate gain can be calculated by subtracting the operator's past utility from its present utility, accordingly,

\begin{equation*}
{{G}^{t+1}}=N\log \left( \frac{{{k}^{t+1}}}{N}{{\log }_{2}}\left( 1+\gamma  \right) \right)-N\log \left( \frac{{{k}^{t}}}{N}{{\log }_{2}}\left( 1+\gamma  \right) \right),
\end{equation*}

\noindent
where ${{k}^{t}}$ is the past orthogonal carrier allocation, ${{k}^{t+1}}$ is the present orthogonal carrier allocation (s.t., ${{k}^{t+1}}>{{k}^{t}}$), $\gamma$ is the SNR of component carriers and $N$ is the load an operator. Simplifying $G$,

\begin{equation} \label{eq:IGGen}
{{G}^{t+1}}=N\log \left( \frac{{{k}^{t+1}}}{{{k}^{t}}} \right).
\end{equation}

\noindent
Similarly, the immediate loss for an operator (with ${{k}^{t+1}}<{{k}^{t}}$) can be shown as

\begin{equation} \label{eq:ILGen}
{{L}^{t+1}}=N\log \left( \frac{{{k}^{t}}}{{{k}^{t+1}}} \right).
\end{equation}

\noindent
It is observable in Eq.~\eqref{eq:IGGen} and~\eqref{eq:ILGen} that both ${{G}^{t+1}}$ and ${{L}^{t+1}}$ will be larger for high load $N$. With $N_{a} > N_{b}$, using initial orthogonal utilities $U_{o,a}$ and $U_{o,b}$ as past utilities, i.e, $k_{a}^{t}=k_{b}^{t}={K}/{2}$, $k_{a}^{t+1}=({K}/{2}) + x$ and $k_{b}^{t+1}=({K}/{2}) - x$, where $0<x<{K}/{2}$, it can be implied that,

\begin{equation} \label{eq:IGILProb}
G_{a}^{t+1}>L_{b}^{t+1}.
\end{equation}

\noindent
Even though ${k_{b}^{t}}/{k_{b}^{t+1}}$ is slightly superior than ${k_{a}^{t+1}}/{k_{a}^{t}}$, but the load's influence is far more pronounced, i.e., due to larger $N_{a}$ in comparison to $N_{b}$, $G_{a}$ gets better. A numerical example may clarify the mechanics of the presented analysis. Using simulation input parameters, $K=8$, $N_{a}=25$ and $N_{b}=5$, the transference of 1 carrier ($x=1$) from low load operator $O_{b}$ to high load operator $O_{a}$ from an initial equal orthogonal carrier allocation fetches values where $\text{log}_{10}(k_{a}^{t+1}/k_{a}^{t})=0.097$ is lower than $\text{log}_{10}(k_b^{t}/k_{b}^{t+1})= 0.125$; on the opposite, $G_{a}^{t+1}=2.423$ exceeds $L_{b}^{t+1}=0.624$.\\

\noindent
From Eq.~\eqref{eq:GainLosInit},~\eqref{eq:BeginGameCaseI} and~\eqref{eq:IGILProb}, we get,

\begin{equation*}
G_{a}^{t+1}>\delta>L_{b}^{t+1},
\end{equation*}
\noindent
and it demonstrates that \textit{Case I} is more likely to prevail over \textit{Case II}. Therefore, at the beginning of the game, the high load operator is preferred over the low load operator for spectrum usage favors, i.e., operator $O_{a}$  starts gaining carriers.

\paragraph{Game Progression}

\noindent\\\\
If, for the time, the state of load conditions remains same, i.e., operator $O_{a}$ continues to bear a high load while operator $O_{b}$ has a low load, in due course, the tendency of low load operator $O_{b}$ to rent out more and more carriers turns highly unlikely. This is due to the fact that it increases its immediate loss ${{L}^{t+1}}$ very much against expected gain $\widehat G^{t}$ (see Eq.~\eqref{eq:ILGen}, ${{\lim }_{{{k}^{t+1}}\to 0}}{{L}^{t+1}}=\infty $). As a result, the carrier transfer ceases at some point where it cannot satisfy ${{L}^{t+1}}<\widehat G^{t}$ further anymore. However, the amount of carriers transferred to the high load operator by the low load operator does not guarantee of an optimal carrier allocation. Therefore, we require a constraint in the form of a surplus which limits the carrier transference resulting in an approximate optimal carrier allocation.

\paragraph{Game: Role of Surplus}

\noindent\\\\
Let us assume, at the end of the game, operator $O_{a}$ has collected $\Delta K$ additional carriers from operator $O_{b}$ from its initial orthogonal carrier allocation. The value of $\Delta K$, for which the game fetches the approximate best sum-utility for the operators is given by the Eq.~\eqref{eq:GamePriceMaxProb} and its solution is exactly the same described by $\Delta K_{limit}$ for the same equation. By appropriately selecting surplus limit $S$\footnote{According to the definition, surplus limit is defined in terms of favors (see Section~\ref{sec:GameExpectationCons}). However, it is possible to translate the surplus limit definition in terms of carriers limit, because $S_{carriers} = 2S_{favors}$.}, we can cap the carrier transference limit to $\Delta K_{limit}$. However, $\Delta K_{limit}$ changes with temporal load variations because operators' loads $N_{a}$ and $N_{b}$ are not fixed. Besides, the game is noncooperative, and the operators are not allowed to convey their load related information to each other. Therefore, the operators at the beginning can optimize $\Delta K_{limit}$ based on load estimations (commonly observed average load, maximum load, minimum load, etc.) and decide the surplus limit parameter.\\
 
\noindent
In the simulation, we have used $K=8$ (\acp{PCC} =  2, \acp{SCC} = 6), ${{N}_{a}}=25$, ${{N}_{b}}=5$ and the users are uniformly distributed in the operator's access area. We observed, $\Delta {{K}_{limit }}=2$\footnote{See Section~\ref{sec:Algo2Analysis}, for $S$ = 4, $O_{a}$'s \acp{SCC} = 5, $O_{b}$'s \ac{SCC} = 1 and both have a single \ac{PCC}; therefore, $\Delta K_{limit}$ = $\left| {8}/{2} - (5+1)\right|$ or $\left| {8}/{2} - (1+1)\right|$.} (for surplus limit $S$=4) in a closer bound to the cooperative solution. For the same inputs, theoretically where many assumptions are regarded, we obtain $\Delta {{K}_{limit }}=2.7$ and lies close to the simulated result.

\paragraph{Game Reversal}

\noindent\\\\
In this section, we would like to show that with a change in relative load status, the game behaviour also changes, i.e., the new high load operator starts collecting spectrum usage favors from the low load operator. Here, we discuss two cases, the one in which the new low load (previously high load) operator gets more favor(s) from the other or vice versa. We will provide suggestive evidence that the operation of the first case becomes predominantly impossible, whereas the second case is plausible with a dependency on input specifications.\\

\noindent
Proceeding with the analysis, we would like to measure first the expected gain/loss for operators $O_{a}$ and $O_{b}$ till time slot $t+1$. Representing operators' previously load status as $N_{a}^{-}$ and $N_{b}^{-}$, s.t., $N_{a}^{-} > N_{b}^{-}$\footnote{\textquoteleft -\textquoteright~denotes the past load status.}, and assuming $x$ component carriers had transferred to operator $O_{a}$ by operator $O_{b}$ during time slot $t+1$, s.t., $x\le \Delta K_{limit}$, we calculate,

\begin{equation} \label{eq:GamRevMeanLG}
\widehat L_{a}^{t+1}\approx \widehat G_{b}^{t+1}\approx \delta,
\end{equation}

\begin{equation*}
\widehat G_{a}^{t+1}=\frac{h_{a} \delta +G_{a}^{t+1}}{h_{a}+1},
\end{equation*}

\begin{equation*}
\widehat L_{b}^{t+1}=\frac{h_{a} \delta +L_{b}^{t+1}}{h_{a}+1},
\end{equation*}

\noindent
where $h_{a}$ is the number of times operator $O_{a}$ had received favors from the other during the game initialization. $G_{a}^{t+1}$ and $L_{b}^{t+1}$ can be calculated using Eq.~\eqref{eq:IGGen} and~\eqref{eq:ILGen} with carrier allocation status $k_a^{t+1}={K}/{2}+x$ and $k_b^{t+1}={K}/{2} - x$ at time slot $t+1$. Therefore, $\widehat G_{a}^{t+1}$ and $\widehat L_{b}^{t+1}$ can be rewritten as

\begin{equation} \label{eq:EGt2}
\widehat G_{a}^{t+1}=\frac{h_{a} \delta+{{N}_{a}^{-}}\log \left( \frac{\frac{K}{2}+x}{\frac{K}{2}} \right)}{h_{a}+1},
\end{equation}

\begin{equation} \label{eq:ELt2}
\widehat L_{b}^{t+1}=\frac{h_{a} \delta +{{N}_{b}^{-}}\log \left( \frac{\frac{K}{2}}{\frac{K}{2}-x} \right)}{h_{a}+1}.
\end{equation}

\noindent
Let us assume at time slot $t+2$, operator $O_{b}$ is experiencing relatively high load conditions than operator $O_{a}$, i.e., $N_{a} < N_{b}$. The operators play the game and ask each other for $y$ additional carriers. Again two cases arises, either low load operator $O_{a}$ gets resources or high load operator $O_{b}$.\\

\noindent
\textit{Case I}: Low load operator $O_{a}$ receives more favor(s) from high load operator $O_{b}$ at time slot $t+2$,

\begin{subequations} \label{eq:GameRevCase1}
\begin{align}
G_{a}^{t+2} & > \widehat L_{a}^{t+1},\\
L_{b}^{t+2} & < \widehat G_{b}^{t+1}.
\end{align}
\end{subequations}

\noindent
Calculating $G_{a}^{t+2}$ and $L_{b}^{t+2}$ using Eq.~\eqref{eq:IGGen} and~\eqref{eq:ILGen}, we get,

\begin{equation} \label{eq:GamRevLowLoadImGain}
G_{a}^{t+2}={{N}_{a}}\log \left( \frac{\frac{K}{2}+x+y}{\frac{K}{2}+x} \right),
\end{equation}

\begin{equation} \label{eq:GamRevLowLoadImLoss}
L_{b}^{t+2}={{N}_{b}}\log \left( \frac{\frac{K}{2}-x}{\frac{K}{2}-x-y} \right).
\end{equation}

\noindent
Let us evalute \textit{Case I} feasibility. Using Eq.~\eqref{eq:GamRevMeanLG}, we rewrite Eq.~\eqref{eq:GameRevCase1} as

\begin{equation} \label{eq:GamRevLowLoadGen}
G_{a}^{t+2}> \delta > L_{b}^{t+2}.
\end{equation}

\noindent
Using the results of Eq.~\eqref{eq:GamRevLowLoadImGain} and~\eqref{eq:GamRevLowLoadImLoss} in Eq.~\eqref{eq:GamRevLowLoadGen}, we get,

\begin{equation} \label{eq:GamRevLowLoad}
{{\left( \frac{\frac{K}{2}+x+y}{\frac{K}{2}+x} \right)}^{{{N}_{a}}}}>{{\left( \frac{\frac{K}{2}-x}{\frac{K}{2}-x-y} \right)}^{{{N}_{b}}}}.
\end{equation}

\noindent
With $x>0$, $y>0$, $x+y<\frac{K}{2}$, and ${{N}_{a}}<{{N}_{b}}$; Eq.~\eqref{eq:GamRevLowLoad} is not satisfied and the prevalence of \textit{Case I} almost becomes impossible.\\

\noindent
\textit{Case II}: High load operator $O_{b}$ receives more favor(s) from low load operator $O_{a}$ at time slot $t+2$,

\begin{equation*}
\begin{aligned}
G_{b}^{t+2} & >\widehat L_{b}^{t+1},\\
L_{a}^{t+2} & <\widehat G_{a}^{t+1}.
\end{aligned}
\end{equation*}

\noindent
Calculating $G_{b}^{t+2}$ and $L_{a}^{t+2}$ using Eq.~\eqref{eq:IGGen} and~\eqref{eq:ILGen}, we get,

\begin{equation} \label{eq:GamRevHigLoadImGain}
G_{b}^{t+2}={{N}_{b}}\log \left( \frac{\frac{K}{2}-x+y}{\frac{K}{2}-x} \right),
\end{equation}

\begin{equation} \label{eq:GamRevHigLoadImLoss}
L_{a}^{t+2}={{N}_{a}}\log \left( \frac{\frac{K}{2}+x}{\frac{K}{2}+x-y} \right).
\end{equation}

\noindent
Comparing $G_{b}^{t+2}$ with $\widehat L_{b}^{t+2}$ (See Eq.~\eqref{eq:GamRevHigLoadImGain} and~\eqref{eq:ELt2}); there is a good amount of probability that $G_{b}^{t+2}$ can exceed $\widehat L_{b}^{t+2}$ because for some values of $N_{b}$ and $y$, it is possible to have,

\begin{equation*}
{{N}_{b}}\log \left( \frac{\frac{K}{2}-x+y}{\frac{K}{2}-x} \right) > \delta + \frac{N_{b}^{-}}{h_{a}+1}\log \left( \frac{\frac{K}{2}}{\frac{K}{2}-x} \right),
\end{equation*}

\noindent
where $N_{b}>N_{a}$, $N_{b}^{-}<N_{a}^{-}$, $0<x<K/2$, $0<y<(K/2) + x$ and $h_{a} > 0$.\\

\noindent
Similarly, there is a room of possibility that $\widehat G_{a}^{t+2}$ can exceed $L_{a}^{t+2}$, i.e.,

\begin{equation*}
\delta + \frac{N_{a}^{-}}{h_{a}+1}\log \left( \frac{\frac{K}{2}+x}{\frac{K}{2}} \right) > {{N}_{a}}\log \left( \frac{\frac{K}{2}+x}{\frac{K}{2}+x-y} \right) 
\end{equation*}

\noindent
for the same inputs.\\

\noindent
Observing the analysis, it can be inferred that the noncooperative spectrum sharing games between the competitive operators benefit the heavily loaded operator for spectrum usage favors and improve spectrum efficiency of the network operators.

\clearpage


\section{Simulation Results and Analysis} \label{chap:Simulation}

\noindent
In this chapter, several system level simulation results of the proposed schemes in Chapters~\ref{chap:GamePrice} and~\ref{chap:GameExpectation} are presented. The \ac{DSA} techniques for inter-operator spectrum sharing are analyzed and investigated against the static allocation schemes - orthogonal and full spread spectrum sharing with varying interference conditions. In addition, as a baseline for comparison, the simulation results are also compared with the Pareto optimal cooperative schemes. We have used Monte Carlo methods for the simulations.


\subsection{Simulation Scenario} \label{sec:SimSce}

\begin{figure}[H]
\begin{subfigure}{1\textwidth}
  \centering
  \includegraphics[scale=0.226, trim = 0mm 0mm 0mm 0mm, clip]{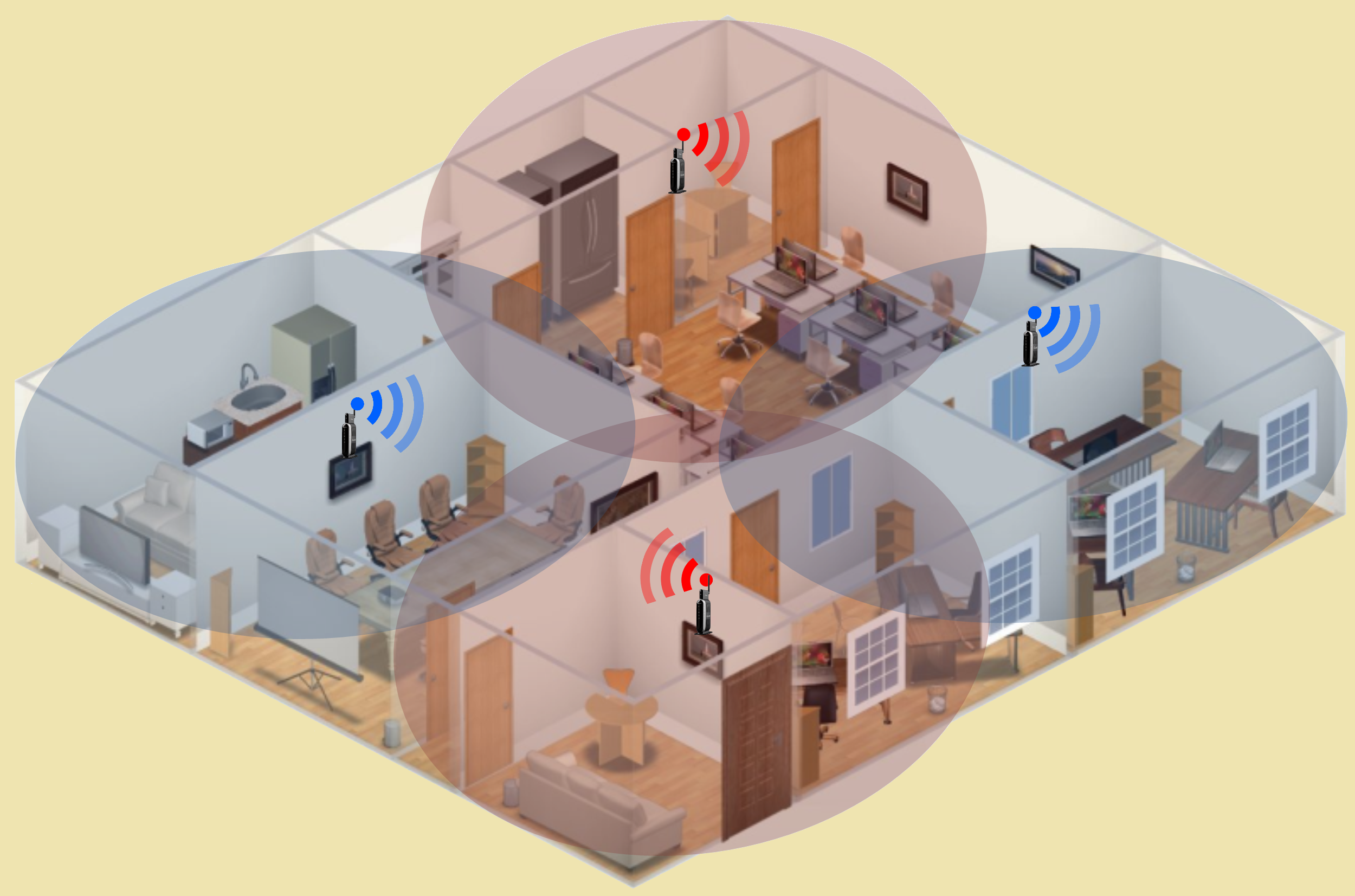}
  \caption{Multi-operator scenario in an office building}
\end{subfigure} \\
\begin{subfigure}{1\textwidth}
  \centering
  \includegraphics[scale=1, trim = 0mm 0mm 0mm 0mm, clip]{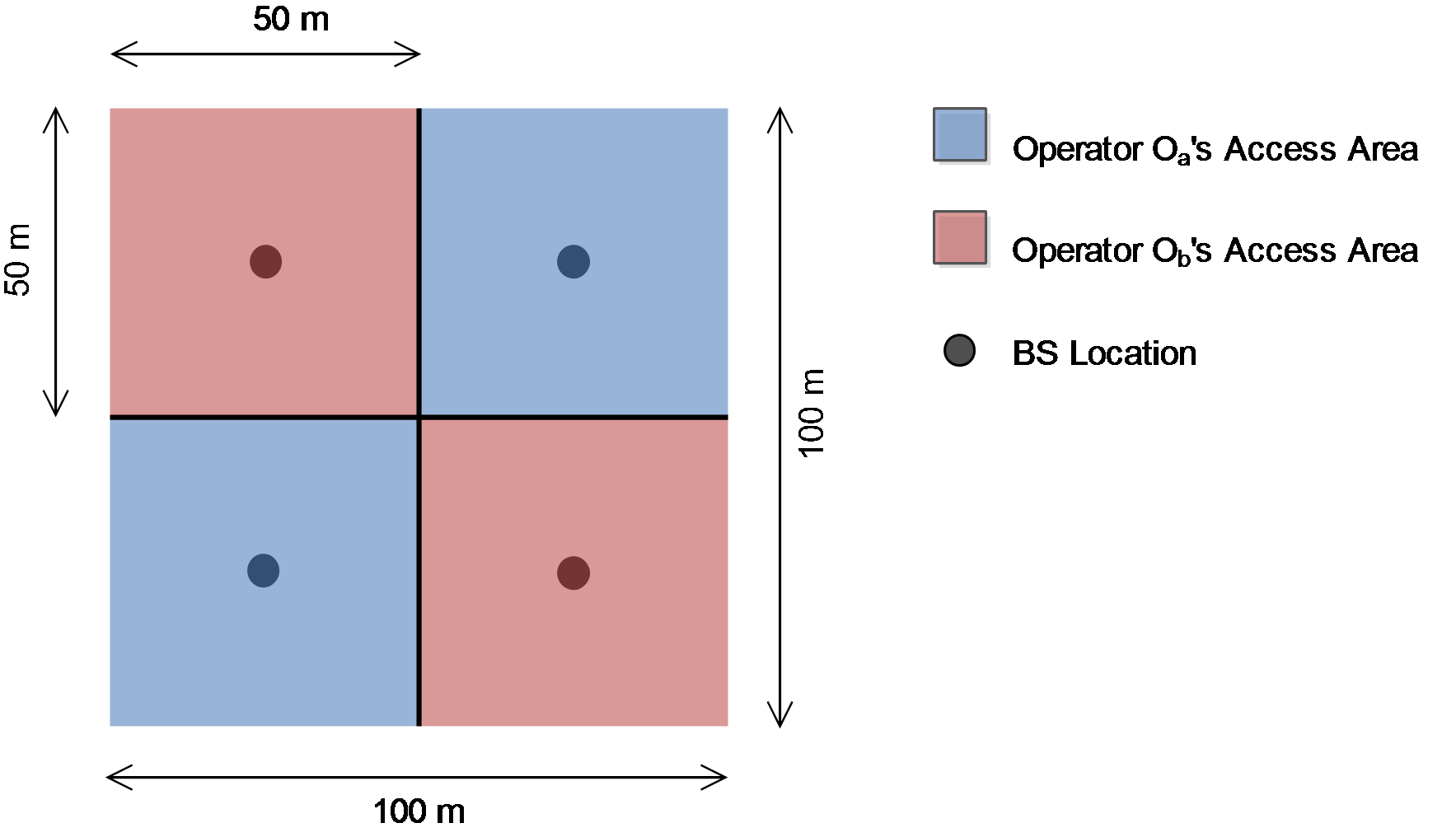}
  \caption{Single floor layout of an office building}
\end{subfigure}
\caption{Indoor inter-operator deployment scenario}
\label{fig:BuildingLayout}
\end{figure}

\noindent
We consider a small cell \ac{LTE} based network comprising of 2 operators each have 2 \acp{BS} and a Poisson distributed mean load of 25 and 5 users in their given access area. The \acp{BS} are deployed in a single storey building separated by walls and the users are uniformly distributed within the operator's access area. The \acp{BS}' locations and coverage areas are illustrated in Fig.~\ref{fig:BuildingLayout} and Tab.~\ref{tab:SimParam}.\\

\noindent
The total bandwidth of the system comprising operators $O_{a}$ and $O_{b}$ is equally divided into 8 component carriers. The centre of frequencies of the operators needs not to be adjacent. For example, if we assume that the \acp{UE} are \ac{LTE} Release-10 \acp{UE}, and then carrier aggregation can be applied to serve the \acp{UE} in the sub-band of shared frequency if the bandwidth of the operators is not contiguous. The additive white Gaussian noise (AWGN) for each component carrier is kept constant, $N_{o}$. Downlink power control is not exercised. The available power budget is divided equally among the used carriers for downlink transmissions. The transmitting \acp{BS} are full buffer in that they always have data to send. \\

\noindent
Regarding the channel modelling, the signal power attenuates according to a power law model for distance-based path loss, i.e., $C{d^{-A}}$ with path loss exponent $A$ = 3.6, attenuation constant $C$ = 1e-4 and distance $d$ between the \ac{BS} and the \ac{UE}. For the sake of simplicity, we do not consider shadow fading, or frequency selective fading, or other indoor channel models, e.g., WINNER II because the protocol's behaviour is independent of fading or attenuation characterizations. The degree of spectrum sharing depends on the inter-operator interference. Therefore, to model different interference environments, the wall attenuation between the neighbouring \acp{BS} is allowed to change. Details of the system parameters are given in Tab.~\ref{tab:SimParam}.

\begin{table}[h]
\centering
\caption{Simulation Parameters}
{\begin{tabular}{ |p{7.0cm}|p{7.0cm}|}
\hline
\multicolumn{2}{|c|}{\bf{Systeml Model}} \\
\hline
Carrier frequency & 2.6 [GHz] \\
Carrier bandwidth & 12.5 [MHz] \\
Total component carriers & 8\\
Primary component carriers (\acp{PCC}) & 2\\
Secondary component carriers (\acp{SCC}) & 6\\
BS transmit power & 30 [dBm]\\
Antenna patterns & Omni directional \\
Noise figure & 15 [dB]\\
Noise thermal power & -174 [dBm/Hz]\\
\hline
\multicolumn{2}{|c|}{\bf{Path Loss Model}} \\
\hline
Power law path loss model & $PL\text{ [dB]}=A*10{{\log }_{10}}\left( d [m] \right)+10{{\log }_{10}}\left( \frac{1}{C} \right)+W$\\
Path loss coefficients & $A = 3.6$\\
& $C=\text{1e-4}$\\
Wall attenuation ($W$) & 0 [dB] (High interference scenario)\\
& 10 [dB] (Low interference scenario)\\
\hline
\multicolumn{2}{|c|}{\bf{Scenario Model}} \\
\hline
Number of operators & 2\\
Number of BSs/operator & 2\\
Number of buildings & 1\\
Number of floors/building & 1\\
Number of rooms/floor & 4\\
Number of BSs/room & 1\\
\hline
\multicolumn{2}{|c|}{\bf{Traffic Model}} \\
\hline
Number of UEs/operator & Poisson distributed load with mean 25 or 5\\
UEs distribution & Uniformly distributed\\
\hline
\multicolumn{2}{|c|}{\bf{Link Level Model}} \\
\hline
Spectral efficiency & $r = \text{BW}_{\text{eff}}*\text{BW}*\text{log}_{2}\left(1+{\text{SINR}} \right)$ \\
Bandwidth efficiency & $\text{BW}_{\text{eff}} = .56$\\
\hline
\multicolumn{2}{|c|}{\bf{Algorithm Parameters}} \\
\hline
Maximum outstanding favors or surplus ($S$) & 2, 4\\
\hline
\end{tabular}}
\label{tab:SimParam}
\end{table}


\subsection{Performance Evaluation} \label{sec:PerfEval}

\noindent
The results are presented for 1000 random network instantiations, generated according to the aforementioned parameters. Scheduling weight per carrier $w_{i,j,k}$ (Eq.~\eqref{eq:GamPricThroughput}) is fixed and inversely proportional to their \ac{BS}'s load. The \acp{BS} within an operator use the same component carriers, however the \acp{BS} of different operators can have different carrier allocations. For each deployment, the repeated games are allowed to run for 30 counters. The operators' favors (refer Section~\ref{sec:GamePriceFavors},~\ref{sec:GameExpectationCons}) and gains/losses (only in the case of Algorithm II, refer Section~\ref{sec:GameHisDecis}) are recorded at the end of the game sequences and fed to the next deployment. \\

\noindent
The data rates are experienced by the individual users tracked after each deployment and collected over 1000 deployments. The histogram is used to generate the user rate probability distribution over all realizations, which then used to plot the user rate \acp{CDF}. The user rate \acp{CDF} are plotted for operators $O_{a}$ and $O_{b}$ with respective Poisson distributed mean load of 25 and 5 users. Load reversal cases also been considered to assess the performance with varying load conditions. The user rate \acp{CDF} are plotted for 2000 instantiations, where in the halfway of the simulation, the loads are reversed with same Poisson distributed mean load, i.e., for the first 1000 deployments, the loads of operators $O_{a}$ and $O_{b}$ are 25 and 5 users respectively, and during the latter half, the respective loads are 5 and 25 users. The effect of temporal load variations depicted in the simulations defines the practicability of the scenarios. Besides, the effect of surplus $S$ has also been analyzed for the mentioned plots. Two cases with different interference conditions are taken into consideration -

\begin{enumerate}
\item High interference scenario (with wall loss of 0 dB),
\item Low interference scenario (with wall loss of 10 dB).
\end{enumerate}

\noindent
Henceforth, we present the analysis of both algorithms achieving about the same outcomes.


\subsubsection{Algorithm I Analysis} \label{sec:Algo1Analysis}

\begin{figure}[H]
\centering
\includegraphics[scale=.73, trim = 1mm 0.6mm 9mm 6.5mm, clip]{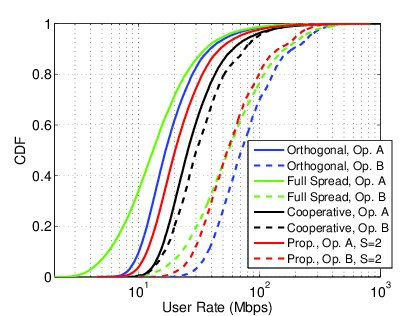}
\caption{Rate distribution for operator $O_{a}$, $N_{a} = 25$ users and operator $O_{b}$, $N_{b} = 5$ users using the traditional orthogonal and full spread spectrum sharing, cooperative algorithm and proposed scheme based on Algorithm I, in a high interference environment (wall loss, 0 dB). The maximum number of outstanding favors, $S = 2$.}
\label{fig:Algo1HighIntfSur2}
\end{figure}

\begin{figure}[H]
\centering
\includegraphics[scale=.73, trim = 1mm 0.6mm 9mm 6.5mm, clip]{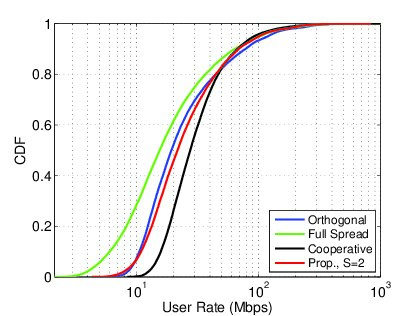}
\caption{Rate distribution for operator $O_{a}$ with temporal load variations using the traditional orthogonal and full spread spectrum sharing, cooperative algorithm and proposed scheme based on Algorithm I, in a high interference environment (wall loss, 0 dB). The maximum number of outstanding favors, $S = 2$.}
\label{fig:Algo1HighIntfSur2LoadRev}
\end{figure}

The Algorithm I considers a carrier pricing based utility function for the repeated games framework. The operators pay the penalty on their carriers usage which forces them to share the bandwidth resources according to their relative needs. In the simulation, the pricing constants $p_1$ and $p_2$ are set as 7 and 0.8 respectively according to the optimization criteria discussed in Section~\ref{sec:GamPriceOpt}. Fig.~\ref{fig:Algo1HighIntfSur2} shows the user rate \acp{CDF} for two operators $O_{a}$ and $O_{b}$ in a high interference scenario (wall loss, 0 dB). The maximum number of outstanding favors is set equal to, $S = 2$. It can be seen, with dynamic spectrum sharing, high load operator $O_{a}$ becomes able to improve its delivered throughput in comparison with the orthogonal static allocation. On the other hand, low load operator $O_{b}$'s throughput falls, but operators do not mind sacrificing their resources during low load conditions if they are anticipating benefits in the demanding circumstances. This behaviour is captured in Fig.~\ref{fig:Algo1HighIntfSur2LoadRev} when the load gets reversed. The figure shows the user rate distribution for operator $O_{a}$ spanning over the temporal load variations be fitted with its initial high load (of 25 users), and latter low load (of 5 users) instances. In the figure, it is observable that operator $O_{a}$'s delivered throughput has improved over time in comparison to the orthogonal sharing even though it has sacrificed the spectrum resources latter when the load was sparse. The similar behaviour has been noticed for operator $O_{b}$ during the simulation because, on an average, the loads are same over time for the both operators in the load reversal scenario.

\begin{figure}[H]
\centering
\includegraphics[scale=.73, trim = 1mm 0.6mm 9mm 6.5mm, clip]{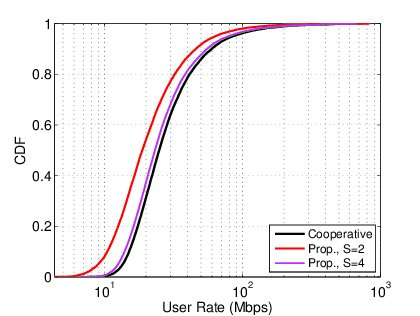}
\caption{Rate distribution for operator $O_{a}$, $N_{a} = 25$ users using the cooperative algorithm and proposed scheme based on Algorithm I, in a high interference environment (wall loss, 0 dB). The maximum number of outstanding favors is varied and rate curves are analysed for, $S = 2$ and 4.}
\label{fig:Algo1HighIntfSur2Sur4}
\end{figure}

\noindent
In the game, surplus parameter plays a crucial role in controlling the trading of spectrum resources. It puts a limit on the number of spectrum usage favors given by the operators to each other as it ensures that the sacrificing operator retains an adequate amount of spectrum for its smoother operation while dynamically sharing the spectrum. In Fig.~\ref{fig:Algo1HighIntfSur2Sur4}, the effect of surplus is analyzed on the user rate distributions for high load operator $O_{a}$. In the simulation, the average carrier utilizations in \acp{SCC} (6 carriers) using a surplus limit of 2 are observed as 4.02 for high load operator $O_{a}$ and 1.99 for low load operator $O_{b}$, whereas with surplus limit 4, the average carrier utilizations are 4.97 and 1.03 respectively. It indicates that with easing off in surplus limit value, the operators' exploitation of the degree of freedom in frequency domain increases and the throughput gain approaches the efficient cooperative solution. Though, operators have \acp{SCC} of 6 carriers, which means, the maximal surplus limit can be fixed at 6. Nevertheless, it has been observed that the large surplus limit value becomes redundant once it reaches the approximate cooperative solution, i.e., surplus 4 to 6 almost give the same performance. The reason is that the carrier pricing component in the utility function keeps the carrier allocations (of both operators) optimal if the surplus limit is larger than the optimal carrier allocation (cooperative carrier allocation) of the high load operator. However,  in the case of high load operator, its carrier allocation reduces to the surplus limit if the surplus limit is lower than its optimal carrier allocation.

\begin{figure}[H]
\centering
\includegraphics[scale=.73, trim = 1mm 0.6mm 9mm 6.5mm, clip]{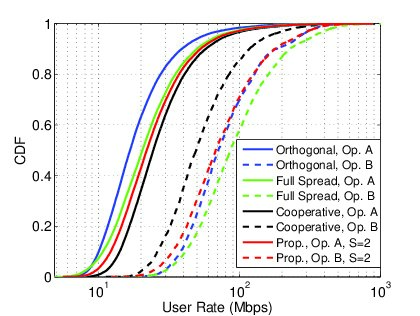}
\caption{Rate distribution for operator $O_{a}$, $N_{a} = 25$ users and operator $O_{b}$, $N_{b} = 5$ users using the traditional orthogonal and full spread spectrum sharing, cooperative algorithm and proposed scheme based on Algorithm I, in a low interference environment (wall loss, 10 dB). The maximum number of outstanding favors, $S = 2$.}
\label{fig:Algo1LowIntfSur2}
\end{figure}

\begin{figure}[H]
\centering
\includegraphics[scale=.73, trim = 1mm 0.6mm 9mm 6.5mm, clip]{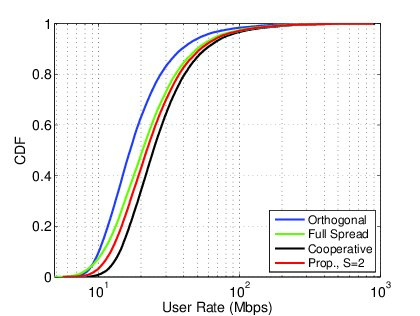}
\caption{Rate distribution for operator $O_{a}$ with temporal load variations using the traditional orthogonal and full spread spectrum sharing, cooperative algorithm and proposed scheme based on Algorithm I, in a low interference environment (wall loss, 10 dB). The maximum number of outstanding favors, $S = 2$.}
\label{fig:Algo1LowIntfSur2LoadRev}
\end{figure}

\begin{figure}[H]
\centering
\includegraphics[scale=.73, trim = 1mm 0.6mm 9mm 6.5mm, clip]{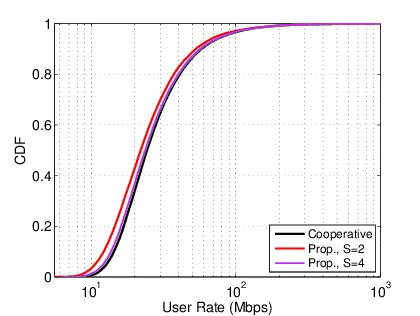}
\caption{Rate distribution for operator $O_{a}$, $N_{a} = 25$ users using the cooperative algorithm and proposed scheme based on Algorithm I, in a low interference environment (wall loss, 10 dB). The maximum number of outstanding favors is varied and rate curves are analysed for, $S = 2$ and 4.}
\label{fig:Algo1LowIntfSur2Sur4}
\end{figure}

\noindent
The algorithm's efficiency is also tested in a low interference environment, where interference is suppressed by increasing the wall attenuation to 10 dB between the \acp{BS}.  The analysis is quite much same as it is discussed for the above case of high interference scenario. The only difference lies here is that now operators have more degree of overlapped carriers. Describing briefly, Fig.~\ref{fig:Algo1LowIntfSur2} shows the user rate distributions for operators $O_{a}$ and $O_{b}$. In the figure, high load operator $O_{a}$ gathers more spectrum resources than low load operator $O_{b}$ according to their load conditions. Similarly, Fig.~\ref{fig:Algo1LowIntfSur2LoadRev} shows the user rate distribution for operator $O_{a}$ over time when it had a high load initially and later a low load. The figure shows that the operators are able to improve their throughput over time through dynamically sharing the spectrum in comparison to the static allocations. The behaviour  of surplus is captured in Fig.~\ref{fig:Algo1LowIntfSur2Sur4}. In the simulation, the average carrier utilizations in \acp{SCC} with surplus limit 2 are observed as 6.0 for high load operator $O_{a}$ and 3.99 for low load operator $O_{b}$, and with surplus limit 4, the average carrier utilizations are 6.0 and 2.14 respectively.

\subsubsection{Algorithm II Analysis} \label{sec:Algo2Analysis}

\noindent
Algorithm II tries to attain the same objective what Algorithm I does. The Algorithm II considers mutual gain/loss history by which the operators play repeated games to share the spectrum dynamically. The working of the algorithm requires an initialization, and for this we have initialized both operators with a load of a single user in the simulation, and let it run for around 100 instants. In the performance curves, we have considered PF based user rates. The analysis of results is similar to what we have presented for Algorithm I in Section~\ref{sec:Algo1Analysis}.\\

\begin{figure}[H]
\centering
\includegraphics[scale=.73, trim = 1mm 0.6mm 9mm 6.5mm, clip]{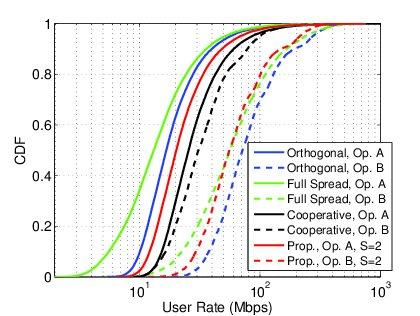}
\caption{Rate distribution for operator $O_{a}$, $N_{a} = 25$ users and operator $O_{b}$, $N_{b} = 5$ users using the traditional orthogonal and full spread spectrum sharing, cooperative algorithm and proposed scheme based on Algorithm II, in a high interference environment (wall loss, 0 dB). The maximum number of outstanding favors $S = 2$.}
\label{fig:Algo2HighIntfSur2}
\end{figure}

\begin{figure}[H]
\centering
\includegraphics[scale=.73, trim = 1mm 0.6mm 9mm 6.5mm, clip]{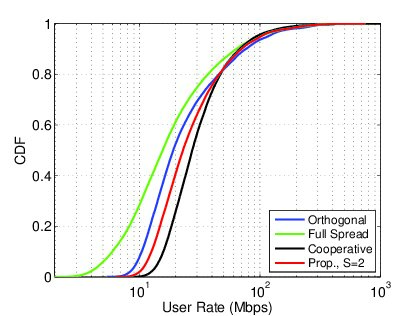}
\caption{Rate distribution for operator $O_{a}$ with temporal load variations using the traditional orthogonal and full spread spectrum sharing, cooperative algorithm and proposed scheme based on Algorithm II, in a high interference environment (wall loss, 0 dB). The maximum number of outstanding favors $S = 2$.}
\label{fig:Algo2HighIntfSur2LoadRev}
\end{figure}

\noindent
Simulation result in Fig.~\ref{fig:Algo2HighIntfSur2} shows the user rate \acp{CDF} of operators $O_{a}$ and $O_{b}$ with respective mean load of 25 and 5 users. It is quite visible from the figure, that high load operator $O_{a}$ has become able to improve its delivered throughput at the expense of low load operator $O_{b}$. Though, operator $O_{b}$ suffers at the moment, but when its load gets high in the near future, it will be getting its rightful share of the spectrum resources. This behaviour  is captured in Fig.~\ref{fig:Algo2HighIntfSur2LoadRev}, where loads get reversed after some time, i.e., now operator $O_{a}$ has a mean load of 5 users while operator $O_{b}$ has a mean load of 25 users. The user rate curves are plotted for operator $O_{a}$ over the time span when it had a mean load of 25 and latter of 5 users. The game is modelled in a high interference scenario (wall loss, 0 dB), and the plotted rate curves affirm that the game based \ac{DSA} scheme provides a clear benefit over the orthogonal sharing with asymmetric loading and improves throughput with time.

\begin{figure}[H]
\centering
\includegraphics[scale=.73, trim = 1mm 0.6mm 9mm 6.5mm, clip]{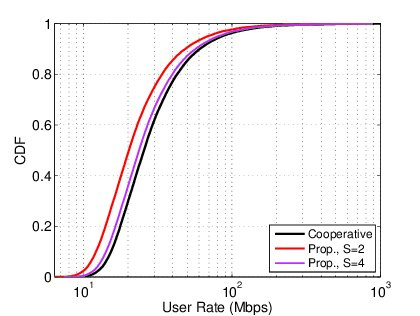}
\caption{Rate distribution for operator $O_{a}$, $N_{a} = 25$ users using the cooperative algorithm and proposed scheme based on Algorithm II, in a high interference environment (wall loss, 0 dB). The maximum number of outstanding favors $S$ is varied and rate curves are analysed for, $S = 2$ and 4.}
\label{fig:Algo2HighIntfSur2Sur4}
\end{figure}

\noindent
Apart from that, in the game, the maximum limit for outstanding favors (surplus limit $S$) is kept at 2, which means neither of the operators can trade spectrum resources more than the limit while dynamically sharing the spectrum. In Fig.~\ref{fig:Algo2HighIntfSur2Sur4}, the surplus limit behaviour on the game is captured. It can be seen that with an increase in the surplus limit, the high load operator's throughput improvement steadily approaches the cooperative solution. In the simulation, the average carrier utilizations in \acp{SCC} (6 carriers) using a surplus limit of 2 are observed as 4.09 for high load operator $O_{a}$ and 2.0 for low load operator $O_{b}$, whereas with surplus limit 4, the average carrier utilizations are 5.0 and 1.0 respectively. It indicates that with easing off in the surplus limit value, the operators' tendency for sharing the spectrum increases and become more pronounced. Though, the operators have \ac{SCC} of 6 carriers, which means, the maximal surplus limit can be fixed at 6. However, with a larger surplus limit (e.g., here, if more than 4), high load operator $O_{a}$'s gain surpasses the cooperative solution. It indicates that the loss of low load operator $O_{b}$ will be more than the gain of high load operator $O_{a}$, and the sum-throughput of operators will diminish. Therefore, it is essential to optimize the surplus limit parameter to have a maximal benefit, which has been discussed in detail, in Section~\ref{sec:GameExpectMath}.\\

\begin{figure}[H]
\centering
\includegraphics[scale=.73, trim = 1mm 0.6mm 9mm 6.5mm, clip]{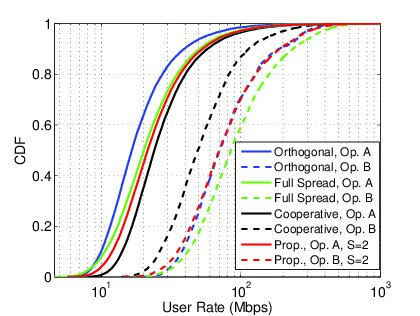}
\caption{Rate distribution for operator $O_{a}$, $N_{a} = 25$ users and operator $O_{b}$, $N_{b} = 5$ users using the traditional orthogonal and full spread spectrum sharing, cooperative algorithm and proposed scheme based on Algorithm II, in a low interference environment (wall loss, 10 dB). The maximum number of outstanding favors $S = 2$.}
\label{fig:Algo2LowIntfSur2}
\end{figure}

\begin{figure}[H]
\centering
\includegraphics[scale=.73, trim = 1mm 0.6mm 9mm 6.5mm, clip]{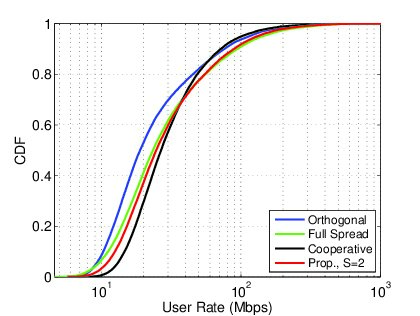}
\caption{Rate distribution for operator $O_{a}$ with temporal load variations using the traditional orthogonal and full spread spectrum sharing, cooperative algorithm and proposed scheme based on Algorithm II, in a low interference environment (wall loss, 10 dB). The maximum number of outstanding favors $S = 2$.}
\label{fig:Algo2LowIntfSur2LoadRev}
\end{figure}

\begin{figure}[H]
\centering
\includegraphics[scale=.73, trim = 1mm 0.6mm 9mm 6.5mm, clip]{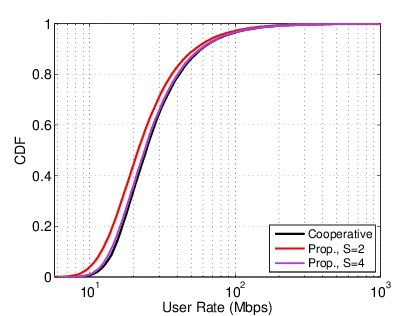}
\caption{Rate distribution for operator $O_{a}$, $N_{a} = 25$ users using the cooperative algorithm and proposed scheme based on Algorithm II, in a low interference environment (wall loss, 10 dB). The maximum number of outstanding favors $S$ is varied and rate curves are analysed for, $S = 2$ and 4.}
\label{fig:Algo2LowIntfSur2Sur4}
\end{figure}

\noindent
Similarly, the rate curves have also been plotted for a low interference environment (with wall loss of 10 dB). In this, the full spread is the dominant static allocation. According to Fig.~\ref{fig:Algo2LowIntfSur2} and~\ref{fig:Algo2LowIntfSur2LoadRev}, it is observable that, with the game based spectrum sharing, the operators' spectrum allocation are now more closely aligned to the full spread rather than the orthogonal. Besides, the rate curves are observed better than the full spread. The reason is that the interference in this scenario is suppressed, but not completely eliminated. However, it has been observed that with a wall loss of over 20 dB, both full spread and game curves converge, and the operators utilize the full spectrum with negligible interference. The surplus behaviour  also been captured in Fig.~\ref{fig:Algo2LowIntfSur2Sur4}, which shows that, with an increase in the surplus limit, the spectrum sharing improves and reaches the cooperative solution for some limit (of 4, as observed in the figure). In the simulation, the average carrier utilizations in \acp{SCC} with surplus limit 2 are observed as 5.9 for high load operator $O_{a}$ and 4.3 for low load operator $O_{b}$, and with surplus limit 4, the average carrier utilizations are 5.9 and 2.3 respectively. With a further increase in the surplus limit parameter, the overall sum-through starts declining and deviating from the optimal gain, same as it is discussed for the high interference scenario.

\clearpage

 
\section{Conclusion and Future Work}

 
\noindent
This thesis reports the findings, whose main objective is to demonstrate how sharing paradigms in wireless networks, in particular spectrum sharing improves the spectral efficiency. We describe and evaluate different scenarios where spectrum sharing is considered in the context of multi-operator cooperation, namely full spread or orthogonal sharing. More specifically, we investigate the impact of noncooperative games between the operators on spectrum sharing. Our numerical results show that properly modelled games may provide a gain in terms of system-level throughput, with respect to full spread and orthogonal spectrum sharing scenarios over the time. The performance of the specific techniques is strongly dependent on several system parameters, such as the number of users, QoS and the rational outlook of the operators. More importantly, the gains are significant when the number of serving users by an operator is relatively large, and the BSs have enough degrees of freedom to efficiently schedule the users.

\subsection{Summary}

\noindent
In this thesis, we have investigated the inter-operator spectrum sharing problem between the self-interested operators. Operators exist in nearby geographical area with neighbouring \acp{RAN}. The problem is modelled via game theoretic approach for an efficient \ac{DSA}. The spectrum resources amongst the operators divided into, (i) \ac{FSA} and (ii) \ac{DSA}. In \ac{FSA}, privately owned orthogonal frequency bands are allocated to the operators where no inter-operator interference exists. Whereas in \ac{DSA}, the operators contend for resources from a common spectrum pool. The operators follow the repeated games framework and devise strategies to fetch spectrum resources based on their requirements (e.g., load congestion, \ac{QoS}, etc.). The games are played entirely on a noncooperative basis as no operational information is revealed to the other. Leveraging this analysis, two different taxonomies of the noncooperative \ac{DSA} algorithm have been proposed.\\

\noindent
Chapter~\ref{chap:GamePrice} discusses the first algorithm, where a carrier pricing based utility function has been considered for the repeated games framework. The utility design penalizes the operators when it comes to their spectrum usage. In the games, the operators aim to maximize their utility at every game sequence. This leads to sharing of spectrum resources between the operators based on their spectrum affordability, eventually favoring the congested operators.\\

\noindent
In Chapter~\ref{chap:GameExpectation}, another coordination algorithm is introduced, where mutual interactions between the operators is recognized as the basis for resource sharing. In this, the operators translate their past gains/losses due to spectrum sharing over the previous games into their future benefits. The operators play noncooperative repeated games, and if they expect promising future gains, they readily sacrifice their resources upon requests and accordingly trade resources.\\

\noindent
To curb the favoritism towards operators(s) in collecting spectrum usage favors in the algorithms, a limit has been imposed in the form of surplus. Surplus establishes a trust mechanism and ensures that the operators sacrifice resources for each other as long as the other was helpful in the past. Setting up this parameter requires an appropriate measure, as too small value does not let the game run effectively or too much of a relaxation in the value might distinctly favor the operators.\\

\noindent
For the purpose of performance analysis, a scenario has been considered, comprising of two operators with their neighboring \acp{RAN} in a single storey building. Each operator has two \acp{BS}, and all the \acp{BS} are geographically separated by the walls. The operator's loads are Poisson distributed, but their locations are uniformly distributed within the operator's access area. A spectrum band of 8 carriers is available to the operators, which is partitioned exclusively into two allocations - \ac{FSA}, and \ac{DSA}. In \ac{FSA}, both operators are allocated a single orthogonal carrier whereas in \ac{DSA}, a spectrum pool of 6 carriers is shared. As a baseline for comparison, the benefits of the algorithms are assessed against the static allocations - orthogonal and full spread, and Pareto efficient cooperative algorithm (reference Chapter~\ref{chap:Coop}).\\

\noindent
In the simulation results, the performance of algorithms achieving dynamic spectrum sharing outperforms the typical static allocation schemes (orthogonal or full spread) under varying interference conditions or load factors. The cooperative algorithm serves best in the scenario. However, such choice is disregarded by the self-interested operators because of the trust issues and significant overhead constraints. Though, it categorically provides an optimal benchmark solution for the study of game algorithms. The operators opportunistically share resources and maximize their sum-throughput noncooperatively with an effort to converge to the ideal cooperative solutions.

 
\subsection{Future Work}

\noindent
We do not expect this work to be complete and there are several areas where the research done in this thesis could be expanded. 

\begin{enumerate}

\item In the simulations, only two operators each have two \acp{BS}, have been considered. Extending this work many \acp{BS} could lead to interesting findings in terms of interference management. One also can implement the cooperative schemes for intra-operator radio resource management. However, this makes the problem of scheduling extremely complicated and requires some implementation to speed up the scheduling of users in the network.

\item The proposed Algorithm II, based on mutual history of gains/losses, could add many facets to its decision making mechanism. For instance, the operators could incorporate outstanding favors in the decision making process rather than using it alone for a hard check. Also, the operators can categorize favors into big and small favors, and formulate policies in granting them, e.g., resorting to leniency in granting small favors.

\item The operators consider the current traffic load only without anticipating the future. The algorithm has a room to be equipped with accurate load modelling. This could save the operators from unnecessary processing time and energy consumption in forwarding the requests and subsequent decision making. With accurate load modelling, the operators convey requests when it is right to do and thus, can make accurate reservations of spectrum resources proactively.

\end{enumerate}

\clearpage
\phantomsection







\addcontentsline{toc}{section}{References}

\begin{thebibliography}{10}
\providecommand{\url}[1]{#1}

\bibitem{WP11NSN}
{Nokia Siemens Networks}, ``Wake-up call: Industry collaboration needed to make
  beyond {4G} networks carry 1000 times more traffic by 2020,'' White Paper,
  Aug. 2011. [Online]. Available:
  \url{http://blogs.nokiasiemensnetworks.com/news/2011/08/24/beyond-4g-networks/}

\bibitem{WP13Huawei}
Huawei, ``The second phase of {LTE-A}dvanced ({LTE-B}): 30-fold capacity
  boosting to {LTE},'' White Paper, Feb. 2013. [Online]. Available:
  \url{http://www.huawei.com/ilink/en/download/HW_259010}

\bibitem{WP11Ericsson}
Ericsson, ``More than 50 billion connected devices,'' White Paper, Feb. 2011.
  [Online]. Available:
  \url{http://www.ericsson.com/res/docs/whitepapers/wp-50-billions.pdf}

\bibitem{WP13Ericsson}
------, ``{LTE} release 12 - taking another step toward the networked
  society,'' White Paper, Jan. 2013. [Online]. Available:
  \url{http://www.ericsson.com/res/docs/whitepapers/wp-lte-release-12.pdf}

\bibitem{RP12FCC}
{Federal Communications Commission (FCC)}, ``Mobile broadband: The benefits of
  additional spectrum,'' Tech. Rep., Oct. 2010.

\bibitem{RP11NTIA}
{National Telecommunications and Information Administration (NTIA)}, ``{U}nited
  {S}tates frequency allocation chart,'' Tech. Rep., 2011. [Online]. Available:
  \url{http://www.ntia.doc.gov/page/2011/united-states-frequency-allocation-chart}

\bibitem{RP06ITU}
{International Telecommunication Union (ITU)}, ``Estimated spectrum bandwidth
  requirements for the future development of {IMT}-2000 and {IMT-A}dvanced,''
  Tech. Rep. M.2078, 2006.

\bibitem{RP02FCC}
{Federal Communications Commission (FCC)}, ``First report and order in the
  matter of revision of part 15 of the commission's rules regarding ultra
  wideband transmission systems,'' FCC 02-48 ET Docket 98-153, Feb. 2002.

\bibitem{CP04Cabric}
D.~\v{C}abri\'{c}, S.~M. Mishra, and R.~W. Brodersen, ``Implementation issues
  in spectrum sensing for cognitive radios,'' in \emph{Proc. Asilomar
  Conference on Signals, Systems and Computers ({ASILOMAR}'04)}, Nov. 2004, pp.
  772--776.

\bibitem{CP10Valenta}
V.~Valenta, R.~Mar\v{s}\'{a}lek, G.~Baudoin, M.~Villegas, M.~Suarez, and
  F.~Robert, ``Survey on spectrum utilization in {E}urope: Measurements,
  analyses and observations,'' in \emph{Proc. {IEEE} International Conference
  on Cognitive Radio Oriented Wireless Networks ({CROWNCOM}'10)}, Jun. 2010,
  pp. 1--5.

\bibitem{RP03McHenry}
M.~McHenry, ``Spectrum white space measurements,'' New America Foundation
  BroadBand Forum, jun 2003. [Online]. Available:
  \url{http://www.ericsson.com/res/docs/whitepapers/wp-50-billions.pdf}

\bibitem{BK03Osborne}
M.~J. Osborne, \emph{An Introduction to Game Theory}.\hskip 1em plus 0.5em
  minus 0.4em\relax Oxford: Oxford University Press, 2003.

\bibitem{CP12Peltomaki}
M.~Peltomaki, J.~Koljonen, O.~Tirkkonen, and M.~Alava, ``Algorithms for
  self-organized resource allocation in wireless networks,'' in \emph{Proc.
  {IEEE} Vehicular Technology Conference ({VTC}'12 Spring)}, vol.~61, no.~1,
  Jan. 2012, pp. 346--359.

\bibitem{JR97Kelly}
F.~Kelly, ``Charging and rate control for elastic traffic,'' \emph{European
  Trans. Telecomm.}, vol.~8, pp. 33--37, 1997.

\bibitem{JR03Sung}
C.~W. Sung and W.~S. Wong, ``A noncooperative power control game for multirate
  {CDMA} data networks,'' \emph{{IEEE} Trans. Wireless Commun.}, vol.~2, no.~1,
  pp. 186--194, Jan. 2003.

\bibitem{JR08Wu}
K.~D. Wu and W.~Liao, ``Flow allocation in multi-hop wireless networks: A
  cross-layer approach,'' \emph{{IEEE} Trans. Wireless Commun.}, vol.~7, no.~1,
  pp. 269--276, Jan. 2008.

\bibitem{CP02Cao}
Y.~Cao and V.~O.~K. Li, ``Utility-oriented adaptive qos and bandwidth
  allocation in wireless networks,'' in \emph{Proc. {IEEE} International
  Conference on Communications ({ICC}'02)}, vol.~5, 2002, pp. 3071--3075.

\bibitem{CH05Navaie}
K.~Navaie, D.~Montuno, and Y.~Q. Zhao, \emph{Resource Allocation in Next
  Generation Wireless Networks}.\hskip 1em plus 0.5em minus 0.4em\relax New
  York: Nova Science Publishers, 2005, ch. Fairness of resource allocation in
  cellular networks: A Survey.

\bibitem{CP01Gao}
X.~Gao, T.~Nandagopal, and V.~Bharghavan, ``Achieving application level
  fairness through utility-based wireless fair scheduling,'' in \emph{Proc.
  {IEEE} Global Communications Conference ({GLOBECOM}'01)}, vol.~6, 2001, pp.
  3257--3261.

\bibitem{CP04Liu}
P.~Liu, R.~Berry, and M.~L. Honig, ``A fluid analysis of utility-based wireless
  scheduling policies,'' in \emph{Proc. {IEEE} Conference on Decision and
  Control ({CDC}'04)}, vol.~3, 2004, pp. 3283--3288.

\bibitem{CP02Siris}
V.~A. Siris, B.~Briscoe, and D.~Songhurst, ``Economic models for resource
  control in wireless networks,'' in \emph{Proc. {IEEE} International Symposium
  on Personal, Indoor and Mobile Radio Communications ({PIMRC}'02)}, vol.~3,
  2002, pp. 1112--1116.

\bibitem{CP02Marbach}
P.~Marbach and R.~Berry, ``Downlink resource allocation and pricing for
  wireless networks,'' in \emph{Proc. {IEEE} International Conference on
  Computer Communications ({INFOCOM}'02)}, vol.~3, 2002, pp. 1470--1479.

\bibitem{CP02Liu}
P.~Liu, R.~Berry, M.~L. Honig, and S.~Jordan, ``Slow-rate utility-based
  resource allocation in wireless networks,'' in \emph{Proc. {IEEE} Global
  Communications Conference ({GLOBECOM}'02)}, vol.~1, 2002, pp. 799--803.

\bibitem{JR00Bianchi}
G.~Bianchi and A.~T. Campbell, ``A programmable {MAC} framework for
  utility-based adaptive quality of service support,'' \emph{{IEEE} J. Sel.
  Areas Commun.}, vol.~18, no.~2, pp. 244--255, Feb. 2000.

\bibitem{JR01Liao}
R.~R.~F. Liao and A.~T. Campbell, ``A utility-based approach for quantitative
  adaptation in wireless packet networks,'' \emph{Springer Wireless Netw.},
  vol.~7, no.~5, pp. 541--557, Sep. 2001.

\bibitem{JR02Saraydar}
C.~U. Saraydar, N.~B. Mandayam, and D.~Goodman, ``Efficient power control via
  pricing in wireless data networks,'' \emph{{IEEE} Trans. Commun.}, vol.~50,
  no.~2, pp. 291--303, Feb. 2002.

\bibitem{JR07Huang}
W.~L. Huang and K.~B. Letaief, ``Cross-layer scheduling and power control
  combined with adaptive modulation for wireless ad hoc networks,''
  \emph{{IEEE} Trans. Commun.}, vol.~55, no.~4, pp. 728--739, Apr. 2007.

\bibitem{JR03Xiao}
M.~Xiao, N.~B. Shroff, and E.~K.~P. Chong, ``A utility-based power-control
  scheme in wireless cellular systems,'' \emph{{IEEE/ACM} Trans. Netw.},
  vol.~11, no.~2, pp. 210--221, Apr. 2003.

\bibitem{BK97Keshav}
S.~Keshav, \emph{An engineering approach to computer networking}.\hskip 1em
  plus 0.5em minus 0.4em\relax Reading, MA: Addison-Wesley, 1997.

\bibitem{JR06HuangACM}
J.~Huang, R.~A. Berry, and M.~L. Honig, ``Auction-based spectrum sharing,''
  \emph{Springer Mobile Netw. Applicat.}, vol.~11, no.~3, pp. 405--418, Jun.
  2006.

\bibitem{JR98Kelly}
F.~P. Kelly, A.~K. Maulloo, and D.~K.~H. Tana, ``Rate control for communication
  networks: Shadow prices, proportional fairness and stability,'' \emph{J.
  Operational Research Soc.}, vol.~49, no.~3, pp. 237--252, 1998.

\bibitem{JR00Mo}
J.~Mo and J.~Walrand, ``Fair end-to-end window-based congestion control,''
  \emph{{IEEE/ACM} Trans. Netw.}, vol.~8, no.~5, pp. 556--567, 2000.

\bibitem{JR81Axelrod}
R.~Axelrod and W.~D. Hamilton, ``The evolution of cooperation,'' \emph{{AAAS}
  Sci.}, vol. 211, no. 4489, pp. 1390--1396, Mar. 1981.

\bibitem{BK07Axelrod}
R.~M. Axelrod, \emph{The Evolution of Cooperation}.\hskip 1em plus 0.5em minus
  0.4em\relax New York, NY: Basic Books, 1984.

\bibitem{JR09Garcia}
L.~G.~U. Garcia, K.~I. Pedersen, and P.~E. Mogensen, ``Autonomous component
  carrier selection: interference management in local area environments for
  {LTE-A}dvanced,'' \emph{{IEEE} Commun. Mag.}, vol.~47, no.~9, pp. 110--116,
  Sep. 2009.

\bibitem{CP11Prasad}
A.~Prasad, K.~Doppler, M.~Moisio, K.~Valkealahti, and O.~Tirkkonen,
  ``Distributed capacity based channel allocation for dense local area
  deployments,'' in \emph{Proc. {IEEE} Vehicular Technology Conference
  ({VTC}'11 Fall)}, Sep. 2011, pp. 1--5.

\bibitem{CP13Amin}
P.~Amin, O.~Tirkkonen, T.~Henttonen, and E.~Pernila, ``Dynamic frequency
  selection based on carrier pricing between cells,'' in \emph{Proc. {IEEE}
  Vehicular Technology Conference ({VTC}'13 Spring)}, Jun. 2013, pp. 1--5.

\bibitem{CP12Ahmed}
F.~Ahmed, A.~A. Dowhuszko, and O.~Tirkkonen, ``Distributed algorithm for
  downlink resource allocation in multicarrier small cell networks,'' in
  \emph{Proc. {IEEE} International Conference on Communications ({ICC}'12)},
  Jun. 2012, pp. 6802--6808.

\bibitem{CP12Anchora}
L.~Anchora, L.~Badia, E.~Karipidis, and M.~Zorzi, ``Capacity gains due to
  orthogonal spectrum sharing in multi-operator {LTE} cellular networks,'' in
  \emph{Proc. International Symposium on Wireless Communication Systems
  ({ISWCS}'12)}, Aug. 2012, pp. 286--290.

\bibitem{CP07Niyato}
D.~Niyato and E.~Hossain, ``A game-theoretic approach to competitive spectrum
  sharing in cognitive radio networks,'' in \emph{Proc. {IEEE} Wireless
  Communications and Networking Conference ({WCNC}'07)}, Mar. 2007, pp. 16--20.

\bibitem{CP06Middleton}
G.~Middleton, K.~Hooli, A.~T{\"o}lli, and J.~Lilleberg, ``Inter-operator
  spectrum sharing in a broadband cellular network,'' in \emph{Proc. {IEEE}
  International Symposium on Spread Spectrum Techniques and Applications
  ({ISSSTA}'06)}, Aug. 2006, pp. 376--380.

\bibitem{CP07Bennis}
M.~Bennis and J.~Lilleberg, ``Inter base station resource sharing and improving
  the overall efficiency of {B3G} systems,'' in \emph{Proc. {IEEE} Vehicular
  Technology Conference ({VTC}'07 Fall)}, Sep. 2007, pp. 1494--1498.

\bibitem{JR09Bennis}
M.~Bennis, S.~Lasaulce, and M.~Debbah, ``Inter-operator spectrum sharing from a
  game theoretical perspective,'' \emph{Springer {EURASIP} J. Advances Signal
  Process.}, vol. 2009, no.~4, Mar. 2009.

\bibitem{CP10Lindblom}
J.~Lindblom and E.~Karipidis, ``Cooperative beamforming for the miso
  interference channel,'' in \emph{Proc. European Wireless Conference
  {({EW}'10)}}, Apr. 2010, pp. 631--638.

\bibitem{CP11Jorsweik}
E.~A. Jorswieck, L.~Badia, T.~Fahldieck, M.~Haardt, E.~Karipidis, J.~Luo,
  R.~Pisz, and C.~Scheunert, ``Resource sharing improves the network efficiency
  for network operators,'' in \emph{Proc. 27\textsuperscript{th} Meeting of the
  Wireless World Research Forum ({WWRF}'11)}, Oct. 2011.

\bibitem{JR06HuangIEEE}
J.~Huang, R.~A. Berry, and M.~L. Honig, ``Distributed interference compensation
  for wireless networks,'' \emph{{IEEE} J. Sel. Areas Commun.}, vol.~24, no.~5,
  pp. 1074--1084, May 2006.

\bibitem{JR07Etkin}
R.~Etkin, A.~Parekh, and D.~Tse, ``Spectrum sharing for unlicensed bands,''
  \emph{{IEEE} J. Sel. Areas Commun.}, vol.~25, no.~3, pp. 517--528, Apr. 2007.

\bibitem{JR09Wu}
Y.~Wu, B.~Wang, K.~Liu, and T.~Clancy, ``Repeated open spectrum sharing game
  with cheat-proof strategies,'' \emph{{IEEE} Trans. Wireless Commun.},
  vol.~20, no.~8, pp. 1922--1933, Nov. 2009.

\bibitem{CP05Ileri}
O.~Ileri, D.~Samardzija, T.~Sizer, and N.~B. Mandayam, ``Demand responsive
  pricing and competitive spectrum allocation via spectrum server,'' in
  \emph{Proc. {IEEE} International Symposium on New Frontiers in Dynamic
  Spectrum Access Networks ({DySPAN}'05)}, Nov. 2005, pp. 194--202.

\bibitem{JR08Niyato}
D.~Niyato and E.~Hossain, ``Competitive pricing for spectrum sharing in
  cognitive radio networks: Dynamic game, inefficiency of nash equilibrium, and
  collusion,'' \emph{{IEEE} J. Sel. Areas Commun.}, vol.~26, no.~1, pp.
  192--202, Jan. 2008.

\bibitem{JR09Niyato}
D.~Niyato, E.~Hossain, and Z.~Han, ``Dynamics of multiple-seller and
  multiple-buyer spectrum trading in cognitive radio networks: A game-theoretic
  modeling approach,'' \emph{{IEEE} Trans. Mobile Comput.}, vol.~8, no.~8, pp.
  1009--1022, Aug. 2009.

\bibitem{JR09Sengupta}
S.~Sengupta and M.~Chatterjee, ``An economic framework for dynamic spectrum
  access and service pricing,'' \emph{{IEEE/ACM} Trans. Netw.}, vol.~17, no.~4,
  pp. 1200--1213, Aug. 2009.

\bibitem{BK02Krishna}
V.~Krishna, \emph{Auction Theory}.\hskip 1em plus 0.5em minus 0.4em\relax
  London, UK: Academic Press, 2002.

\bibitem{JR06Sun}
J.~Sun, E.~Modiano, and L.~Zheng, ``Wireless channel allocation using an
  auction algorithm,'' \emph{{IEEE} J. Sel. Areas Commun.}, vol.~24, no.~5, pp.
  1085--1096, May 2006.

\bibitem{STD80211}
IEEE, ``{Std. 802.11 - Part 11: Wireless LAN Medium Access Control (MAC) and
  Physical Layer (PHY) Specifications},'' Aug. 1999.

\bibitem{WP11Qualcomm}
{Qualcomm}, ``A comparison of {LTE A}dvanced {HetNets} and {Wi-Fi},'' White
  Paper, Oct. 2011. [Online]. Available:
  \url{http://www.qualcomm.com/media/documents/comparison-lte-advanced-hetnets-and-wifi}

\bibitem{STD80211h}
IEEE, ``{Std. 802.11h - Part 11: Wireless LAN Medium Access Control (MAC) and
  Physical Layer (PHY) Specifications - Amendment 5: Spectrum and Transmit
  Power Management Extensions in the 5GHz band in {E}urope},'' Dec. 2003.

\bibitem{STD80216h}
------, ``{Std. 802.16h - Part 16: Air Interface for Broadband Wireless Access
  Systems - Amendment 2: Improved Coexistence Mechanisms for License-Exempt
  Operation},'' Jul. 2010.

\bibitem{STD80222}
------, ``{Std. 802.22 - Part 22: Cognitive Wireless RAN Medium Access Control
  (MAC) and Physical Layer (PHY) Specifications - Policies and Procedures for
  Operation in the TV Bands},'' Jul. 2011.

\bibitem{JR08Simeone}
O.~Simeone, I.~Stanojev, S.~Savazzi, Y.~Bar-Ness, U.~Spagnolini, and
  R.~Pickholtz, ``Spectrum leasing to cooperating secondary ad hoc networks,''
  \emph{{IEEE} J. Sel. Areas Commun.}, vol.~26, no.~1, pp. 203--213, Jan. 2008.

\bibitem{JR08Leshem}
A.~Leshem and E.~Zehavi, ``Cooperative game theory and the gaussian
  interference channel,'' \emph{{IEEE} J. Sel. Areas Commun.}, vol.~26, no.~7,
  pp. 1078--1088, Sep. 2008.

\bibitem{JR12Su}
W.~Su, J.~D. Matyjas, and S.~Batalama, ``Active cooperation between primary
  users and cognitive radio users in heterogeneous ad-hoc networks,''
  \emph{{IEEE} Trans. Signal Process.}, vol.~60, no.~4, pp. 1796--1805, Apr.
  2012.

\bibitem{JR09Suris}
J.~E. Suris, L.~A. DaSilva, Z.~Han, A.~B. MacKenzie, and R.~S. Komali,
  ``Asymptotic optimality for distributed spectrum sharing using bargaining
  solutions,'' \emph{{IEEE} Trans. Wireless Commun.}, vol.~8, no.~10, pp.
  5225--5237, Oct. 2009.

\bibitem{JR10Yang}
C.~G. Yang, J.~D. Li, and Z.~Tian, ``Optimal power control for cognitive radio
  networks under coupled interference constraints: A cooperative game-theoretic
  perspective,'' \emph{{IEEE} Trans. Veh. Technol.}, vol.~59, no.~4, pp.
  1696--1706, May 2010.

\bibitem{JR11Liu}
X.~Liu and E.~Erkip, ``A game-theoretic view of the interference channel:
  Impact of coordination and bargaining,'' \emph{{IEEE} Trans. Inf. Theory},
  vol.~57, no.~5, pp. 2805--2820, May 2011.

\bibitem{CP12Zhai}
C.~Zhai, W.~Zhang, and P.~C. Ching, ``Spectrum sharing based on two-path
  successive relaying,'' in \emph{Proc. {IEEE} International Conference on
  Acoustics, Speech and Signal Process ({ICASSP}'12)}, Mar. 2012, pp.
  2909--2912.

\bibitem{JR04Laneman}
J.~N. Laneman, D.~N.~C. Tse, and G.~W. Wornell, ``Cooperative diversity in
  wireless networks: Efficient protocols and outage behavior,'' \emph{{IEEE}
  Trans. Inf. Theory}, vol.~50, no.~12, pp. 3062--3080, Dec. 2004.

\bibitem{JR11Saad}
W.~Saad, Z.~Han, T.~Basar, M.~Debbah, and A.~Hjorungnes, ``Network formation
  games among relay stations in next generation wireless networks,''
  \emph{{IEEE} Trans. Commun.}, vol.~59, no.~9, pp. 2528--2542, Sep. 2011.

\bibitem{JR08Huang}
J.~Huang, Z.~Han, M.~Chiang, and H.~V. Poor, ``Auction-based resource
  allocation for cooperative communications,'' \emph{{IEEE} J. Sel. Areas
  Commun.}, vol.~26, no.~7, pp. 1226--1237, Sep. 2008.

\bibitem{JR99Myerson}
R.~B. Myerson, ``Nash equilibrium and the history of economic theory,''
  \emph{{AEA} J. Econ. Literature}, vol.~37, pp. 1067--1082, 1999.

\bibitem{CP07Gandhi}
S.~Gandhi, C.~Buragohain, L.~Cao, H.~Zheng, and S.~Suri, ``A new pricing model
  for next generation spectrum access,'' in \emph{Proc. {IEEE} International
  Symposium on New Frontiers in Dynamic Spectrum Access Networks
  ({DySPAN}'07)}, Apr. 2007, pp. 22--33.

\bibitem{CP06Ryan}
K.~Ryan, E.~Aravantinos, and M.~M. Buddhikot, ``A general framework for
  wireless spectrum auctions,'' in \emph{Proc. International Workshop on
  Technology and Policy for Accessing Spectrum ({TAPAS}'06)}, Aug. 2006.

\bibitem{CP04Huang}
J.~Huang, R.~A. Berry, and M.~L. Honig, ``Auction mechanisms for distributed
  spectrum sharing,'' in \emph{Proc. Allerton Conference'04}, Sep. 2004.

\bibitem{JR05Juliusson}
E.~\a'{A}. Juliusson, N.~Karlsson, and T.~Gr\"{a}ling, ``Weighing the past and
  the future in decision making,'' \emph{European J. Cognitive Psychology},
  vol.~17, no.~4, pp. 561--575, 2005.

\bibitem{JR08Stanovich}
K.~E. Stanovich and R.~F. West, ``On the relative independence of thinking
  biases and cognitive ability,'' \emph{{APA} J. Personality, Social
  Psychology}, vol.~94, no.~4, pp. 672--695, Apr. 2008.

\bibitem{BK94Snowdon}
B.~Snowdon, H.~Vane, and P.~Wynarczyk, \emph{A modern guide to
  macroeconomics}.\hskip 1em plus 0.5em minus 0.4em\relax Aldershot, UK: Edward
  Elgar Publishing Limited, 1994.
	
	\end{thebibliography}
\end{document}